\documentclass[pra,aps,superscriptaddress,notitlepage,longbibliography,twocolumn,nofootinbib,superscriptaddress]{revtex4-2}

\usepackage{amsmath}
\usepackage{amsfonts}
\usepackage{graphicx}
\usepackage{bm}
\usepackage{bbold}
\usepackage{color}
\usepackage{braket}
\usepackage[normalem]{ulem}
\newcommand{\stkout}[1]{\ifmmode\text{\sout{\ensuremath{#1}}}\else\sout{#1}\fi}
\usepackage[dvipsnames]{xcolor}

\newcommand{\blue}{}

\usepackage[colorlinks]{hyperref}
\long\def\comment#1{}
\usepackage{soul}

\newcommand{\ahatbt}{\skew{6}{\hat}{\bm{\mathcal{A}}}} %truncated A with boldface and properly aligned hat
\hypersetup{
    colorlinks=true,
    linkcolor=blue,
    filecolor=magenta,      
    urlcolor=magenta,
    citecolor={blue},
    }

\DeclareMathAlphabet{\mathcallc}{U}{dutchcal}{m}{n}
\SetMathAlphabet{\mathcallc}{bold}{U}{dutchcal}{b}{n}
\DeclareMathAlphabet{\mathbcallc}{U}{dutchcal}{b}{n}

\begin{document}

% change equation skips for align
\abovedisplayskip=6pt
\abovedisplayshortskip=6pt
\belowdisplayskip=6pt
\belowdisplayshortskip=6pt

\title{Gauge-invariant theory of truncated quantum light-matter interactions in arbitrary media}
\author{Chris Gustin}
\email{cgustin@stanford.edu}
\affiliation{Edward L.\ Ginzton Laboratory, Stanford University, Stanford, California 94305, USA}
\author{Sebastian Franke}
\affiliation{\hspace{0pt}Department of Physics, Engineering Physics, and Astronomy, Queen's University, Kingston, Ontario K7L 3N6, Canada\hspace{0pt}}
\affiliation{Technische Universit\"at Berlin, Institut f\"ur Theoretische Physik,
Nichtlineare Optik und Quantenelektronik, Hardenbergstra{\ss}e 36, 10623 Berlin, Germany}
\author{Stephen Hughes}
\affiliation{\hspace{0pt}Department of Physics, Engineering Physics, and Astronomy, Queen's University, Kingston, Ontario K7L 3N6, Canada\hspace{0pt}}

\begin{abstract}
The loss of gauge invariance in models of light-matter interaction which arises from material and photonic space truncation can pose significant challenges to conventional quantum optical models when matter and light strongly hybridize. 
In structured photonic environments, necessary in practice to achieve strong light-matter coupling, a rigorous model of field quantization within the medium is also needed.
Here, we use the framework of macroscopic QED by quantizing the fields in an arbitrary material system, with a 
spatially-dependent dispersive and absorptive dielectric, starting from a fundamental light-matter action. We truncate the material and mode degrees of freedom  while respecting the gauge principle by imposing a partial gauge fixing constraint during canonical quantization, which admits a large number of gauges including the Coulomb and multipolar gauges commonly
used in quantum optics. We also consider gauge conditions with explicit time-dependence, enabling us to unambiguously introduce additional phenomenologically time-dependent light-matter interactions in any gauge. Our results allow one to derive rigorous non-relativistic models of ultrastrong light-matter interactions in structured photonic environments with no gauge ambiguity. 
Results for two-level systems and the dipole approximation are discussed, as well as how to go beyond the dipole approximation for effective single-particle models. By comparing with the limiting case of an inhomogeneous dielectric, where dispersion and absorption can be neglected and the fields can be expanded in terms of the generalized transverse eigenfunctions of the dielectric, we show how lossy systems can introduce an additional gauge ambiguity, which we resolve and predict to have fundamental implications for open quantum system models. Finally, we show how observables in mode-truncated systems can be calculated without ambiguity by using a simple gauge-invariant model of photodetection.

\end{abstract}

\maketitle
\section{introduction}

In nanophotonics, one often would like to describe the interaction of a small number of emitters,
 treated as microscopic degrees of freedom, interfacing via the electromagnetic field with a macroscopic medium, wherein the different degrees of freedom are not tracked explicitly. In classical electromagnetism, this is accomplished by the macroscopic Maxwell's equations, where the medium is, assuming a linear response of the medium to applied fields, ascribed a dielectric function, which in general can be frequency-dependent (allowing for dispersion) and complex (allowing for energy losses via absorption). In fact, the requirement of the constituent medium response to the applied field to follow a causal relationship implies, via the Kramers-Kronig relations, that in general a frequency-dependent dielectric function is complex, and vice-versa.

In quantum mechanics, the direct quantization of the macroscopic Maxwell's equations is complicated by the fact that under an imaginary permittivity, the operators describing the electromagnetic field would in general decay to zero amplitude, in violation of the fundamental commutation relations~\cite{grunwel}.
One particularly successful method to quantize the electromagnetic field in a dispersive and absorbing medium is 
the macroscopic quantum electrodynamics (QED) approach~\cite{PhysRevA.11.230,Huttner,grunwel,Dung,scheel1998qed,suttorp2004field,Philbin2010Dec,PhysRevA.12.1475}, where the field is expanded in terms of the photonic Green's function of the medium (obtained from the impulse response of the electric field to a localized dipole source) and a bosonic polariton field. As the Green's function can be obtained by purely classical calculations---including analytic solutions for simple geometries, and general numerical techniques (e.g., finite-difference time-domain simulations) for general cases---this powerful approach provides a quantum mechanical framework for studying the dynamics of light-matter systems in practical nanophotonic settings. The Green's function quantization can be regarded as a generalization of the usual normal mode expansion from lossless systems. Moreover, the Green's functions can also be obtained
through mode expansion techniques, even for complex geometries and in the presence of photon loss~\cite{NormKristHughes,Kristensen:20}.

The macroscopic QED formalism can be justified on purely phenomenological grounds, by virtue of its simultaneous fulfillment of the macroscopic Maxwell's equations and Lorentz equations of motion, the fundamental quantization commutation relations of the electromagnetic field, and the dissipation-fluctuation theorem~\cite{Dung}. Microscopic derivations can also be performed, wherein a material medium ``reservoir'' field is coupled via the fundamental minimal coupling prescription of QED to the vacuum electromagnetic fields, and a process of ``Fano diagonalization'' is used to diagonalize the total Hamiltonian (medium plus electromagnetic field) in terms of bosonic polariton operators which can then couple to microscopic material particles~\cite{Huttner,Suttorp2004Sep}. In this latter microsopic derivation, the coupling can be written as a quantized Hamiltonian interaction, or more generally quantization can be performed at the level of an initial Lagrangian~\cite{Philbin2010Dec} by identifying canonical variables and quantizing in accordance with Dirac's prescription for quantization with constraints~\cite{Weinberg1995Jun}. These  results are also in accordance with the phenomenological quantization approach.

Macroscopic QED, already having proved a powerful theoretical tool for modelling light-matter interactions in a medium (e.g.,
spontaneous emission in arbitrary environments~\cite{PhysRevA.62.053804,PhysRevA.60.2534,PhysRevB.80.195106}, dipole-dipole interactions~\cite{Dung2002,Kristensen2011}, Casimir-Polder forces~\cite{Buhmann2004Nov}, and surface-enhanced Raman spectroscopy~\cite{KamandarDezfouli2017May}), is an excellent candidate for improving models and providing  fundamental analysis of light-matter interactions in the so-called ultrastrong coupling
(USC)
regime~\cite{frisk_kockum_ultrastrong_2019,RevModPhys.91.025005}. In the USC regime, the parameters which characterize the coupling between the material and field degrees of freedom become substantial compared to the bare resonances of the subsystems,
which strongly hybridizes the field and matter degrees of freedom, and common frameworks for understanding the dynamics of the interaction break down. These changes can be dynamical, as a consequence of having to forgo the widely used rotating-wave approximation, but 
%more 
recently it has come to be more widely understood that the fundamental Hamiltonian used to describe the coupling between the subsystems itself can become questionable in any situation where the field or material degrees of freedom are expressed in a {\it truncated basis}. In particular, the gauge invariance of the theory is broken when the minimal coupling Hamiltonian is expressed in terms of variables which are truncated to a finite energy basis for the material degrees of freedom, or the number of modes (and presumably Fock number states) for the field~\cite{DeBernardis2018Apr,DeBernardis2018Nov,Taylor2022Mar}. Notably, the loss of gauge invariance remains highly significant in interaction regimes where the truncation process retains all energy levels near-resonant with the ultrastrong interaction, where a truncated model should, in principle, be accurate and independent of choice of gauge.

The breakdown of gauge-invariance in material systems \cite{Starace1971Apr,Keeling2007Jun,DeBernardis2018Nov,DiStefano2019Aug,Stokes2019Jan,Taylor2020Sep} can be understood by noting that a truncation in an energy basis (e.g., to the widely-used two-level system (TLS) model) implies a truncation in a position basis (which is continuous without truncation). As a U(1) gauge theory, QED promotes the global symmetry of the Schr{\"o}dinger equation's invariance under a total change in the phase of the state vector to a \emph{local} symmetry.
The presence of this new symmetry induces a minimal coupling to a gauge boson field. In the case of nonrelativistic QED, a transformation of a wavefunction of a particle with charge $q$, through
\begin{equation}\label{eq:eq1}
    \psi(\mathbf{x}) \rightarrow \exp{\left[iq\Lambda(\mathbf{x})/\hbar\right]}\psi(\mathbf{x}),
\end{equation}
can be compensated by a corresponding gauge transformation of the potential fields,  $\mathbf{A} \rightarrow \mathbf{A} + {\bm \nabla} \Lambda$ and $\phi \rightarrow \phi - \dot{\Lambda}$.
All physical results must be invariant under local U(1) transformations.

Critically, if the model is to be implemented in a truncated basis, the truncation must be carried out in a way which is consistent with the gauge transformation in Eq.~\eqref{eq:eq1}; that is, a gauge transformation of the field must be able to be compensated with an appropriate dimensional unitary transformation of the state vector. In a different context, these insights form the basis of lattice gauge theory, introduced by Wilson~\cite{PhysRevD.10.2445} to study quantum chromodynamics on a lattice, where the continuous representation of position is truncated to a finite basis in a manner which respects the gauge symmetry of the theory. \blue{Recent work has shown that gauge invariance can be restored in truncated material systems~\cite{DiStefano2019Aug,Savasta2021May,Taylor2020Sep}, by using a generalized minimal coupling replacement in the form of a unitary transformation, which correctly constrains the light-matter interaction within the truncated subspace.}

Equivalent insights can be used to show that truncation of photonic degrees of freedom (e.g., the number of photonic modes) also yields gauge-dependent predictions~\cite{DeBernardis2018Apr,Taylor2022Mar}\blue{, and that gauge invariance can be restored by a similar unitary transformation.} \blue{It has also} been noted in different contexts, such as high-order harmonic generation~\cite{Han2010Jun}, and tight-binding models~\cite{Li2020May}, that truncation in the Coulomb gauge (and its analogues) allows for converging results with less modes, while the multipolar gauge (and its analogues) allow for less material states. In the context of media with loss, discrete mode expansions of the electromagnetic fields can take the form of,  e.g., quasi-modes of various types~\cite{Dalton1999Jul,Fussell2008May,Vendromin2022Oct}, or quasinormal modes (QNMs)~\cite{2ndquant2,Lalanne_review,Kristensen:20}---where for the latter,  a fully quantized theory has been developed recently~\cite{frankequantization} and applied to plasmonic single-photon sources~\cite{Hughes_SPS_2019}, and coupled resonators~\cite{franke2020quantized}, including gain-loss systems~\cite{PhysRevA.105.023702,PhysRevX.11.041020}, and appears to be an excellent candidate for modelling ultrastrong cavity-QED interactions in realistic photonic media.

In this work, we take the view that any rigorous picture of open system quantum optics that involves a few discrete modes~\cite{Lentrodt2020Jan} manifests in some degree of mode truncation, and thus understanding gauge invariance in these models is essential.

In the quantization of systems with matter (that is, \blue{degrees of freedom} beyond the ``passive'' material medium considered in the macroscopic QED approach) an additional gauge symmetry beyond that of the field potentials  arises due to the polarization field of the matter component. In this manner, for light-matter interactions it is therefore necessary to formulate a theoretical framework which allows for gauge transformations consistent with this greater class of symmetry under material truncation.  Previous work has shown these methods to maintain gauge invariance on the basis of semiclassical arguments~\cite{Starace1971Apr,DiStefano2019Aug,Savasta2021May}, which is generally sufficient, as after quantization, the gauge is invariably at least partially fixed by the requirement of constraints to quantize the electromagnetic field (and the same approach can be justified without reference to gauge transformations by instead appealing to the notion of constraining interactions within the correctly-truncated subspace~\cite{Taylor2020Sep}). However, this approach does not shed insight into what exactly the realizable gauges are and what a gauge transformation consists of in the truncated space. The usual approach is to quantize in the Coulomb gauge and construct the multipolar gauge by means of the unitary Power-Zienau-Woolley (PZW) transformation~\cite{Babiker1983Feb,Woolley2020Feb}; however, 
because a unitary transformation from a fixed gauge cannot implement a \emph{generic} gauge transformation~\cite{Woolley1999Jan}, this has raised questions~\cite{Rousseau2017Sep} (and the resolutions to these questions~\cite{Vukics2021Aug,Andrews2018Jan,Stokes2021Sep}) about the validity of such a procedure recently. To be consistent with macroscopic QED, any approach must start from a Lagrangian which respects this gauge symmetry of both the electromagnetic and material polarization fields, and contains the material reservoir fields which describe the medium.

In this paper, we accomplish this task by quantizing the electromagnetic fields in the presence of both a material medium reservoir field (the ``passive'' component), and free charged particles (the ``active'' component), the latter of which can interact with the electromagnetic fields with arbitrary strength, allowing one to study USC effects. Using a $c$-number quantization function method by Woolley~\cite{Woolley1999Jan}, {\it we quantize in a way which does not require choosing a specific gauge, allowing us to derive manifestly gauge-invariant models} which incorporate a broad class of gauges, including the most commonly used gauges in quantum optics: the Coulomb and multipolar (or dipole, when using a dipole approximation) gauges. We show explicitly the validity of previous theoretical works on restoring gauge invariance, and shed light on the non-relativistic limit of their application. 

Our results provide a rigorous and gauge-invariant framework for describing light-matter interactions in the USC regime from a first-principles approach,
for arbitrary media, and one that
%one that 
can still take advantage of a reduced description of the medium in terms of a linear susceptibility function. We stress that from a theoretical perspective,  our results need not be implemented in the context of a macroscopically quantized medium, and indeed the formalism also applies for free space quantization, but the presence of a medium (generally one that supports resonant modes) is necessary to reach the USC regime in practice in  optical systems.

Our work also lays the necessary groundwork for the future development of first-principles models of loss from cavity-QED systems in the USC regime. This is timely and highly desired,  as it as been shown recently that the nearly universally-used phenomenological model of dissipation (standard input-output theory~\cite{Gardiner1985Jun}) is insufficient in the USC regime~\cite{Salmon2022Mar}. We expect our work to be useful and applicable to, in addition to quantized QNMs, studies based on, for example, pseudomodes~\cite{Medina2021Mar,Sanchez-Barquilla2022Aug}, or simulations involving matrix product states~\cite{delPino2018Nov,Regidor2021Apr}.

In addition to laying out the fundamental theory of gauge-invariant interactions in quantum light-matter systems in a quantized and arbitrary medium, our work also contributes three additional main findings:

\blue{(i) Any open quantum systems approach to photon loss (e.g., a master equation) in a system interacting ultra-strongly with matter, from a rigorous theoretical perspective, should be derived in the Coulomb gauge, as it is the unique gauge in which the reservoir can be described by a subspace \emph{unentangled} with the light-matter system.
However, a reduced notion of a gauge transformation can be defined only with respect to a truncated field (e.g., a cavity mode in cavity-QED models), which has allowed for previous developments of gauge-invariant models~\cite{Salmon2022Mar,Mercurio2022Apr} (in these cases, assuming phenomenological models of system-reservoir coupling). This necessarily requires a mode-truncated description of the reduced gauge transformation, and thus we propose that there exists a potential intrinsic gauge ambiguity due to mode truncation in rigorous open quantum system models of photon loss, for which the techniques described in this work to retain gauge invariance are important.}

(ii) Contrasting previous claims~\cite{Stokes2021Feb}, we show that it is possible to introduce unambiguous phenomenological time-dependent light-matter interactions in \emph{any gauge}, provided the time-dependence of the gauge condition is consistently accounted for in the quantization.

(iii) By considering an explicit model of photodetection in a truncated mode system, we resolve a gauge ambiguity regarding observables, and provide justification for the recent approach that has been used in previous works, for simple model systems of cavity-QED~\cite{PhysRevResearch.3.023079,Savasta2021Jan,Salmon2022Mar,Mercurio2022Apr}. \blue{By identifying the \emph{correctly mode-truncated electric field operator}, we also refute recent claims~\cite{Stokes2022Nov} that the modal expansion operators of the transverse electric field are not the correct operators to couple to external reservoir modes in the case of open quantum systems.}

The rest of our paper is organized as follows.
In Sec.~\ref{sec:action}, we present the fundamental {\it action} from which we derive Maxwell's equations and the Lorentz force law in a dispersive and absorbing dielectric, where we treat the medium degrees of freedom explicitly as a frequency and spatially dependent reservoir with an oscillator field.

In Sec.~\ref{sec:quantization}, we show how this general system can be quantized using Dirac's method of canonical quantization with constraints, using Woolley's ~\cite{Woolley1999Jan} quantization function approach. Following previous works ~\cite{Suttorp2004Sep,Philbin2010Dec}, we then perform a Fano diagonalization to express part of the quantum Hamiltonian for this (arbitrary-gauge) system as a bosonic polariton harmonic oscillator field, which removes any explicit reference to the medium oscillator degrees of freedom, and express the electromagnetic fields in terms of the photonic Green's function of the medium and the polariton operators. 

\begin{figure}[ht]
    \centering
    \includegraphics[width=0.94\columnwidth]{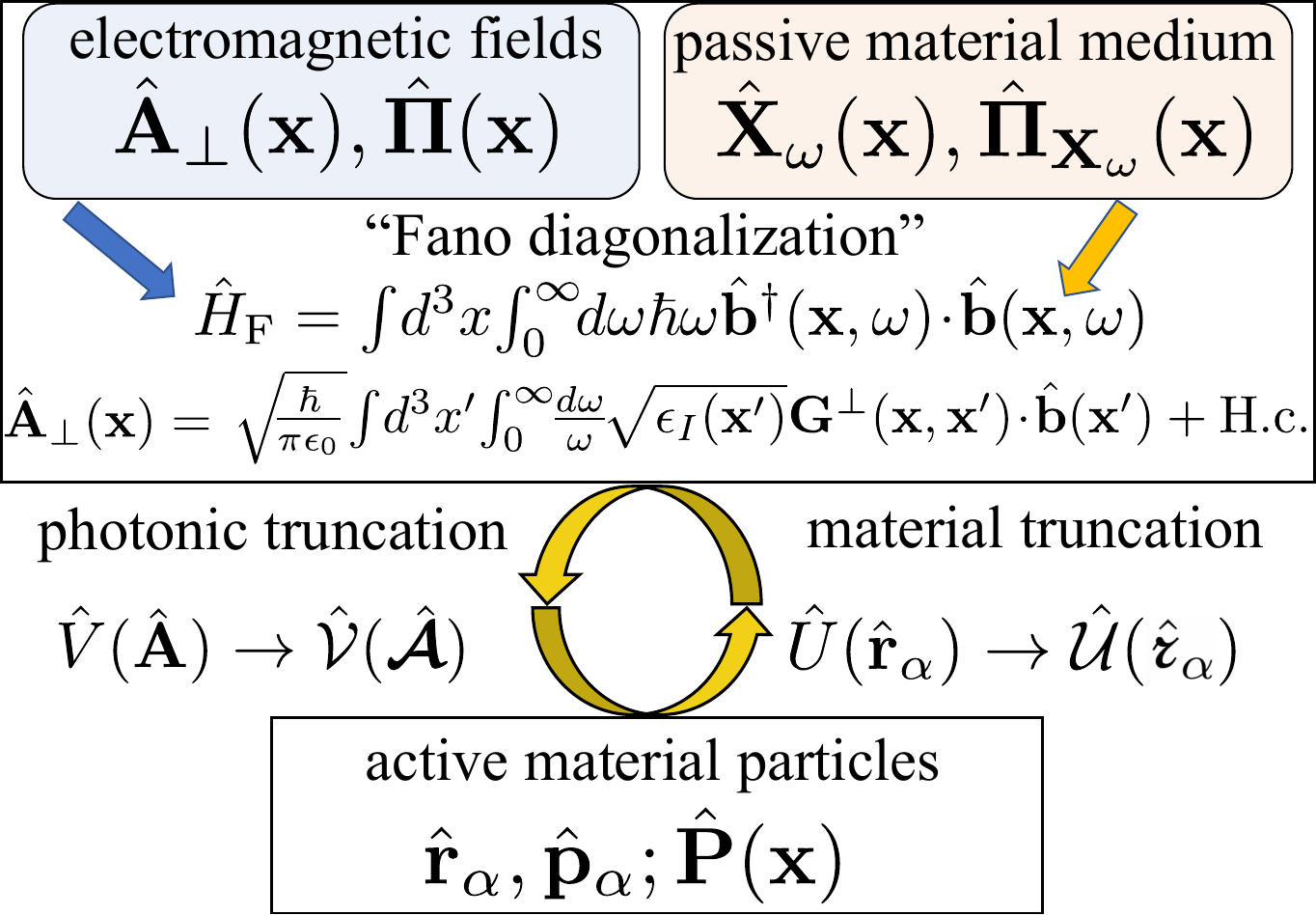}
    \caption{Schematic representation of the gauge-invariant approach to light-matter interactions within a linear medium (see text for definition of mathematical symbols). The electromagnetic fields interact weakly with a passive material reservoir representing the dispersive and absorbing dielectric medium, allowing for the fields to be expressed as a linear functional of fundamental polariton operators $\hat{\mathbf{b}}(\mathbf{x},\omega)$, $\hat{\mathbf{b}}^{\dagger}(\mathbf{x},\omega)$ (we have suppressed the $\omega$ functional dependence in the dielectric function, Green's function, and annihilation operator in the schematic for brevity). Light-matter interactions are introduced by a minimal coupling scheme of free ``active'' material particles with the (vacuum) electromagnetic fields. Gauge invariance under active material truncation is retained by expressing the Hamiltonian in terms of the unitary operators $\hat{\mathcal{U}}(\hat{\mathbcallc{r}}_{\alpha})$, where $\hat{\mathbcallc{r}}_{\alpha}$ are the directly truncated position operators of the active material component. Similarly, gauge invariance is retained under photonic truncation by expressing the Hamiltonian in terms of $\hat{\mathcal{V}}(\ahatbt)$, where $\ahatbt$ is the truncated vector potential. The approach is manifestly gauge invariant as the longitudinal component of the vector potential $\ahatbt_{\parallel}$ and transverse component of the polarization $\hat{\bm{\mathcal{P}}}_{\perp}$  are determined by the arbitrary transverse quantization function $\mathbf{K}_{\perp}(\mathbf{x},\mathbf{x'})$. 
    }\label{fig:1}
\end{figure}

In Sec.~\ref{sec:GI}, we show how gauge invariance manifests in the quantum theory with the quantization function approach, and show how material and mode truncation can be introduced in a manner that respects the gauge principle. We contrast the form of  discrete mode expansion that can be done in a lossy system with the more commonly employed normal mode expansion (namely, using lossless eigenmodes), which is generally only strictly possible in lossless and dispersionless media. Figure~\ref{fig:1} shows a schematic conceptual representation of our approach. 

In Sec.~\ref{sec:time-dep}, we generalize the quantization function approach to quantizing in an arbitrary gauge by allowing the gauge condition to have explicit time-dependence. This permits for a broader set of gauge transformations to be considered than have previously appeared in the literature. We then use this formalism to show how phenomenological time-dependent interactions in ultrastrong transverse light-matter interactions can be introduced \emph{unambiguously} in any gauge.

In Sec.~\ref{sec:dtls}, we apply our results to some common approximations in quantum optics, specifically the dipole and material TLS approximations, and show how to go beyond the dipole approximation for the case of an effective single-particle model.

In Sec.~\ref{sec:res}, we identify and resolve a potential gauge ambiguity regarding observables of the electromagnetic field, by introducing an explicit simple model of photodetection in a truncated mode system. Using our formalism, we show that for mode truncation to be correctly performed, it must be done with respect to the \emph{vector potential}, and that the correctly mode-truncated form of the electric field operator subsequently takes a modified form. 

Finally, in Sec.~\ref{sec:conc}, we conclude. 
In addition, we also include \blue{four} appendices. \blue{Appendices~\ref{app:action} and~\ref{app:dirac} give extra details on the fundamental light-matter interaction action and canonical quantization procedure in the presence of constraints.} In Appendix~\ref{sec:appA}, we show in more detail how a discrete mode expansion can be generally constructed from the continuous polariton operators, and connect to the important case of quantized QNMs. We consider, in Appendix~\ref{sec:appB}, the case of an inhomogeneous but nondispersive and lossless dielectric, with a real dielectric permittivity that is independent of frequency $\epsilon(\mathbf{x})$. In this important special case (justifiable in a limited frequency regime), the field variables can be expressed in terms of the so-called generalized transverse eigenfunctions of the system, which are normal modes of Maxwell's equations in the dielectric medium with closed or periodic boundary conditions, by choosing the generalized Coulomb gauge, which satisfies ${\bm \nabla} \cdot \left[\epsilon(\mathbf{x})\mathbf{A}(\mathbf{x})\right] =0$, or the generalized multipolar gauge. We quantize the electromagnetic field in this medium using Dirac's constrained quantization technique for the generalized Coulomb gauge condition, which allows us to recover previously known results (e.g., Refs.~\cite{motu3,Dalton1997Jul,Wubs2003Jul}), now using   a systematic method. We then discuss how gauge-invariant truncated models can be obtained, similarly to the procedure in the main text. 

\section{Electromagnetic fields, arbitrary dielectric passive medium, and active material particles}
\label{sec:action}

\blue{We take our model to consist of the electromagnetic field, a passive medium reservoir field $\mathbf{X}_{\omega}(\mathbf{x},t)$ (corresponding to ``bound'' charges), as well as active free particles, indexed by $\alpha$, with charge $q_{\alpha}$, mass $m_{\alpha}$, and position coordinate $\mathbf{r}_{\alpha}(t)$. For simplicity we assume no net charge,  such that $\sum_{\alpha}q_{\alpha}=0$. We would like to ultimately model the interaction of the free charges with the electromagnetic field using a {\it macroscopic} approach, where the field can be expressed in a quantized form, using the photonic Green's function of the medium, which is determined by its dielectric function $\mathbf{\epsilon}(\mathbf{x},\omega)$. In general, the dielectric constant is a complex-valued function that
depends on space and frequency with real and imaginary parts $\epsilon_{R}(\mathbf{x},\omega)$ and $\epsilon_I(\mathbf{x},\omega)$, respectively. It may also be anisotropic, but we will consider this function as
a scalar.

We consider the action $\mathcal{S} = \mathcal{S}[A_\mu;  \mathbf{r}_\alpha; \mathbf{X}_{\omega}]$.
This action is a functional of the electromagnetic potential fields $A^{\mu}(x) = (\phi(\mathbf{x},t)/c,\mathbf{A}(\mathbf{x},t))$, the medium field excitations $\mathbf{X}_{\omega}(\mathbf{x},t)$, and the material particle coordinates $\mathbf{r}_{\alpha}(t)$, as well as the time derivatives of each of these respective quantities. The medium is described by a dielectric function with real and imaginary parts $\epsilon_R(\mathbf{x},\omega)$ and $\epsilon_I(\mathbf{x},\omega)$, describing dispersion and absorption of the medium, respectively, and which satisfy the Kramers-Kronig relations. We assume non-magnetic media, such that $\mu(\mathbf{x},\omega) \approx 1$, although the theory can be generalized to also incorporate a general complex magnetic susceptibility~\cite{Philbin2010Dec}. Note that here and throughout, $\omega$ refers to a continuous modal index, and not the argument of a Fourier transform of time. 

The full action is given in Appendix~\ref{app:action}, as well as the equations of motion it generates. These are the Amp{\`e}re and Lorentz equations~\eqref{eq:eom2} and~\eqref{eq:eom3}, as well as an equation of motion for the passive medium reservoir oscillator field~\eqref{eq:eom4}, and Gauss's law~\eqref{eq:eom1}, which acts as a constraint.

}

%Clearly, Eqs.~\eqref{eq:eom1} and~\eqref{eq:eom2} are the Gauss and Amp{\`e}re-Maxwell laws with active material sources, as well as a polarization field $\mathbf{P}_{\rm M} = \int d\omega \alpha(\mathbf{x},\omega) \mathbf{X}_{\omega}$ arising from the passive medium reservoir which also acts as a source---the ``passivity'' of the medium will ultimately be demonstrated by the ``Fano diagonalization'' of Sec.~\ref{sec:Fano_diagnonalization}, which allows the fields to be expressed in terms of bosonic polariton operators and the photonic Green's function without explicit reference to the medium interaction. Equation~\eqref{eq:eom3} is the Lorentz force equation of motion for the active free particles, and Eq.~\eqref{eq:eom4} is the equation of motion for the passive medium reservoir field. 

\section{Quantization}
\label{sec:quantization}
In order to perform canonical quantization and promote field coordinates to operators on a Hilbert space, we first move to a Hamiltonian picture. To do so, we  choose the Lagrangian $\mathcal{L}$, which satisfies $\mathcal{S} = \int \! dt \mathcal{L}$, to take the form of the time integrands of the action in Eqs.~\eqref{eq:A1}-\eqref{eq:A4}. A total time derivative can be added to the Lagrangian without changing the resulting equations of motion, which is related to gauge symmetry of the theory, discussed in Sec.~\ref{sec:gauge}.
We then identify the conjugate momenta to the field and particle coordinates:
\begin{subequations}
\begin{equation}\label{eq:p1}
    \Pi_{\phi} = \frac{\delta \mathcal{L}}{\delta \dot{\phi}} = 0,
\end{equation}
\begin{equation}\label{eq:p2}
    \mathbf{\Pi}_{\mathbf{A}} = \frac{\delta \mathcal{L}}{\delta \dot{\mathbf{A}}}=-\epsilon_0 \mathbf{E} - \int_0^{\infty}d\omega \alpha(\mathbf{x},\omega)\mathbf{X}_{\omega},
\end{equation}
\begin{equation}\label{eq:p3}
    \mathbf{p}_{\alpha} = \frac{\partial \mathcal{L}}{\partial \mathbf{r}_{\alpha}} = m_{\alpha}\dot{\mathbf{r}}_{\alpha} + q_{\alpha}\mathbf{A}(\mathbf{r}_{\alpha}),
\end{equation}
\begin{equation}\label{eq:p4}
    \mathbf{\Pi}_{\mathbf{X}_{\omega}} =\frac{\delta \mathcal{L}}{\delta \dot{\mathbf{X}}_{\omega}}= \dot{\mathbf{X}}_{\omega},
\end{equation}
\end{subequations}
\blue{and $\alpha(\mathbf{x},\omega)$ is defined in Eq.~\eqref{eq:alpha}.}

The vanishing of $\Pi_{\phi}$ is indicative of the fact that the equations of motion have redundant degrees of freedom; while the four-potential contains four degrees of freedom, only three dynamical equations of motion are given by Eq.~\eqref{eq:eom2}. In the process of canonical quantization, the usual approach of promoting Poisson brackets to commutators of operators requires a one-to-one correspondence for dynamical equations of motion and unconstrained degrees of freedom. Instead, here we have a constrained system, where the longitudinal component of $\mathbf{\Pi}_{\mathbf{A}}$ is constrained by Gauss's law (Eq.~\eqref{eq:eom1}), and the canonical momentum $\Pi_{\phi}$ is constrained to vanish. These constraints restrict the phase space manifold in which the system is to be quantized, and describe a so-called ``singular'' system, where the Hessian of the Lagrangian does not have full rank~\cite{Weinberg1995Jun,Cohen-Tannoudji1997Mar,sundermeyer1982constrained}. 

In addition to the two constraints described above, which we can write as
\begin{subequations}
\begin{equation}\label{eq:c0}
    \chi_0 = \Pi_{\phi} =0, 
\end{equation}
\begin{equation}\label{eq:c1}
   \chi_1 = {\bm \nabla} \cdot \mathbf{\Pi}_{\mathbf{A}} + \rho_{\rm A} = 0,
\end{equation}
\end{subequations}
\blue{where $\rho_{\rm A}$ is the active particle charge density given by Eq.~\eqref{eq:charged},}
a third constraint on the four-potential is required to ensure that one dynamical equation of motion exists for each of the unconstrained quantization variables. The reason for this is that typically in constrained systems, one can quantize by using Dirac's prescription, which modifies the Poisson brackets as to include the effect of the constraints in a manner that implies there exist unconstrained variables which can be quantized in accordance with the usual Poisson bracket prescription~\cite{Weinberg1995Jun}. The difficulty here is that we find that the Poisson brackets of the above constraints vanishes, which precludes one from directly applying this procedure. This is ultimately because Eq.~\eqref{eq:c1} cannot be solved to uniquely eliminate one of the field degrees of freedom (e.g., $\phi$) by expressing it in terms of the others, which is a consequence of the gauge freedom of the theory (that is, Eq.~\eqref{eq:c1} only allows one to solve for the longitudinal part of $\mathbf{\Pi}_{\mathbf{A}}$)~\cite{Weinberg1995Jun}. To remedy this, we will apply another constraint, which we will choose to manifestly preserve the gauge symmetry of the Lagrangian, allowing us to quantize in a (mostly) \emph{arbitrary gauge}.

\subsection{Gauge Symmetry}\label{sec:gauge}

Note that under a general gauge transformation:
\begin{subequations}
\begin{equation}\label{eq:gauge1}
    \mathbf{A}(\mathbf{x},t) \rightarrow \mathbf{A}(\mathbf{x},t) + {\bm \nabla}\Lambda(\mathbf{x},t),
\end{equation}
\begin{equation}\label{eq:gauge2}
    \phi(\mathbf{x},t) \rightarrow \phi(\mathbf{x},t) - \dot{\Lambda}(\mathbf{x},t),
\end{equation}
\end{subequations}
the Lagrangian is not invariant, while the equations of motion~\eqref{eq:eom} are. However, if we add the additional term
\begin{equation}\label{eq:ls}
    \mathcal{L} \rightarrow \mathcal{L} - \frac{d}{dt} \int d^3x \mathbf{A}(\mathbf{x},t) \cdot \mathbf{P}(\mathbf{x},t),
\end{equation}
where the auxiliary polarization field $\mathbf{P}$ satisfies ${\bm \nabla} \cdot \mathbf{P}(\mathbf{x},t) = -\rho_{\rm A}(\mathbf{x},t)$ (which fixes its longitudinal part, but leaves the transverse part arbitrary), then the Lagrangian becomes invariant. Note here we have used the relation
$\nabla \cdot \mathbf{J}_{\rm A} + \dot{\rho}_{\rm A} = 0$\blue{, where $\mathbf{J}_{\rm A}$ is the active particle current density given by Eq.~\eqref{eq:currentd}.}
Also note that the action only involves the vector potential in the free particle part, whereas the medium is coupled to the (manifestly gauge invariant) electric field. More fundamentally, this is a consequence of the fact the medium part of the action is chosen to give the macroscopic Maxwell's equations, in the form of a dielectric function.

Since we know how to quantize in the Coulomb gauge (where the vector potential is transverse), it would be useful to have a representation of the vector potential in terms of its transverse part $\mathbf{A}_{\perp}$, which is gauge-invariant. To do this, we use a method devised by Woolley, in which the longitudinal component of the vector potential is determined by a c-number function projected onto the transverse vector potential~\cite{Woolley1999Jan,Stokes2021Sep}. We call this method the \emph{quantization function} approach. 

Note that we can write, following Helmholtz's theorem, the vector potential as $\mathbf{A} = \mathbf{A}_{\perp} + \mathbf{A}_{\parallel}$, where
\begin{equation}\label{eq:apar}
   \mathbf{A}_{\parallel}(\mathbf{x},t) = {\bm \nabla^{\mathbf{x}}} \int d^3 x'  \mathbf{K}_{\parallel}(\mathbf{x},\mathbf{x'}) \cdot  \mathbf{A}(\mathbf{x'},t) ,
\end{equation}
and $\mathbf{K}_{\parallel}(\mathbf{x'},\mathbf{x})$ is the Green's function for the divergence operator
\begin{equation}
    \mathbf{K}_{\parallel}(\mathbf{x},\mathbf{x'}) = -{\bm \nabla}^{\mathbf{x}} \frac{1}{4\pi|\mathbf{x}-\mathbf{x'}|},
\end{equation}
which satisfies ${\bm \nabla}^{\mathbf{x}} \cdot \mathbf{K}_{\blue{\parallel}}(\mathbf{x},\mathbf{x'}) = \delta(\mathbf{x}-\mathbf{x'})$, and we have let the $\parallel$ ($\perp$) subscript denote the longitudinal (transverse) part of a function with respect to its first spatial argument. Equation~\eqref{eq:apar} is verified easily by noting that the gradient of $\mathbf{K}_{\parallel}$ gives the longitudinal Dirac delta function
\begin{equation}
    {\bm \nabla}^{\mathbf{x}} \mathbf{K}_{\parallel}(\mathbf{x},\mathbf{x'}) = {\bm \delta}^{\parallel}(\mathbf{x}-\mathbf{x'}) = \int \frac{d^3k}{(2\pi)^3} \hat{\mathbf{k}}\hat{\mathbf{k}}e^{i \mathbf{k} \cdot (\mathbf{x}-\mathbf{x'})}.
\end{equation}

Next, by imposing the constraint
\begin{equation}\label{eq:c2}
    \chi_2 = \int d^3x'  \mathbf{A}(\mathbf{x'},t) \cdot \mathbf{K}(\mathbf{x'},\mathbf{x}) = 0,
\end{equation}
and noting that $\mathbf{K}_{\parallel}(\mathbf{x},\mathbf{x'})$ is antisymmetric with respect to exchange of its arguments,
it follows that one can write
\begin{equation}\label{eq:Afull}
    \mathbf{A}(\mathbf{x},t) = \mathbf{A}_{\perp}(\mathbf{x},t) + {\bm \nabla^{\mathbf{x}}} \int d^3x' \mathbf{A}_{\perp}(\mathbf{x'},t) \cdot \mathbf{K}_{\perp}(\mathbf{x'},\mathbf{x}),
\end{equation}
such that the longitudinal part of the vector potential becomes $\mathbf{A}_{\parallel}(\mathbf{x},t) = {\bm \nabla} \Lambda(\mathbf{x})$, with the gauge function
\begin{equation}\label{eq:gaugef}
    \Lambda(\mathbf{x}) = \int d^3x' \mathbf{A}_{\perp}(\mathbf{x'},t) \cdot \mathbf{K}_{\perp}(\mathbf{x'},\mathbf{x}).
\end{equation}
The transverse part of the quantization function $\mathbf{K}_{\perp}\!\!\!:(\mathbb{R}^3,\mathbb{R}^3) \rightarrow \mathbb{R}^3$ is nearly completely arbitrary~\cite{Woolley1999Jan}, and uniquely fixes the gauge with respect to the fields. \blue{Choosing $\chi_2$ to take the form in Eq.~\eqref{eq:c2} allows one to write the entire vector potential in terms of its gauge-invariant transverse component, and a function which remains a c-number after quantization.}
All that remains is to specify the transverse part of the polarization, to describe a unique Lagrangian as per the prescription in Eq.~\eqref{eq:ls}. Noting that since we can write the longitudinal part as 
\begin{equation}\label{eq:PL}
\mathbf{P}_{\parallel}(\mathbf{x},t) = -\int d^3x' \mathbf{K}_{\parallel}(\mathbf{x},\mathbf{x'})\rho_{\rm A}(\mathbf{x'},t),
\end{equation}
we see that the additional term in the Lagrangian becomes (suppressing the time index)
\begin{equation}
    -\frac{d}{dt} \! \int \! \! d^3x\left \{ \mathbf{A}_{\perp}(\mathbf{x}) \! \cdot \! \left[\mathbf{P}_{\perp}(\mathbf{x}) \! + \!  \int \! \! d^3x' \mathbf{K}_{\perp}(\mathbf{x},\mathbf{x'})\rho_{\rm A}(\mathbf{x'})\right]\right\}.
\end{equation}

A convenient definition for the polarization is thus
\begin{equation}\label{eq:pdef}
\mathbf{P}(\mathbf{x},t) = -\int d^3x\, \mathbf{K}(\mathbf{x},\mathbf{x'})\rho_{\rm A}(\mathbf{x'},t),
\end{equation}
such that the total time derivative vanishes, \blue{as}
\begin{equation}
\blue{\int d^3x \mathbf{A}(\mathbf{x},t) \cdot \mathbf{P}(\mathbf{x},t) =0,}
\end{equation}
and the polarization no longer appears at all in the Lagrangian; this  allows one to quantize entirely on the basis of the gauge symmetry associated with the electromagnetic field (albeit in  a subspace with the reduced gauge symmetry associated with the quantization constraint). Note that with this definition, a \emph{gauge transformation} consists of the simultaneous change of the gauge function $\mathbf{K}(\mathbf{x},\mathbf{x'}) \rightarrow \mathbf{K'}(\mathbf{x},\mathbf{x'})$, with an associated change in the vector potential and transverse polarization $\mathbf{A}_{\parallel} \rightarrow \mathbf{A}_{\parallel}'$, $\mathbf{P}_{\perp} \rightarrow \mathbf{P}_{\perp}'$ as determined by Eqs.~\eqref{eq:Afull} and~\eqref{eq:pdef}. In Sec.~\ref{sec:time-dep}, we extend this scheme to allow for \emph{time-dependent} gauge conditions, by letting $\mathbf{K}_{\perp}$ be an explicit function of time, such that a gauge transformation also changes $\phi$. The quantization function approach can also be used for quantization of systems with no explicit medium reservoir degrees of freedom, when dissipation and dispersion can be neglected in the dielectric function $\epsilon$, as we show in Appendix~\ref{sec:appB}.

Two important choices of gauge in the theory of light-matter interactions are the Coulomb and multipolar gauges; for the Coulomb gauge
\begin{equation}\label{eq:kc}
    \mathbf{K}^{\rm C}_{\perp}(\mathbf{x},\mathbf{x'})=0,
\end{equation}
and for the multipolar gauge,
\begin{equation}\label{eq:kmp}
    \mathbf{K}^{\rm mp}_{\perp}(\mathbf{x},\mathbf{x'})=-(\mathbf{x'}- \mathbf{r}_{\rm A})\cdot \! \int_0^{1} ds {\bm \delta}^{\perp}(\mathbf{x}-\mathbf{r}_{\rm A} - s(\mathbf{x'} - \mathbf{r}_{\rm A})),
\end{equation}
    where $\mathbf{r}_{\rm A}$ is an arbitary c-number position, and ${\bm \delta}^{\perp} = \mathbf{I}\delta - {\bm \delta}^{\parallel}$ is the transverse delta function (using dyadic notation with $I_{ij} = \delta_{ij}$). This form is particularly useful for computing multipolar expansions for charge distributions localized around a position $\mathbf{r}_{\rm A}$ (e.g., a molecular center), and is closely related to the PZW transformation from the Coulomb to multipolar gauges, as  discussed more in Sec.~\ref{ssec:matT}.

\subsection{Canonical Quantization with Constraints}
\label{sec:can_quantization}
\blue{ Having specified the gauge of the theory by means of the quantization function $\mathbf{K}(\mathbf{x},\mathbf{x'})$ (in particular, its transverse component), we can now apply Dirac's constrained quantization procedure using the constraints $\chi_0 = \chi_1 = \chi_2 = 0$. The details of this procedure are given in Appendix~\ref{app:dirac}. 

After quantization, we find that the theory can be expressed in terms of transverse canonical field variables $\hat{\mathbf{A}}_{\perp}$ and $\hat{\mathbf{\Pi}}$, where
\begin{equation}
\hat{\mathbf{\Pi}} = \hat{\mathbf{\Pi}}_{\mathbf{A}} - \hat{\mathbf{P}},
\end{equation}
as well as $\hat{\mathbf{r}}_{\alpha}$, $\hat{\mathbf{p}}_{\alpha}$, $\hat{\mathbf{X}}_{\omega}$, and $\hat{\mathbf{\Pi}}_{\mathbf{X}_{\omega}}$. These canonical coordinates satisfy the usual canonical commutation relations:
    \begin{subequations}
    \begin{equation}\label{eq:commA}
        [\mathbf{\hat{A}}_{\perp}(\mathbf{x}),\mathbf{\hat{\Pi}}(\mathbf{x'})] = i\hbar{\bm \delta}^{\perp}(\mathbf{x}-\mathbf{x'})
    \end{equation}
    \begin{equation}\label{eq:commr_maintext}
    [\mathbf{\hat{r}}_{\alpha},\mathbf{\hat{p}}_{\alpha'}] = i\hbar \mathbf{I}\delta_{\alpha \alpha'},
\end{equation}
\begin{equation}\label{eq:commX}
    [\mathbf{\hat{X}}_{\omega}(\mathbf{x}),\mathbf{\hat{\Pi}}_{\omega'}(\mathbf{x'})] = i\hbar\mathbf{I}\delta(\mathbf{x}-\mathbf{x'})\delta(\omega-\omega')
\end{equation}
\end{subequations}
with all other commutators vanishing.

}

\subsection{Fano diagonalization and Green's function expansion}
\label{sec:Fano_diagnonalization}
As a result of the canonical commutation relations in Eqs.~\eqref{eq:commA},~\eqref{eq:commr_maintext}, and~\eqref{eq:commX}, the fields $\hat{\mathbf{A}}_{\perp}$ and $\hat{\mathbf{X}}_{\omega}$, as well as their conjugates $\hat{\mathbf{\Pi}}$ and $\hat{\mathbf{\Pi}}_{\mathbf{X}_{\omega}}$ can be expressed as a sum over bosonic creation and annihilation operators. Furthermore, the total Hamiltonian can be decomposed into a component which only consists of these operators, $\hat{H}_{\rm F}$,
as well as a component which involves the polarization operators associated with the $\hat{\mathbf{r}}_{\alpha}$ operators and their conjugates $\hat{H}_{\rm P}$: 
\begin{align}\label{eq:hp}
    \hat{H}_{\rm F} = \frac{1}{2\epsilon_0}&\int d^3x\left[\hat{\mathbf{\Pi}} + \int d\omega \alpha(\mathbf{x},\omega) \hat{\mathbf{X}}_{\omega}\right]^2 
    \nonumber \\ &  + \frac{1}{2\mu_0}\left({\bm \nabla} \times \hat{\mathbf{A}}_{\perp}\right)^2 + \hat{H}_{\mathbf{X}_{\omega}} ,
\end{align}
with 
\begin{equation}
    \hat{H}_{\mathbf{X}_{\omega}} = \frac{1}{2}\int d^3x\int_0^{\infty} d\omega\left(\hat{\mathbf{\Pi}}_{\omega}^2 + \omega^2 \hat{\mathbf{X}}_{\omega}^2\right),
\end{equation}
and
\begin{align}
    \hat{H}_{\rm P} =& \sum_{\alpha}\frac{\left[\hat{\mathbf{p}}_{\alpha} - q_{\alpha}  \hat{\mathbf{A}}(\hat{\mathbf{r}}_{\alpha})\right]^2}{2m_{\alpha}} - \int d^3x \hat{\mathbf{P}}(\mathbf{x}) \cdot \hat{\mathbf{E}}(\mathbf{x}) \nonumber \\ & - \frac{1}{2\epsilon_0}\int d^3x \hat{\mathbf{P}}^2(\mathbf{x}),
\end{align}
where Eq.~\eqref{eq:p2} was used to express the electric field operator in the quantization variables as
\begin{equation}
    \hat{\mathbf{E}}(\mathbf{x}) = -\frac{1}{\epsilon_0}\left[\hat{\mathbf{\Pi}}(\mathbf{x}) + \hat{\mathbf{P}}(\mathbf{x}) + \int_0^{\infty} d\omega \alpha(\mathbf{x},\omega) \hat{\mathbf{X}}_{\omega}(\mathbf{x})\right].
\end{equation}
The total Hamiltonian is then $\hat{H} = \hat{H}_{\rm F}  + \hat{H}_{\rm P}$.

The term $\hat{H}_{\rm F}$ is quadratic in the quantization variables over bosonic fields, and as such, one should be able to, by a process of Fano diagonalization, express it in the form
\begin{equation}
    \hat{H}_{\rm F} = \int d^3x \int_0^{\infty}d \omega \hbar \omega \hat{\mathbf{b}}^{\dagger}(\mathbf{x},\omega) \cdot \hat{\mathbf{b}}(\mathbf{x},\omega),
\end{equation}
where $\hat{\mathbf{b}}(\mathbf{x},\omega)$ are bosonic excitation operators which combine medium harmonic oscillator fields $\hat{\mathbf{X}}_{\omega}$ and electromagnetic degrees of freedom, and satisfy 
\begin{subequations}
\begin{align}
    [\mathbf{\hat{b}}(\mathbf{x},\omega),\mathbf{\hat{b}}^{\dagger}(\mathbf{x'},\omega')] &= \mathbf{I}\delta(\mathbf{x}-\mathbf{x'})\delta(\omega-\omega') \\
     [\mathbf{\hat{b}}(\mathbf{x},\omega),\mathbf{\hat{b}}(\mathbf{x'},\omega')] &= {\bf 0}\\
    [\mathbf{\hat{b}}^{\dagger}(\mathbf{x},\omega),\mathbf{\hat{b}}^{\dagger}(\mathbf{x'},\omega')] &= {\bf 0}.
\end{align}
\end{subequations}

For the Coulomb gauge, this is done precisely in the derivation by (e.g.) Philbin~\cite{Philbin2010Dec}. Here, we  note that the results of this derivation can be applied directly with the substitution $(\hat{\mathbf{A}},\hat{\mathbf{\Pi}}_{\mathbf{A}}) \rightarrow (\hat{\mathbf{A}}_{\perp}, \hat{\mathbf{\Pi}})$. The important result is that the fields can be expressed as
\begin{subequations}\label{eq:fieldGf}
\begin{equation}
\hat{\mathbf{E}}_{\blue{{\rm F}}}(\mathbf{x}) = i\int \! d^3x' \! \int_0^{\infty}\! \! \!  d\omega \frac{1}{\epsilon_0 \omega} \mathbf{G}(\mathbf{x},\mathbf{x'},\omega) \cdot \hat{\mathbf{J}}_{\rm N}(\mathbf{x'},\omega) + \text{H.c.}
\end{equation}
\begin{equation}
\hat{\mathbf{A}}_{\perp}(\mathbf{x}) = \int \! d^3x' \int_0^{\infty}\! \! \! d\omega \frac{1}{\epsilon_0 \omega^2} \mathbf{G}^{\perp}(\mathbf{x},\mathbf{x'},\omega) \cdot \hat{\mathbf{J}}_{\rm N}(\mathbf{x'},\omega) + \text{H.c.},
\end{equation}
\end{subequations}
where
\begin{equation}
    \hat{\mathbf{J}}_{\rm N }(\mathbf{x},\omega) = \sqrt{\frac{\hbar\omega}{2}}\alpha(\mathbf{x},\omega) \hat{\mathbf{b}}(\mathbf{x},\omega),
\end{equation}
and $\hat{\mathbf{E}}_{\blue{{\rm F}}}$ is the part of the total electric field operator $\hat{\mathbf{E}} = \hat{\mathbf{E}}_{\blue{{\rm F}}} - \hat{\mathbf{P}}/\epsilon_0$ which can be expressed in terms of the bosonic operators that diagonalize $\hat{H}_{\rm F}$. 
In accordance with our assumption of non-magnetic media,
the photonic Green's function $\mathbf{G}(\mathbf{x},\mathbf{x}',\omega)$ is a tensor (or dyad) which satisfies the Helmholtz equation for a dipole source:
    \begin{equation}
    \left[\boldsymbol{\nabla}^{\mathbf{x}}\times\boldsymbol{\nabla}^{\mathbf{x}}\times-\frac{\omega^2}{c^2}\epsilon(\mathbf{x},\omega)\right]\mathbf{G}(\mathbf{x},\mathbf{x'},\omega)=\frac{\omega^2}{c^2}\mathbf{I}\delta(\mathbf{x-\mathbf{x'}}),\label{eq: GreenHelm}
\end{equation}
together with the corresponding retarded boundary conditions. For example, for open dielectrics (i.e., cavity resonators surrounded by a homogeneous medium with index of refraction $n_B$), one can use the Silver-M{\"u}ller radiation condition:
\begin{equation}
    \frac{\mathbf{x}}{|\mathbf{x}|}\times\boldsymbol{\nabla}^{\mathbf{x}}\times\mathbf{G}(\mathbf{x},\mathbf{x'},\omega)\rightarrow in_{\rm B}\frac{\omega}{c}\mathbf{G}(\mathbf{x},\mathbf{x'},\omega) \label{eq: SM_Cond_GF},
\end{equation}
which holds as $|\mathbf{x}| \rightarrow \infty$. Here we let the notation $\mathbf{G}^{\perp}(\mathbf{x},\mathbf{x'},\omega)$ refer to the transverse part of $\mathbf{G}(\mathbf{x},\mathbf{x'},\omega)$ with respect to the left-hand side of the dyad, and spatial argument $\mathbf{x}$.

\section{Gauge Invariance and Hilbert Space Truncation}
\label{sec:GI}
In this section, we first show in Sec.~\ref{ssec:GI} how gauge invariance manifests in the quantized theory, and how gauge transformations can be implemented as unitary transformations within the general gauge function quantization method. We then discuss how material and mode truncation can potentially break this gauge invariance, and how this can be avoided, in Secs.~\ref{ssec:matT} and~\ref{ssec:modeT}. Our main contribution in this section is the arbitrary-gauge Hamiltonian under material and mode truncation $\hat{\tilde{\mathcal{H}}}$, given by Eq.~\eqref{eq:hmode}, from which we resolve gauge ambiguities and derive simplified models in later sections.

\subsection{Gauge Invariance in the Quantum Theory}\label{ssec:GI}
Prior to quantization, gauge invariance manifests as the invariance of the Lagrangian under gauge transformations of the four-potential. 
%In general, this is the only way to explicitly demonstrate the full gauge symmetry of the theory.
After quantization using the arbitrary gauge approach (in terms of the quantization function $\mathbf{K}(\mathbf{x},\mathbf{x'})$), however, the hallmark of gauge invariance is the invariance of the Schr{\" o}dinger equation under a local phase change of the state vector simultaneous with a gauge transformation of the potentials. Such a process can be implemented as a unitary transformation which transforms the system from one \emph{fixed} gauge to another~\cite{Stokes2021Sep}.

As mentioned in the previous section, a gauge transformation consists of the change $\mathbf{K}(\mathbf{x},\mathbf{x'}) \rightarrow \mathbf{K'}(\mathbf{x},\mathbf{x'})$ and the associated change in the longitudinal component of the vector potential and the transverse component of the polarization. Thus, for the theory to be gauge invariant, this change should be compensated by a local phase change in the state vector, implemented as a unitary operator.

Specifically, for the quantization scheme developed here and the set of gauge transformations allowed therein, this phase variation can be expressed as a unitary transformation $\ket{\psi} \rightarrow \hat{W}\ket{\psi}$, with an accompanying change of the Hamiltonian $\hat{W}\hat{H}\hat{W}^{\dagger}$, in order to preserve the form of the Schr{\"o}dinger equation evolution. To be concrete, consider two gauges indexed by $g$ and $g'$, with quantization functions $\mathbf{K}^g(\mathbf{x},\mathbf{x'})$ and $\mathbf{K}^{g'}(\mathbf{x},\mathbf{x'})$, respectively. Then, a gauge transformation from gauge representation $g$ to $g'$ can be found as 
\begin{align}\label{eq:wdef}
    \hat{W}_{g'g} &= \exp{\left[\frac{i}{\hbar}\sum_{\alpha} q_{\alpha}\left[\hat{\Lambda}_{g'}(\hat{\mathbf{r}}_{\alpha}) -\hat{\Lambda}_{g}(\hat{\mathbf{r}}_{\alpha})\right] \right]} \nonumber \\ 
    & = \exp{\left[\frac{i}{\hbar}\int d^3x \left[ \hat{\Lambda}_{g'}(\mathbf{x})-\hat{\Lambda}_g(\mathbf{x}) \right]\hat{\rho}_{\rm A}(\mathbf{x})\right]} \nonumber \\ 
    & =\exp{\left[-\frac{i}{\hbar} \!\int \! \! d^3x \left[\hat{\mathbf{P}}^{g'}_{\perp}(\mathbf{x})-\hat{\mathbf{P}}^{g}_{\perp}(\mathbf{x})\right] \! \cdot \! \hat{\mathbf{A}}_{\perp}(\mathbf{x})\right]},
\end{align}
and $\hat{\Lambda}_{g}(\mathbf{x})$ is the quantized version of the gauge function expressed in Eq.~\eqref{eq:gaugef} for a gauge indexed by $g$.  The state vector transforms as $\ket{\psi_{g'}} = \hat{W}_{g'g}\ket{\psi_{g}}$. To determine the transformation effect on the variables constituting the Hamiltonian, it is useful to note the following relations (suppressing frequency indices):
\begin{align}\label{eq:aperpcomm}
    &[\hat{A}_{\perp,i}(\mathbf{x}),\hat{A}_{\perp,j}(\mathbf{x'})]  \nonumber \\ & = \frac{\hbar}{2\epsilon_0^2}  \int \!  d^3r  \! \int_0^{\infty}\! \! \! d\omega  \frac{\alpha^2(\mathbf{r})}{\omega^3}G_{ik}^{\perp}(\mathbf{x},\mathbf{r}) G^{\perp *}_{kj}(\mathbf{r},\mathbf{x'}) - \text{H.c.} \nonumber \\
    &=\frac{\hbar}{\pi \epsilon_0} \int_0^{\infty}\frac{d\omega}{\omega^2}\text{Im}\{G^{\perp}_{ij}(\mathbf{x},\mathbf{x'})\} - \text{H.c.} \nonumber \\
    &=0,
\end{align}
where we have used the Green's function relations~\cite{Dung}
\begin{equation}\label{eq:gf1}
    G_{ij}(\mathbf{x},\mathbf{x'},\omega) = G_{ji}(\mathbf{x'},\mathbf{x},\omega),
\end{equation}
and
\begin{equation}\label{eq:gf2}
    \int \! \! d^3r \epsilon_I(\mathbf{r},\omega)\mathbf{G}(\mathbf{x},\mathbf{r},\omega)\cdot \mathbf{G}(\mathbf{r},\mathbf{x'},\omega) = \text{Im}\{\mathbf{G}(\mathbf{x},\mathbf{x'},\omega)\}.
\end{equation}

Equations~\eqref{eq:aperpcomm} and~\eqref{eq:Afull} together imply that $[\hat{A}_i(\mathbf{x}),\hat{A}_{j}(\mathbf{x'})]=0$ as well.

Of particular use is the transformation of the canonical momenta:
\begin{equation}\label{eq:ptP}
    \hat{W}_{g'g}\left[\hat{\mathbf{\Pi}}(\mathbf{x}) + \hat{\mathbf{P}}^{g}(\mathbf{x})\right]\hat{W}_{g'g}^{\dagger} = \hat{\mathbf{\Pi}}(\mathbf{x}) +\hat{\mathbf{P}}^{g'}(\mathbf{x}),
\end{equation}
\begin{equation}\label{eq:ptp}
    \hat{W}_{g'g}\left[\hat{\mathbf{p}}_{\alpha} - q_{\alpha} \hat{\mathbf{A}}^g(\hat{\mathbf{r}}_{\alpha})\right]\hat{W}^{\dagger}_{g'g} = \hat{\mathbf{p}}_{\alpha} - q_{\alpha} \hat{\mathbf{A}}^{g'}(\hat{\mathbf{r}}_{\alpha}).
\end{equation}
To show the effect of this transformation on the Hamiltonian, it is useful to write the arbitrary gauge Hamiltonian \blue{$\hat H^{g}$} in the form 
\begin{align}
    \hat{H}^{\blue{g}} = \int d^3x & \left[\frac{\epsilon_0}{2}\blue{\left[\hat{\mathbf{E}}^g(\mathbf{x})\right]^2} + \frac{1}{2\mu_0}\hat{\mathbf{B}}^2(\mathbf{x})\right] \nonumber \\ &+ \hat{H}_{\mathbf{X}_{\omega}} + \sum_{\alpha}\frac{\left[\hat{\mathbf{p}}_{\alpha} - q_{\alpha} \hat{\mathbf{A}}^{\blue{g}}(\hat{\mathbf{r}}_{\alpha})\right]^2}{2m_{\alpha}},
\end{align}
where
\begin{equation}\label{eq:Eg}
    \hat{\mathbf{E}}^{\blue{g}}(\mathbf{x}) =- \frac{1}{\epsilon_0}\left[\hat{\mathbf{\Pi}}(\mathbf{x}) + \hat{\mathbf{P}}^g(\mathbf{x}) + \! \int_0^{\infty}\! \! d\omega \alpha(\mathbf{x},\omega) \hat{\mathbf{X}}_{\omega}(\mathbf{x})\right].
\end{equation}
The unitary transformation has no effect on $\hat{\mathbf{B}}$ (as a result of Eq.~\eqref{eq:aperpcomm}) or the reservoir operators $\hat{\mathbf{X}}_{\omega}$, $\hat{\mathbf{\Pi}}_{\mathbf{X}_\omega}$, and clearly $\hat{W}_{g'g}\hat{\mathbf{E}}^{\blue{g}}(\mathbf{x})\hat{W}^{\dagger}_{g'g} = \hat{\mathbf{E}}^{\blue{g'}}(\mathbf{x})$, as a consequence of Eq.~\eqref{eq:ptP}. Using this fact and Eq.~\eqref{eq:ptp}, it is easy to see that the unitary transform has the effect of performing the gauge transform, and thus we confirm the quantum theory is indeed gauge invariant.

\subsection{Material truncation}
\label{ssec:matT}
The issue with truncation is that the transformation given by Eq.~\eqref{eq:ptp} necessarily requires the full infinite dimensional operator algebra to be implemented, and thus any method of truncating the operator $\hat{W}_{g'g}$ (which must be truncated to operate on the reduced dimensionality state vector) will fail to give the necessary gauge transformation~\cite{Starace1971Apr,Keeling2007Jun,DeBernardis2018Apr,DeBernardis2018Nov,Stokes2019Jan,Rouse2021Feb}. This results in a theory which does not respect the gauge principle and gives ambiguous results, especially in the USC regime. This can also be seen as the truncation creating a non-local potential which can be expressed as a function of truncated momentum operators, to which the minimal coupling replacement that ensures gauge invariance has not been applied~\cite{Starace1971Apr,Savasta1995Nov,DiStefano2019Aug}, or inconsistent constraining of interactions to a specific subspace~\cite{Taylor2020Sep}.

To circumvent this, we can instead write the Hamiltonian, prior to truncation, in terms of unitary operators which implement the minimal coupling replacement on a Hamiltonian which, in the absence of transverse coupling to the electric field (e.g., the eigenstates of a molecular system), has a discrete set of energy levels which are near-resonant with relevant medium-assisted interactions with the electric field. The truncation can then be applied directly to the position operators, which ensures the gauge transformation is consistent with the reduced Hilbert space dimensionality and preserves gauge invariance~\cite{Graf1995Feb, DiStefano2019Aug, Taylor2020Sep}. This procedure is also consistent with a lattice gauge theory perspective, where the local phase transformation acts on the state vector only at discrete ``lattice'' points in space equal in number to the number of states left after truncation~\cite{Savasta2021May}, as well as the Peierls substitution for introducing electromagnetic interactions within tight-binding models~\cite{Graf1995Feb}.

To identify the ``bare'' matter Hamiltonian, containing only the interparticle Coulomb interactions between the constituent particles, note that in the absence of coupling with the transverse field, the material Hamiltonian becomes
\begin{equation}
    \hat{H}_0 = \sum_{\alpha}\frac{\hat{\mathbf{p}}_{\alpha}^2}{2m_{\alpha}} + \sum_{\alpha,\alpha'}\hat{V}_{\rm Coul}(\hat{\mathbf{r}}_{\alpha},\hat{\mathbf{r}}_{\alpha'}),
\end{equation}
where
\begin{align}
    \sum_{\alpha,\alpha'}\hat{V}_{\rm Coul}(\hat{\mathbf{r}}_{\alpha},\hat{\mathbf{r}}_{\alpha'}) &= \int d^3x \frac{\hat{\mathbf{P}}_{\parallel}^2(\mathbf{x})}{2\epsilon_0} \nonumber \\ &= \frac{1}{2} \sum_{\alpha,\alpha'} \frac{q_{\alpha} q_{\alpha'}}{4\pi\epsilon_0|\hat{\mathbf{r}}_{\alpha}-\hat{\mathbf{r}}_{\alpha'}|},
\end{align}
and we are assuming that the medium-assisted longitudinal field is sufficiently weak that it suffices to use the unscreened Coulomb potential to calculate the unperturbed material eigenstates---alternatively, we can simply phenomenlogically use the eigenstates which are corrected by the medium-assisted longitudinal field as the basis $\hat{H}_0$ (see Appendix~\ref{sec:appB} and Ref.~\cite{Wubs2003Jul} for an analogous discussion in the case of a nondispersive and nonabsorbing medium).

In this manner, we can write the entire arbitrary-gauge Hamiltonian as \blue{(for a gauge indexed by `$g$'),}
\begin{equation}
    \hat{H}^{\blue{g}} = \hat{H}_{\rm F} + \hat{H}_0 + \hat{H}^{\blue{g}}_{\rm int},
\end{equation}
where
\begin{align}
    \hat{H}_{\rm int}^{\blue{g}} =& -\sum_{\alpha}\frac{q_{\alpha}}{m_{\alpha}}\hat{\mathbf{p}}_{\alpha} \cdot \hat{\mathbf{A}}^{\blue{g}}(\hat{\mathbf{r}}_{\alpha}) + \sum_{\alpha}\frac{q_{\alpha}^2}{2m_{\alpha}}\blue{\left[\hat{\mathbf{A}}^{g}(\hat{\mathbf{r}}_{\alpha})\right]^2} \nonumber \\ & - \int d^3x \hat{\mathbf{P}}(\mathbf{x}) \cdot \hat{\mathbf{E}}^{\blue{{g}}}_{\rm F}(\mathbf{x}) + \frac{1}{2\epsilon_0}\int d^3x \blue{\left[\hat{\mathbf{P}}^{g}_{\perp}(\mathbf{x})\right]^2}.
\end{align}

Next, we introduce the unitary operator
\begin{equation}
    \hat{U}_{\blue{g}} = \exp{\left[\frac{i}{\hbar}\int d^3x \hat{\mathbf{A}}^{\blue{g}}(\mathbf{x}) \cdot \hat{\mathbf{Z}}_{\rm A}(\mathbf{x})\right]},
\end{equation}
where
\begin{equation}\label{eq:z_def}
    \hat{\mathbf{Z}}_{\rm A}(\mathbf{x}) = \sum_{\alpha}q_{\alpha}\hat{\mathbf{r}}_{\alpha} \int_0^{1}ds \delta(\mathbf{x} - s\hat{\mathbf{r}}_{\alpha})
\end{equation}
is an operator chosen to implement, approximately, the minimal coupling transformation $\hat{\mathbf{p}}_{\rm \alpha} \rightarrow \hat{\mathbf{p}}_{\rm \alpha} - q_{\alpha}\hat{\mathbf{A}}^{\blue{g}}(\hat{\mathbf{r}}_{\alpha})$, as appears in the $\hat{H}_{\rm P}$ part of the full Hamiltonian $\hat{H}^{\blue{g}}$ (Eq.~\eqref{eq:hp}). In actuality, the full transformation of the particle momenta under $\hat{U}_{\blue{g}}$ is
\begin{align}\label{eq:fullt}
    \hat{U}_{\blue{g}}\hat{\mathbf{p}}_{\alpha}&\hat{U}_{\blue{g}}^{\dagger} = \hat{\mathbf{p}}_{\rm \alpha} - q_{\alpha}\hat{\mathbf{A}}^{\blue{g}}(\hat{\mathbf{r}}_{\alpha}) 
    %\nonumber \\ & 
    -q_\alpha \hat{\mathbf{r}}_{\alpha}\times \! \int_0^1\! \! ds~s \hat{\mathbf{B}}(s\hat{\mathbf{r}}_\alpha).
\end{align}
Note that we also have $\hat{U}_{\blue{g}} = \hat{W}_{g,\text{mp}}$; this transformation takes an operator from the multipolar gauge to gauge $g$. \blue{We give this operator as it appears in this context its own symbol $\hat{U}_{g}$ to emphasize its role in restoring gauge invariance under material truncation, described in the following.}

Using  \blue{the transformation of Eq.~\eqref{eq:fullt}}, we can write the full Hamiltonian as
\begin{align}\label{eq:Htot}
    \hat{H}^{\blue{g}} = &\hat{H}_{\rm F} + \hat{U}_{\blue{g}}\hat{H}_0\hat{U}_{\blue{g}}^{\dagger} - \int d^3x \hat{\mathbf{P}}(\mathbf{x}) \cdot \hat{\mathbf{E}}\blue{^{g}_{\rm F}}(\mathbf{x}) \nonumber \\ &+ \frac{1}{2\epsilon_0}\int d^3x \blue{\left[\hat{\mathbf{P}}^{g}_{\perp}(\mathbf{x})\right]^2} \blue{+} \hat{H}^{\blue{g}}_{\rm mag},
\end{align}
where
\begin{align}
&\hat{H}_{\rm mag}^{\blue{g}}  = \sum_{\alpha} \frac{q_{\alpha}}{2m_{\alpha}}\int_0^1 ds s\Bigg[ \hat{\mathbf{p}}_{\alpha} \cdot \hat{\mathbf{r}}_{\alpha} \times \hat{\mathbf{B}}(s\mathbf{\hat{r}}_{\alpha}) \nonumber \\ & +  \hat{\mathbf{r}}_{\alpha} \times \hat{\mathbf{B}}(s\mathbf{\hat{r}}_{\alpha}) \cdot \hat{\mathbf{p}}_{\alpha} - 2q_{\alpha}\hat{\mathbf{A}}^{\blue{g}}(\hat{\mathbf{r}}_{\alpha}) \cdot \hat{\mathbf{r}}_{\alpha} \times \hat{\mathbf{B}}(s\hat{\mathbf{r}}_{\alpha}) \nonumber \\ & - q_{\alpha}\int_0^{1}ds' s'\left[\hat{\mathbf{r}}_{\alpha} \times \hat{\mathbf{B}}(s\hat{\mathbf{r}}_{\alpha})\right] \cdot \left[\hat{\mathbf{r}}_{\alpha} \times \hat{\mathbf{B}}(s'\hat{\mathbf{r}}_{\alpha})\right]\Bigg].
\end{align}
The magnetic terms were analyzed in detail in Ref.~\cite{Buhmann2004Nov} (in that case, as they appear in the multipolar gauge), with the conclusion that, for atomic systems, the scaling of these terms relative to the electric dipole interaction is proportional to $\sim (Z_{\rm eff}\alpha_0)^2$, where $Z_{\rm eff}$ is the effective (screened) charge of the nucleus, and $\alpha_0$ is the fine-structure constant. For non-relativistic systems, $Z_{\rm eff} \alpha_0 \ll 1$, and so going forward we will neglect the influence of $\hat{H}^{\blue{g}}_{\rm mag}$. We note that this is an inherent approximation of the theory (although one that is very well-founded in most circumstances), which has to our knowledge has not been acknowledged to date in the literature on restoring gauge invariance in truncated material systems.

The Hamiltonian of Eq.~\eqref{eq:Htot}, after neglecting the magnetic terms, is gauge invariant under material truncation \emph{provided} we truncate the position operators $\mathbcallc{\hat r}_{\alpha} \equiv \hat{P}\hat{\mathbf{r}}_{\alpha}\hat{P}$, where 
\begin{equation}
    \hat{P} = \sum_{i=1}^{N} \ket{\phi_i}\bra{\phi_i}
\end{equation}
is a projector operator onto a finite set of $N$ eigenstates of $\hat{H}_0$, such that in the truncated space, we have
\begin{equation}
    \hat{P}\hat{H}_0\hat{P} = \hat{\mathcal{H}}_0=\sum_{i=1}^{N} \hbar \omega_i\ket{\phi_i}\bra{\phi_i},
\end{equation}
where $\ket{\phi_i}$ denotes the $i^{\rm th}$ eigenstate of $\hat{H}_0$ with energy $\hbar\omega_i$. We use throughout this work calligraphic characters to denote operators which act on the truncated space. We stress that, except for $\mathcal{\hat{H}}_0$, the  (correctly) truncated operators are those that are expressed in terms of the projected position operators $\hat{\mathbcallc{r}}_{\alpha}$, and \emph{not} those with the projector operator $\hat{P}$ directly applied, which generally will violate gauge invariance.

We then take $\hat{\mathbf{P}}^{\blue{g}} \rightarrow \hat{\bm{\mathcal{P}}}^{\blue{g}}$, where
\begin{equation}
    \hat{\bm{\mathcal{P}}}^{\blue{g}} = -\sum_{\alpha} q_{\alpha}\mathbf{K}^{\blue{g}}(\mathbf{x},\mathbcallc{\hat r}_{\alpha}),
\end{equation}
and $\hat{U}_{\blue{g}} \rightarrow \hat{\mathcal{U}}_{\blue{g}}$:
\begin{align}
\hat{\mathcal{U}}_{\blue{g}}  &= \exp{\left[\frac{i}{\hbar}\int d^3x \hat{\mathbf{A}}^{\blue{g}} (\mathbf{x}) \cdot \hat{\bm{\mathcal{Z}}}_{\rm A}(\mathbf{x})\right]}   \nonumber \\
& = \exp{\left[\frac{i}{\hbar}\sum_{\alpha} q_{\alpha}  \hat{\mathbcallc{r}}_{\alpha} \cdot \int_0^{1}ds \hat{\mathbf{A}}^{\blue{g}} (s\hat{\mathbcallc{r}}_{\alpha})\right]}, 
\end{align}
where $\hat{\bm{\mathcal{Z}}}_{\rm A}$ is $\hat{\mathbf{Z}}_{\rm A}$ expressed in terms of the truncated position operators $\mathbcallc{\hat r}_{\alpha}$.
Thus, the truncated material basis arbitrary-gauge Hamiltonian $\hat{\mathcal{H}}^{\blue{g}}$ can be written as
\begin{align}\label{eq:arbHt}
    \mathcal{\hat{H}}^{\blue{g}} = &\hat{H}_{\rm F} + \hat{\mathcal{U}}_{\blue{g}}\hat{\mathcal{H}}_0\hat{\mathcal{U}}_{\blue{g}}^{\dagger} \nonumber \\ &- \int d^3x \hat{\bm{\mathcal{P}}}^{g}(\mathbf{x}) \cdot \hat{\mathbf{E}}\blue{_{\rm F}^{g}}(\mathbf{x}) + \frac{1}{2\epsilon_0}\int d^3x \blue{\left[\hat{\bm{\mathcal{P}}}_{\perp}^g(\mathbf{x})\right]^2}.
\end{align}

In the Coulomb gauge, $\mathbf{K}^{\rm C}_{\perp}= \hat{\mathbf{P}}^{\rm C}_{\perp} = \hat{\mathbf{A}}^{\rm C}_{\parallel} =0$, and the Hamiltonian, $\hat{\mathcal{H}}^{\blue{{\rm C}}}$, can be expressed as 
\begin{equation}\label{eq:Hc}
    \hat{\mathcal{H}}^{\blue{{\rm C}}} = \hat{H}_{\rm F} + \hat{\mathcal{U}}_{\rm C}\hat{\mathcal{H}}_0\hat{\mathcal{U}}_{\rm C}^{\dagger} - \int d^3x \hat{\bm{\mathcal{P}}}_{\parallel}(\mathbf{x}) \cdot \blue{\left[\hat{\mathbf{E}}_{\rm F}(\mathbf{x})\right]_{\parallel}}.
\end{equation}
In the multipolar gauge, $\mathbf{K}^{\rm mp}$ is given by Eq.~\eqref{eq:kmp}, which implies (again taking $\mathbf{r}_{\rm A} =0$)
\begin{equation}
   \hat{\bm{\mathcal{P}}}^{\rm mp}_{\perp}(\mathbf{x}) = \sum_{\alpha}q_\alpha\hat{\mathbcallc{r}}_{\alpha}\cdot \int_0^{1} ds {\bm \delta}^{\perp}(\mathbf{x} - s\hat{\mathbcallc{r}}_{\alpha}),
\end{equation}
and
\begin{equation}
    \hat{\mathbf{A}}^{\rm mp}(\mathbf{x}) = -\int_0^{1}ds \mathbf{x} \times \hat{\mathbf{B}}(s\mathbf{x}).
\end{equation}
In the multipolar gauge, $\mathbf{x} \cdot \hat{\mathbf{A}}^{\rm mp}(\mathbf{x}) =0$, and from this it is easy to show that $\hat{\mathcal{U}}_{\rm mp} = 1$. Applying this result, we thus find,
\begin{align}
    \hat{\mathcal{H}}^{\blue{{\rm mp}}} &= \hat{H}_{\rm F} + \hat{\mathcal{H}}_0 -\sum_{\alpha} q_{\alpha} \mathbcallc{\hat r}_{\alpha} \cdot \int_0^{1}ds \hat{\mathbf{E}}\blue{_{\rm F}^{\rm mp}}(s\mathbcallc{\hat r}_{\alpha}) \nonumber \\ & + \sum_{\alpha,\alpha'}\frac{q_{\alpha}q_{\alpha'}}{2\epsilon_0} \! \int_0^{1} \! ds \!\int_{0}^1 \! ds'  \mathbcallc{\hat r}_{\alpha} \cdot \mathbf{\delta}^{\perp}(s'\mathbcallc{\hat r}_{\alpha'}-s\mathbcallc{\hat r}_{\alpha}) \cdot \mathbcallc{\hat r}_{\alpha'}.
\end{align}

As in the untruncated theory, we can implement a gauge change from one fixed gauge to another by means of a unitary transformation:
\begin{equation}\label{eq:g2g'}
    \hat{\mathcal{W}}_{g'g} \hat{\mathcal{H}}^\blue{{g}}\hat{\mathcal{W}}_{g'g},
\end{equation}
where $\hat{\mathcal{W}}_{g'g}$, defined from  Eq.~\eqref{eq:wdef}, is expressed in terms of the truncated position operators $\mathbcallc{\hat r}_{\alpha}$, i.e., 
\begin{equation}\label{eq:gauge_tm}
    \hat{\mathcal{W}}_{g'g} = \exp{\left[-\frac{i}{\hbar} \!\int \! \! d^3x \left[\hat{\bm{\mathcal{P}}}^{g'}_{\perp}(\mathbf{x})-\hat{\bm{\mathcal{P}}}^{g}_{\perp}(\mathbf{x})\right] \! \cdot \! \hat{\mathbf{A}}_{\perp}(\mathbf{x})\right]}.
\end{equation}
It is straightforward to verify that the transformation in Eq.~\eqref{eq:g2g'} is equivalent to replacing $\hat{\mathbf{A}}^g$ and $\hat{\bm{\mathcal{P}}}^{g}$ in the arbitrary-gauge Hamiltonian $\hat{\mathcal{H}}$ in Eq.~\eqref{eq:arbHt} with $\hat{\mathbf{A}}^{g'}$ and $\hat{\bm{\mathcal{P}}}^{g'}$ (or equivalently, replacing $\mathbf{K}^g$ with $\mathbf{K}^{g'}$), respectively;  thus we see that gauge invariance is preserved under material truncation in the fully quantized theory.

It is worth noting that $\hat{\mathbf{Z}}_{\rm A}$ is precisely the multipolar polarization $\hat{\mathbf{P}}^{\rm mp}(\mathbf{x})$, and as such, $\hat{U}_{\blue{g}} = \hat{W}_{g, \text{mp}}$, as previously noted. Moreover, if it is evaluated in the Coulomb gauge, then
$\hat{U}_{\rm C}$ is the unitary operator which implements the well-known PZW transformation~\cite{Babiker1983Feb,Woolley2020Feb,Andrews2018Jan}. The PZW transformation $\hat{U}_{\blue{\rm C}}^{\dagger}\hat{H}^{\blue{{\rm C}}} \hat{U}_{\blue{\rm C}} = \hat{H}^{\rm mp}$  removes the transformation that generates minimal coupling from $\hat{H}_0$, which allows one to truncate the energy levels of the bare system without needing to rely on the infinite dimensional operator algebra required to transform the $\hat{\mathbf{p}}_{\alpha}$ operators. This is why naive truncation (in the sense of $\hat{H} \rightarrow \hat{P}\hat{H}\hat{P}$) in the multipolar gauge gives much more accurate results than the Coulomb gauge~\cite{DeBernardis2018Nov}, and is in fact generally assumed to not break gauge invariance~\cite{Salmon2022Mar}. It is worth noting that this argument relies on the neglect of the magnetic terms, however, and should be understood as a non-relativistic approximation. As discussed in the following section, naive truncation in the multipolar gauge also fails in general when anything less than a complete set of modes is used to expand the electromagnetic fields, which is often the case in, for example, cavity-QED.

\subsection{Mode Truncation}
\label{ssec:modeT}
In addition to material truncation, we can also consider truncation of the transverse electrodynamic degrees of freedom: for example, a mode truncation, where the ``modes'' are typically solutions to Maxwell's equations subject to a certain boundary condition (e.g., fixed, periodic, or open). To do so, note that the arbitrary gauge Hamiltonian from Eq.~\eqref{eq:arbHt} can also be written as
\begin{equation}\label{eq:genH}
    \hat{\mathcal{H}}^{\blue{g}} = \hat{V}_{\blue{g}}\hat{H}_{\rm F} \hat{V}^{\dagger}_{\blue{g}} + \hat{\mathcal{U}}_{\blue{g}}\hat{\mathcal{H}}_0 \hat{\mathcal{U}}^{\dagger}_{\blue{g}} - \int d^3x \blue{\left[\hat{\mathbf{E}}_{\rm F}(\mathbf{x})\right]_{\parallel}}\cdot \hat{\bm{\mathcal{P}}}_{\parallel}(\mathbf{x}),
\end{equation}
where
\begin{equation}
    \hat{V}_{\blue{g}} = \exp{\left[-\frac{i}{\hbar}\int d^3x \hat{\bm{\mathcal{P}}}^{\blue{g}}(\mathbf{x}) \cdot \hat{\mathbf{A}}_{\perp}(\mathbf{x})\right]}.
\end{equation}
Note that $\hat{V}_{\blue{g}} = \hat{W}_{g, \text{C}}$; this transformation takes an operator from its Coulomb gauge representation to a generic one.

Similar to the analysis in the case of material truncation, one can show that the transformation induced by $\hat{V}_{\blue{g}}$ relies on the operator relationship $[\hat{\mathbf{A}}_{\perp}(\mathbf{x}),\hat{\mathbf{\Pi}}(\mathbf{x'})] = i\hbar {\bm \delta}^{\perp}(\mathbf{x}-\mathbf{x'})$, which requires a \emph{complete set of transverse modes} to expand the photonic operators in. As such, if the number of modes included in the system Hamiltonian is to be truncated naively, a gauge tranformation in the reduced space can no longer be implemented as a unitary evolution, violating gauge invariance. Equivalently, this can be seen as not properly introducing coupling between the truncated subspaces of the system consistently~\cite{Taylor2022Mar}. Truncation of the Fock space photon number also breaks gauge invariance in this manner, although we focus on the case of mode truncation in this work. 

To be explicit, consider a mode projection operator $\hat{P}_{\rm M}$, that satisfies
\begin{equation}
    \hat{P}_{\rm M}\hat{\mathbf{A}}_{\perp}(\mathbf{x})\hat{P}_{\rm M} = \ahatbt_{\perp}(\mathbf{x}).
\end{equation}
The Hamiltonian which retains gauge invariance under mode truncation is then simply Eq.~\eqref{eq:genH}, but with $\hat{V}_{\blue{g}} \rightarrow \hat{\mathcal{V}}_{\blue{g}}$, and $\hat{\mathcal{V}}_{\blue{g}}$ is $\hat{V}_{\blue{g}}$ evaluated in terms of $\ahatbt_{\perp}$ instead of $\hat{\mathbf{A}}_{\perp}$.

To give a concrete example, let us consider a modal expansion for the transverse vector potential:
\begin{equation}\label{eq:aexp}
    \ahatbt_{\perp}(\mathbf{x}) = \sum_{\mu} \sqrt{\frac{\hbar}{2\epsilon_0 \chi_{\mu \mu}}} \mathbf{f}_{\mu}(\mathbf{x}) \hat{a}_{\mu} + \text{H.c.},
\end{equation}
where we denote the mode expansion over a \emph{finite} sum of ``relevant'' modes with transverse mode profiles $\mathbf{f}_{\mu}(\mathbf{x})$ and annihilation (creation) operators $\hat{a}_{\mu}$ ($\hat{a}^{\dagger}_{\mu}$). 
We can then define the projection operator as
\begin{equation}\label{eq:pmdef}
\hat{P}_{\rm M} = \bigotimes_{\mu}\sum_{n_{\mu}=0}^{\infty}\ket{n_{\mu}}\bra{n_{\mu}}.
\end{equation}
 Subsequently, the bosonic Hamiltonian is
\begin{align}
    \hat{\mathcal{H}}_{\rm F}&= \int d^3x \! \int_0^{\infty} \! \! \! d\omega \omega \hat{P}_{\rm M} \hat{\mathbf{b}}^{\dagger}(\mathbf{x},\omega) \hat{P}_{\rm M} \cdot \hat{P}_{\rm M} \hat{\mathbf{b}}(\mathbf{x},\omega)\hat{P}_{\rm M} \nonumber \\ &= \sum_{\mu\nu} \hbar \chi_{\mu\nu} \hat{a}^{\dagger}_{\mu}\hat{a}_{\nu}.
\end{align}

In Appendix~\ref{sec:appA}, we give more details on the construction of these discrete modes from the continuum, and their relationship to the Hermitian matrix $\chi_{\mu\nu}$. It is important to note that the mode functions $\mathbf{f}_{\mu}$ are not the usual normal mode solutions to the Helmholtz equation, but rather nonorthogonal transverse modal expansion functions which satisfy, if the truncation is not applied,
\begin{equation}\label{eq:complete}
\sum_{\mu\nu} \frac{\chi_{\mu \nu}}{\sqrt{\chi_{\mu \mu} \chi_{\nu \nu}}} \mathbf{f}_{\mu}(\mathbf{x})\mathbf{f}^*_{\nu}(\mathbf{x'}) = {\bm \delta}^{\perp}(\mathbf{x}-\mathbf{x'}),
\end{equation}
as shown in Appendix~\ref{sec:appA}. 

Note that while throughout we refer to these as ``modes'', they are, more generally, a truncation of the spatial and frequency-dependent degrees of freedom of the electromagnetic fields and passive medium reservoir fields. Specifically, the truncation process involves a projection of a spatial and frequency-dependent orthonormal basis onto the polariton operators $\hat{\mathbf{b}}(\mathbf{x},\omega)$, $\hat{\mathbf{b}}^{\dagger}(\mathbf{x},\omega)$, and need not necessarily satisfy the Helmholtz equation with appropriate boundary conditions---although for truncation to be a useful approximation technique, this is presumed to be the case.
A consequence of this is that $\hat{\mathcal{H}}_{\rm F}$ is not diagonal with respect to the finite mode basis (an effect known from, e.g., quantized QNMs~\cite{frankequantization, franke2020quantized}, quasi-modes~\cite{Dalton1999Jul}, as well as supermodes in quantum nonlinear optics~\cite{Onodera2022Mar}).
This is why, in contrast to the case of material truncation, we define the mode truncation with respect to the field expansion itself, and not the field Hamiltonian; in the case of a nondispersive and nonabsorbing medium, it is possible to truncate with respect to the true normal modes of the medium, and both approaches are then equivalent. In Appendix~\ref{sec:appB}, we discuss this case in more detail.

As an important example, in Appendix~\ref{sec:appA}, we show how the discrete modes can be chosen to correspond over a restricted region of space to QNMs---although for this case the completeness relation~\eqref{eq:complete} does not apply directly, as the expansion is only valid over a  spatial region where the QNMs form a well-behaved basis for the transverse Green's function. Also note that even in a dielectric medium with permittivity that is real and independent of frequency, it is often useful to use mode expansions which are not the exact ``true modes'' of the entire system. For example, quasi-modes~\cite{Dalton1999Jul}, which use an artificial permittivity to obtain mode functions which represent an idealized version of the system of interest, and QNMs, where the open-boundary conditions lead to non-Hermitian eigenvalues even without dispersion or absorption.

\begin{figure*}[ht]
    \centering
    \includegraphics[width=1.64\columnwidth]
    {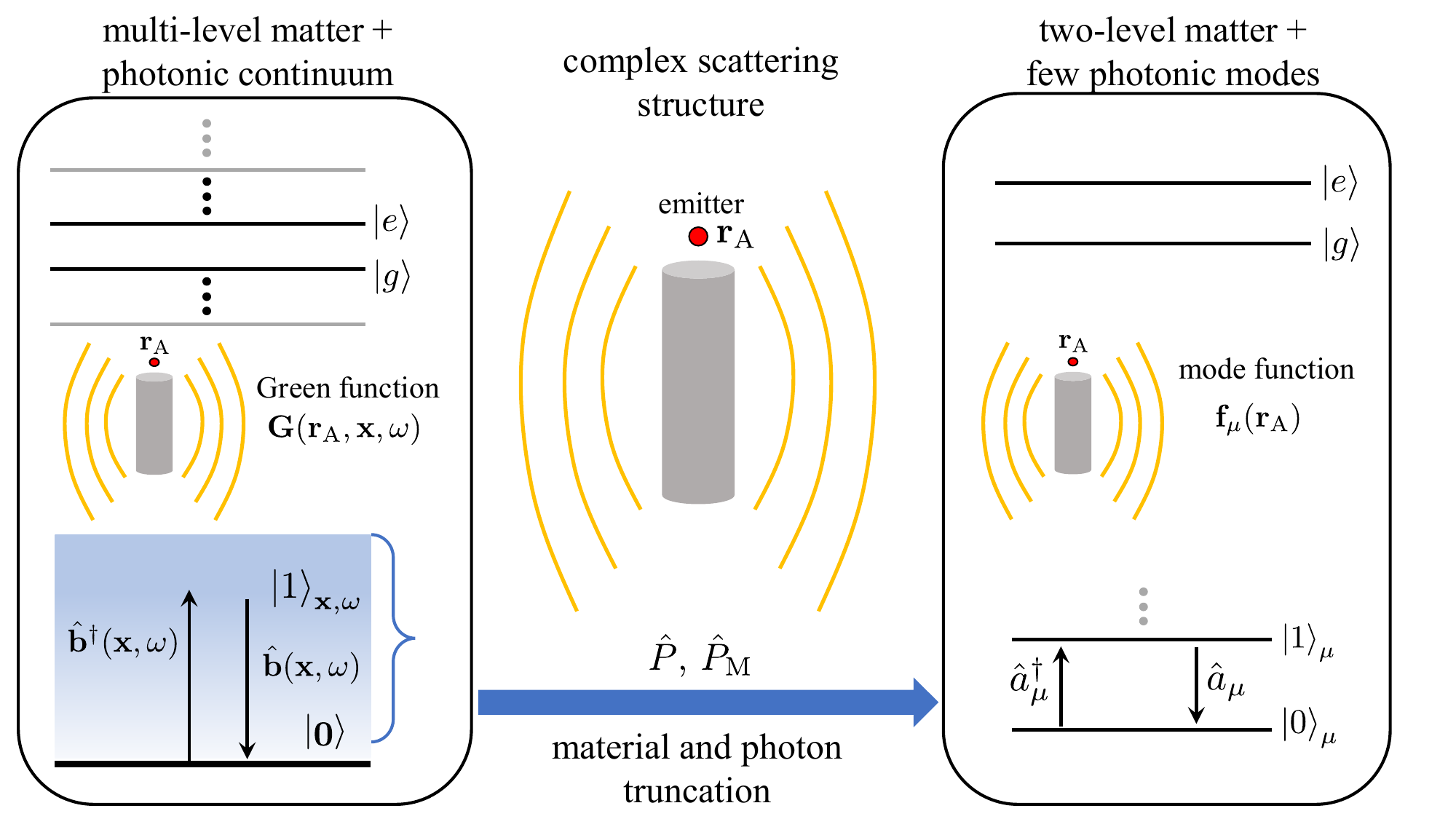}
    \caption{Visualization of the photon and material truncation. An exemplary system consisting of a dielectric rod and a point-like emitter can generally be described quantum mechanically by a multi-level electronic system interacting with a continuum of photon modes (left), which reflect hybridized states of the vacuum electric field and the passive dielectric material. The electromagnetic field depends on a three-dimensional integration over the photon Green's function. The proper truncation of the material and photon degrees of freedom allows for a description of the quantum system with few energy levels (right), while still preserving gauge invariance. The electromagnetic field is obtained from the optical mode functions.
    }\label{fig:2}
\end{figure*}

Applying the correctly-truncated unitary transform to the bosonic Hamiltonian, $\hat{\mathcal{V}}_{\blue{g}}\hat{a}_{\mu}\hat{\mathcal{V}}_{\blue{g}}^{\dagger} = \hat{a}_{\mu} - i \hat{\xi}^{\blue{g}}_{\mu}$, we obtain
\begin{align}\label{eq:80}
    \hat{\mathcal{V}}_{\blue{g}}\hat{\mathcal{H}}_{\rm F}\hat{\mathcal{V}}^{\dagger}_{\blue{g}} =& \hat{\mathcal{H}}_{\rm F} + \left(i\sum_{\mu \nu} \hbar\chi_{\mu\nu}^*\hat{a}_{\mu}\hat{\xi}^{\blue{g}\dagger}_{\nu} + \text{H.c.}\right) \nonumber \\ & + \sum_{\mu\nu}\hbar \chi_{\mu \nu} \hat{\xi}^{\blue{g}\dagger}_{\mu}\hat{\xi}^{\blue{g}}_{\nu},
\end{align}
where 
\begin{equation}
    \hat{\xi}^{\blue{g}}_{\mu} = \sum_{\alpha}\frac{q_{\alpha}}{\sqrt{2\epsilon_0 \hbar \chi_{\mu\mu}}} \int d^3x \mathbf{K}^{\blue{g}}_{\perp}(\mathbf{x},\hat{\mathbcallc{r}}_{\alpha}) \cdot \mathbf{f}^{*}_{\mu}(\mathbf{x}).
\end{equation}
The second term in Eq.~\eqref{eq:80} gives the transverse coupling between photonic and material subspaces, and can also be written as 
\begin{equation}
     \left(i\sum_{\mu \nu} \hbar\chi_{\mu\nu}^*\hat{a}_{\mu}\hat{\xi}^{\blue{g}\dagger}_{\nu} + \text{H.c.}\right) = -\int d^3x \blue{\left[\hat{\bm{\mathcal{E}}}_{\rm F}(\mathbf{x})\right]_{\perp}} \cdot \hat{\bm{\mathcal{P}}}^{\blue{g}}_{\perp}(\mathbf{x}),
\end{equation}
where $\blue{\left[\hat{\bm{\mathcal{E}}}_{\rm F}\right]_{\perp}}$ is the part of the \emph{correctly mode-truncated} transverse electric field operator that can be expressed in terms of the bosonic operators:
\begin{equation}\label{eq:eTcorrect}
   \blue{\left[\hat{\bm{\mathcal{E}}}_{\rm F}(\mathbf{x})\right]_{\perp}}= i\sum_{\mu} \sqrt{\frac{\hbar\chi_{\mu\mu}}{2\epsilon_0}}\mathbf{f}'_{\mu}(\mathbf{x}) \hat{a}_{\mu} + \text{H.c.},
\end{equation}
and $\mathbf{f}'_{\mu} = \sum_{\nu}\frac{\chi_{\mu\nu}^*}{\sqrt{\chi_{\mu\mu}\chi_{\nu\nu}}}\mathbf{f}_{\nu}$. The quantum system after material and photon truncation is visualized in Fig.~\ref{fig:2}. 

In a general gauge, the full correctly-truncated transverse electric field operator can be found from $\hat{\bm{\mathcal{E}}}^{\blue{g}}_{\perp} = \hat{\mathcal{V}}_{\blue{g}}[\hat{\bm{\mathcal{E}}}_{\rm F}]_{\perp}\hat{\mathcal{V}}^{\dagger}_{\blue{g}}$. One can show that this is the correctly-truncated form of the transverse electric field operator 
%can be found 
by enforcing $\hat{\bm{\mathcal{E}}}^{\blue{g}}_{\perp} = -\frac{\partial}{\partial t}\ahatbt_{\perp}$, and applying the Heisenberg equation of motion. Of course, we could also define a mode expansion initially with respect to the transverse electric field, however, \blue{this would violate gauge invariance under mode truncation and not properly constrain interactions to the few-mode subspace.}

The fact that the correctly-truncated transverse electric field operator is expanded in terms of mode profiles $\mathbf{f}_{\mu}'(\mathbf{x})$ which are a linear combination of the mode profiles (in the truncated basis) for the vector potential $\mathbf{f}_{\mu}(\mathbf{x})$ is a fundamental feature of dissipation, and stands in contrast to the case of a normal mode expansion, presented in Appendix~\ref{sec:appB}. Nonetheless, under an assumption of well-separated discrete modes (e.g., high $Q$-factor resonators), the different modal expansions can be related to each other, allowing the transverse electric field to be expanded in the usual form, which we discuss in Appendix~\ref{sec:appA}.

The most general arbitrary-gauge Hamiltonian is then
\begin{equation}\label{eq:hmode}
    \hat{\tilde{\mathcal{H}}}^\blue{g} = \blue{\hat{\tilde{\mathcal{V}}}_g}\hat{\mathcal{H}}_{\rm F}\blue{\hat{\tilde{\mathcal{V}}}_g^{\dagger}} + \blue{\hat{\tilde{\mathcal{U}}}_g}\hat{\mathcal{H}}_0\blue{\hat{\tilde{\mathcal{U}}}_g^{\dagger}} - \int d^3x \blue{\left[\hat{\mathbf{E}}_{\rm F}(\mathbf{x})\right]_{\parallel}} \cdot \hat{\bm{\mathcal{P}}}_{\parallel}(\mathbf{x}),
\end{equation}
where we use a tilde to denote explicitly quantities with both photonic space truncation as well as material truncation. \blue{Equation~\eqref{eq:hmode} is, as expected, consistent with previous work on restoring gauge invariance under material and mode truncation~\cite{DiStefano2019Aug, Taylor2020Sep,Taylor2022Mar}.} Note that no loss of gauge invariance occurs if the longitudinal part of the medium-assisted electric field $\blue{\left[\hat{\mathbf{E}}_{\rm F}(\mathbf{x})\right]_{\parallel}}$ is truncated (i.e., the part that belongs to the photonic subspace). If the transverse mode expansion is complete, in that it satisfies Eq.~\eqref{eq:complete}, the untruncated result is recovered. Note also that we can write the first two terms on the right hand side of Eq.~\eqref{eq:hmode} as $\hat{\tilde{\mathcal{W}}}_{g, \text{C}}\hat{\mathcal{H}}_{\rm F} \hat{\tilde{\mathcal{W}}}^{\dagger}_{g, \text{C}} +\hat{\tilde{\mathcal{W}}}_{g, \text{mp}}\hat{\mathcal{H}}_0 \hat{\tilde{\mathcal{W}}}^{\dagger}_{g, \text{mp}}$, which clearly indicates the special role that the Coulomb and multipolar gauges
play in the the theory of gauge invariance in truncated optical systems.

These expressions can be simplified in the Coulomb and multipolar gauges by noting that for the Coulomb gauge, we have $\hat{\mathcal{V}}_{\blue{\rm C}} = 1$, and in the multipolar gauge, 
\begin{equation}
 \int d^3x  \mathbf{K}^{\blue{\rm mp}}_{\perp}(\mathbf{x},\mathbcallc{\hat r}_{\alpha}) \cdot \mathbf{f}_{\mu}(\mathbf{x}) = - \mathbcallc{\hat r}_{\alpha} \cdot \int_0^{1} ds \mathbf{f}_{\mu}(s\mathbcallc{\hat r}_{\alpha}).
\end{equation}
A gauge transformation in the truncated photonic space is, similar to Eq.~\eqref{eq:gauge_tm},
\begin{equation}
    \hat{\tilde{\mathcal{W}}}_{g'g} = \exp{\left[-\frac{i}{\hbar} \!\int \! \! d^3x \left[\hat{\bm{\mathcal{P}}}^{g'}_{\perp}(\mathbf{x})-\hat{\bm{\mathcal{P}}}^{g}_{\perp}(\mathbf{x})\right] \! \cdot \! \ahatbt_{\perp}(\mathbf{x})\right]},
\end{equation}
which again can be shown to be equivalent to a replacement of the truncated vector potential and transverse polarization.
Of course, all the results in this section can easily be generalized to consider the case of no material truncation by taking $\mathbcallc{\hat r}_{\alpha} \rightarrow \hat{\mathbf{r}}_{\alpha}$.

The difference between Eq.~\eqref{eq:hmode}, the correctly mode-truncated arbitrary-gauge Hamiltonian, and a naively truncated Hamiltonian $\hat{P}_{\rm M}\hat{\mathcal{H}}^{\blue{g}}\hat{P}_{\rm M}$ is twofold: Firstly, the second term in Eq.~\eqref{eq:80}, the interaction term between photonic and material subspaces, is only expressed in terms of the correctly-truncated electric field if the Hamiltonian is properly (not directly) truncated, by instead truncating directly the vector potential as the fundamental field coordinate.

Secondly, the final term in Eq.~\eqref{eq:80} differs from the final term in Eq.~\eqref{eq:arbHt} in that it contains explicit reference to the electromagnetic field via the mode functions $\mathbf{f}_{\mu}$~\cite{Taylor2022Mar}. In the case where truncation is applied after calculating the unitary transformation induced by $\hat{V}$, the transverse delta function in the commutator~\eqref{eq:commA} (which requires a complete set of transverse modes) is used to remove any reference to the field modes. The additional integration present in the properly truncated case resolves issues with the $\hat{\bm{\mathcal{P}}}_{\perp}^2$ term when  \blue{using} the multipolar gauge, where $\hat{\bm{\mathcal{P}}}^{\blue{\rm mp}}_{\perp}$ is expressed in terms of a Dirac delta functions, which can cause problems due to the presence of a product of distributions~\cite{Woolley2020Feb}.

It is instructive to compare our modal expansion and truncation in a fundamentally lossy and dispersive system, with the more commonly employed case in quantum optics of a \emph{normal mode} expansion. To focus on the case of light-matter interactions in a medium, we can consider an inhomogeneous dielectric with a real and independent of frequency permittivity $\epsilon(\mathbf{x})$. In this case, the Helmholtz equation can be used to calculate the true normal modes of this system as the eigenfunctions of ${\bm \nabla} \times {\bm \nabla} \times \mathbf{h}_{\mu} -\frac{\omega_{\mu}^2}{c^2}\epsilon(\mathbf{x})\mathbf{h}_{\mu} =0$, with appropriate closed or periodic boundary conditions. The modal eigenfunctions here are $\mathbf{h}_{\mu}(\mathbf{x})$ with eigenvalue $\omega_{\mu}$, and they are generalized transverse in that they satisfy ${\bm \nabla} \cdot (\epsilon \mathbf{h}_{\mu}) =0$. In this case, an approximate (macroscopically-averaged) Lagrangian can be used with no reference to the medium oscillator fields. In Appendix~\ref{sec:appB}, we quantize this system using Dirac's constrained quantization procedure for the important case of the generalized Coulomb gauge, which satisfies ${\bm \nabla} \cdot (\epsilon \mathbf{A}) =0$, and the generalized multipolar gauge which we obtain by PZW transformation. The main result is that the generalized multipolar Hamiltonian $\hat{\mathcal{H}}_{\rm gmp}$ can be written as 
\begin{align}\label{eq:ghmp}
&    \hat{\mathcal{H}}_{\rm gmp} = \hat{\mathcal{H}}_0 + \sum_{\mu} \hbar\omega_{\mu}\hat{a}^{\dagger}_{\mu}\hat{a}_{\mu}\nonumber \\ & - \int \! \!  d^3x  \hat{\bm{\mathcal{{E}}}}_{\blue{{\rm F}}}(\mathbf{x}) \cdot\hat{\bm{\mathcal{Z}}}_{\rm A}(\mathbf{x})  + \sum_{\mu}\frac{\left[\int d^3x \hat{\bm{\mathcal{Z}}}_{\rm A}(\mathbf{x}) \cdot \mathbf{h}_{\mu}(\mathbf{x})\right]^2}{2\epsilon_0},
\end{align}
where the part of the electric field operator which can be expressed in terms of bosonic normal mode operators is $\hat{\bm{\mathcal{{E}}}}_{\blue{{\rm F}}}(\mathbf{x})=i\sum_{\mu}\sqrt{\frac{\hbar\omega_{\mu}}{2\epsilon_0}}\mathbf{h}_{\mu}(\mathbf{x})\hat{a}_{\mu} + \text{H.c.}$, and for each mode with profile $\mathbf{h}_{\mu}$, we can associate creation and annihilation operators which satisfy $[\hat{a}_{\mu},\hat{a}^{\dagger}_{\nu}] = \delta_{\mu\nu}$. It is important to note that the full electric field operator also contains contributions from the material system operators, which must be considered when calculating physical observables; for details, see Appendix~\ref{sec:appB}.

One  important difference between the generalized multipolar Hamiltonian~\eqref{eq:ghmp} \blue{for a system supporting normal modes} and the multipolar Hamiltonian~\eqref{eq:hmode}, \blue{which can be formulated also for systems with lossy and nondiagonal mode expansions}, is that the interaction term between photonic and material subspaces takes a form which only includes direct mode-polarization couplings in the former case, where loss can be neglected. As a result, a naive truncation of the generalized multipolar Hamiltonian (by applying $\hat{P}_{\rm M}$ operators directly to $\blue{\hat{H}^{\rm mp}}$) only differs from the correct result in the polarization-squared term, which does not couple to the photonic subspace. In many cases (e.g., a fermionic TLS; see Sec.~\ref{ssec:tls}), this term is irrelevant, or otherwise neglected. The ambiguity associated with mode truncation thus does not always play a significant role in light-matter interactions, at least with respect to the Hamiltonian. Note however, that the generalized transverse eigenmodes are found with respect to the entire dielectric system, and thus are generally de-localized and not appropriate for a discrete resonant mode truncation, despite this procedure being common in the literature (i.e., taking a ``single-mode'' limit). 

In contrast, when considering a truncation of lossy-modes, due to the cross-mode nature of the coupling terms in the second term of Eq.~\eqref{eq:80}, truncation in the multipolar gauge will invariably break gauge invariance in a way which can have non-negligible consequences. \blue{Our work generalizes previous results by Ref.~\cite{Taylor2022Mar} on restoring gauge invariance in truncated normal mode systems to consider these more general lossy mode expansions.} One \blue{important example} of this can be seen by considering open quantum systems. 

\blue{To be concrete, consider the case of a cavity resonator that has discrete mode operators which couple to external fields and thus exhibits photon loss with some decay rate. Since there is only one set of electromagnetic fields (i.e., of the universe), we propose that a rigorous model for the dynamics of the cavity field, where the other ``reservoir'' degrees of freedom are traced out (e.g., in the form of a master equation, using the well-known input-output formalism~\cite{Gardiner1985Jun}), requires the use of the Coulomb gauge. This is because, in a generic gauge, the ``reservoir'' field which photons decay into is described by an entangled state of bosonic and fermionic subspaces, which prevents the use of standard open quantum system techniques to trace out the reservoir subsystem. However, the Coulomb gauge is \emph{unique} in that it is the only gauge that satisfies $\hat{\mathcal{V}} = 1$. Thus, the unitary transformation which transforms the field Hamiltonian $\hat{H}_{\rm F}$, and mixes up field and material degrees of freedom, becomes trivial in this gauge.  The Coulomb gauge is thus the unique gauge (within the gauges realizable in the quantization function method) wherein the reservoir degrees of freedom are constrained entirely within a bosonic subspace, which is necessary to perform a Born-Markov approximation and derive a master equation.

It should of course be noted that one can  perform, for example, a PZW-like transformation by \emph{only} using the cavity mode degrees of freedom, to get something resembling the multipolar gauge for a truncated mode system. This is equivalent to defining a new class of gauge transformations which act only on the reduced system degrees of freedom.

In this case, the reduced cavity system is by definition a mode-truncated system. Thus, the potential gauge ambiguity related to mode truncation, and the techniques described in this work to restore gauge invariance under mode truncation, are \emph{intrinsic} to open quantum systems. The Hamiltonian in Eq.~\eqref{eq:hmode} \blue{(or an analogous construction)} is generally required as a starting point to derive the correct model of open quantum system dynamics involving lossy cavities, if the system-reservoir coupling is to be derived rigorously (e.g., as in quantized QNM theory~\cite{franke2020quantized}). 

Recently, it has been argued~\cite{Stokes2022Nov} that is is incorrect to assume that to model an open quantum system (e.g., a lossy cavity), the 
system operator that should couple to external ``reservoir'' modes is the transverse electric field operator, and implied that previous work by some of us~\cite{Salmon2022Mar} is incorrect as a result of this. This argument relies on the fact that, when matter degrees of freedom (e.g., ``atoms'' in the cavity) are localized far from the boundary of the resonator system, they should not play a role in the dissipation process of the cavity. Indeed, this is the case, and this feature is generally observed in rigorous models of loss from quantized cavity systems (e.g.,~\cite{dutra2000quantized,Viviescas2003Jan,Lentrodt2020Jan,frankequantization}). It is a straightforward consequence of this that the electric field operator (or specifically, its modal excitation and de-excitation operators) is in fact the correct operator to use in describing coupling to a reservoir system, as it only consists of optical degrees of freedom. Other works rigorously treating dissipation in open cavities (albeit neglecting dispersion and absorption) have come to the same conclusion~\cite{Viviescas2003Jan,Lentrodt2020Jan}.

Where the potential confusion arises in Ref.~\cite{Stokes2022Nov}, is that, when a complete set of modes is involved (no truncation), the arbitrary-gauge expression for the untruncated electric field operator is equivalent to that of its component which only consists of bosonic operators, for positions far away from the matter degrees of freedom (i.e., the cavity boundary). This led the authors to suggest that the electric field evaluated away from the material degrees of freedom should be used to couple to the external reservoir, and that this field should be expanded using only boson operators.  In a \emph{correctly mode-truncated} picture, however, the electric field operator becomes $\hat{\bm{\mathcal{E}}}^{\blue{g}}_{\perp} = \hat{\mathcal{V}}_{\blue{g}}[\hat{\bm{\mathcal{E}}}_{\rm F]}]_{\perp}\hat{\mathcal{V}}^{\dagger}_{\blue{g}}$, which can not exclusively be described by bosonic degrees of freedom away from the material particle locations unless a complete set of modes is used to describe the cavity field. Using a single mode necessarily collapses the spatial degrees of freedom of the description of the electric field to a single coordinate, and thus the field at the boundary of the cavity and at the location of the material degrees of freedom are described by the same modal expansion operators. 
This
was already pointed out in~\cite{PhysRevResearch.3.023079} (especially see Appendix A
of same paper). Thus, by considering the correctly truncated
form of the field operator, the intuition behind the argument of
Ref.~\cite{Stokes2022Nov} does not hold, and consequently the gauge-invariant
observables reported in~\cite{Salmon2022Mar} are correct within the assumptions of the model considered.

\section{Time-dependent gauge transformations}
\label{sec:time-dep}
In this section, we generalize our approach to consider time-dependent gauge conditions and transformations, as well as how to construct gauge-invariant phenomenological time-dependent models of light-matter interaction.

We can extend the previously developed theory of arbitrary gauge quantization by allowing the transverse part of the quantization function to be an explicit function of time, such that \blue{(dropping the explicit $g$ index for now)} $\mathbf{K} = \mathbf{K}(\mathbf{x},\mathbf{x'},t)$. In this case, one can follow the development in prior sections in the exact same way by taking the system to be quantized at a definite time $t_0$, and then determining the explicit time dependence of any observables when calculating expectation values~\cite{Gitman1990,sundermeyer1982constrained}.

Alternatively, we can account for the explicit time dependence of observables that arises from the time-dependent gauge condition by means of an additional time-dependent Hamiltonian term, generated by a canonical transformation prior to quantization\blue{, and then treating all operators as having no explicit time-dependence.} This is analogous to the approach in Ref.~\cite{PhysRevResearch.3.023079}, where, \emph{after} quantization, a unitary transformation with explicit time-dependence is introduced to define a time-dependent gauge transformation. Here, we generalize this approach by introducing it before quantization, and for arbitrary gauges.

Specifically, a modification to the Hamiltonian arises due to the canonical transformation that takes the coordinates $(\mathbf{A},\mathbf{\Pi}_{\mathbf{A}})$ to $(\mathbf{A}_{\perp},\mathbf{\Pi})$. With time-independent constraints, this transformation does not alter the Hamiltonian. With time-dependent constraints, however, $\mathbf{P}(\mathbf{x})$ becomes explicitly time-dependent, and more care is needed to perform the canonical transformation to the unconstrained variables~\cite{Gitman1990}.

A canonical transformation from the set of constrained variables  $(\mathbf{A},\mathbf{\Pi}_{\mathbf{A}})$ to the unconstrained variables $(\mathbf{A}_{\perp},\mathbf{\Pi})$ can be implemented by writing the Lagrangian expressed in terms of the constrained variables as equal to a new Lagrangian expressed in terms of the unconstrained variables, up to a total time derivative which does not affect the extremization of the action:
\begin{align}
    \int d^3x &\dot{\mathbf{A}} \cdot \mathbf{\Pi}_{\mathbf{A}} - H(\mathbf{A},\mathbf{\Pi}_{\mathbf{A}},t) = \nonumber \\ &  \int d^3x \dot{\mathbf{A}}_{\perp} \cdot \mathbf{\Pi} - H'(\mathbf{A}_{\perp},\mathbf{\Pi},t) + \frac{d G}{dt}.
\end{align}
Here, we have used a condensed notation by suppressing functional dependencies of the fields as well as the dependence of the Hamiltonians $H$ and $H'$ on the variables unaffected by the desired transformation.
One way to implement this canonical transformation is by using a type-2 generating function~\cite{goldstein:mechanics}:
\begin{equation}
    G = -\int d^3x \mathbf{A}_{\perp} \cdot \mathbf{\Pi} + \int d^3x G_2(\mathbf{A},\mathbf{\Pi},t).
\end{equation}

For Hamilton's equations to be preserved in the unconstrained variables, one requires then $\frac{ \partial G_2}{\partial \mathbf{A}} = \mathbf{\Pi}_{\mathbf{A}}$, $\frac{\partial G_2}{\partial \mathbf{\Pi}} = \mathbf{A}_{\perp}$, and $H' = H + \int d^3x\frac{\partial G_2}{\partial t}$. Clearly, this can be accomplished by a form $G_2 = \mathbf{A} \cdot \mathbf{\Pi}_{\mathbf{A}} + \mathbf{\Pi} \cdot \mathbf{A}_{\perp}$, which we can express as a function of $\mathbf{A}$ and $\mathbf{\Pi}$ as
\begin{equation}
G_2(\mathbf{A},\mathbf{\Pi},t) = \mathbf{A} \cdot \left[ \mathbf{\Pi} + \mathbf{P}(t)\right] + \mathbf{\Pi} \cdot \int d^3x' {\bm \delta}^{\perp}(\mathbf{x}-\mathbf{x'}) \cdot \mathbf{A}(\mathbf{x'}),
\end{equation}
where we have emphasized that the explicit time dependence comes from (the transverse part of) $\mathbf{P}$. From this, it is easy to determine the relation
\begin{align}
    &H'(t) - H(t)  \nonumber \\ &= \int d^3x \frac{\partial G_2(\mathbf{A},\mathbf{\Pi},t)}{\partial t} \nonumber \\ 
    &=-\int d^3x \int d^3x' \mathbf{A}_{\perp}(\mathbf{x},t) \cdot\left[ \frac{\partial}{\partial t} \mathbf{K}_{\perp}(\mathbf{x},\mathbf{x'},t)\right]\rho_{\rm A}(\mathbf{x'},t).
\end{align}

Upon quantization, this term becomes simply $\int \hat{\mathbf{A}}_{\perp} \cdot \dot{\hat{\mathbf{P}}}_{\perp}$, and so the only change in the case of time-dependent gauge function is
\begin{equation}\label{eq:Hadditional}
    \hat{H} \rightarrow \hat{H}(t) = \hat{H} + \int d^3x \hat{\mathbf{A}}_{\perp}(\mathbf{x}) \cdot \dot{\hat{\mathbf{P}}}_{\perp}(\mathbf{x},t).
\end{equation}
This additional term added to the Hamiltonian is equivalent to the gauge transformation $\hat{\phi} \rightarrow \hat{\phi} - \dot{\hat{\Lambda}}$, since this transforms $\blue{\left[\hat{\mathbf{E}}_{\rm F}\right]_{\parallel}} \rightarrow \blue{\left[\hat{\mathbf{E}}_{\rm F}\right]_{\parallel}}+ \dot{\hat{\mathbf{A}}}_{\parallel}$ \blue{(i.e., if $\hat{\mathbf{A}}_{\parallel} = {\bm \nabla} \hat{\Lambda}$)}, which when applied to the longitudinal interaction term in Eq.~\eqref{eq:hmode}, and employing the quantization constraint relation in Eq.~\eqref{eq:c2}, generates the term in~\blue{Eq.~\eqref{eq:Hadditional}}. Thus, a broad class of gauge transformations as defined in Eqs.~\eqref{eq:gauge1} and~\eqref{eq:gauge2} can be implemented within the arbitrary gauge quantization theory.

The notion of gauge \emph{transformations} also must be modified to account for the explicit time dependence of the gauge transformation function, $\hat{W}_{g'g}(t)$. Here we shall consider the untruncated case for notational simplicity, although the procedure is identical with truncation of the material and/or photonic subspaces.

It is simple to show that the form of the Schr\"odinger evolution is conserved under a gauge transformation $\ket{\psi_g'(t)} = \hat{W}_{g'g}(t)\ket{\psi_g(t)}$, provided the Hamiltonian changes as \blue{(restoring gauge indices)}
\begin{equation}\label{eq:td1}
    \hat{H}^{\blue{g'}}(t) = \hat{W}_{g'g}(t)\hat{H}^{\blue{g}}(t)\hat{W}^{\dagger}_{g'g}(t) - i\hbar \hat{W}_{g'g}(t) \frac{\partial}{\partial t}\hat{W}^{\dagger}_{g'g}(t).
\end{equation}
Using the definition of $\hat{W}_{g'g}(t)$ (Eq.~\eqref{eq:wdef}, but with $\hat{\Lambda}_g(\mathbf{x}) \rightarrow \hat{\Lambda}_g(\mathbf{x},t)$), one can show the second term in the above equation then takes the form
\begin{align}\label{eq:td2}
    - i\hbar &\hat{W}_{g'g}(t) \frac{\partial}{\partial t}\hat{W}^{\dagger}_{g'g}(t) = \nonumber \\ & \int d^3x \left[\dot{\hat{\mathbf{P}}}^{g'}_{\perp}(\mathbf{x},t) - \dot{\hat{\mathbf{P}}}^{g}_{\perp}(\mathbf{x},t)\right] \cdot \hat{\mathbf{A}}_{\perp}(\mathbf{x}).
\end{align}

Reference~\cite{PhysRevResearch.3.023079} suggests that the Coulomb gauge is somehow more fundamental than the multipolar gauge, and must be used when time-dependent interactions are to be considered. Here, we take the view that gauge symmetry is a fundamental property of the QED Lagrangian, and thus any gauge should give consistent results, provided any approximations to the theory are implemented consistently across gauges.

When describing time-dependent interactions and/or gauge conditions (as in Refs.~\cite{PhysRevResearch.3.023079,Stokes2021Feb}), however,  the Coulomb gauge is potentially \emph{unique} in that it is defined, conveniently, by $\mathbf{K}^{\blue{\rm C}}_{\perp}(\mathbf{x},\mathbf{x'},t) =0$ for all $t$, and thus is a time-invariant gauge. In constrast, when the light-matter interaction strength is given an explicit time-dependence, the multipolar gauge constraint condition itself should also be given this explicit time dependence to get something similar to the usual form of the multipolar Hamiltonian. Thus, the additional term (Eq.~\eqref{eq:Hadditional}) arises, which does not exist in the  time-independent multipolar gauge. 

It should also be noted that an explicitly time-dependent Hamiltonian is always an approximation, since the energy non-conserving nature of a time-dependent Hamiltonian means some external system has dynamics which are not explicitly modelled. Thus, choosing to impose time-dependence in a specific gauge can potentially lead to different results than if it is imposed in other gauges, and analysis of the physical origin of the time-dependence may need to be taken to resolve this potential ambiguity~\cite{Stokes2021Feb}.

For the case of time-dependent coupling between the transverse field and matter, we can introduce this in a way which is unambiguous and equivalent in both the Coulomb and multipolar gauges as follows: in the Coulomb gauge, the minimal coupling is introduced via the transverse part of $\hat{\mathbf{Z}}_{\rm A}$ (see Eq.~\eqref{eq:z_def}). Thus, we can\blue{, for example,}  modulate the transverse field's interaction strength with matter by phenomenologically modulating this parameter by a time-dependent factor $\mu(t)$ such that $\hat{\mathbf{Z}}_{\rm A}(\mathbf{x}) \rightarrow \mu(t)\hat{\mathbf{Z}}_{\rm A}(\mathbf{x})$.
In the multipolar gauge, the coupling is entirely mediated by the transverse polarization $\hat{\mathbf{P}}^{\rm mp}_{\perp}(\mathbf{x})$. Thus, we can make the equivalent approximation by taking $\hat{\mathbf{P}}^{\rm mp}_{\perp}(\mathbf{x}) \rightarrow \mu(t)\hat{\mathbf{P}}^{\rm mp}_{\perp}(\mathbf{x})$.

Now, in the Coulomb gauge, $\hat{\mathbf{Z}}_{\rm A}(\mathbf{x})$ is simply a function which generates (neglecting the magnetic interaction terms) the minimal coupling replacements $\hat{\mathbf{p}}_{\alpha} \rightarrow \hat{\mathbf{p}}_{\alpha} - q_{\alpha} \hat{\mathbf{A}}_{\blue{\perp}}(\hat{\mathbf{r}}_{\alpha})$. In contrast, in the multipolar gauge, $\hat{\mathbf{P}}^{\rm mp}_{\perp}(\mathbf{x})$ is a parameter which appears in the Lagrangian. As such, by making this quantity time-dependent, we must add to the Hamiltonian the additional term given by Eq.~\eqref{eq:Hadditional}, and understand the gauge condition as depending explicitly on time. \blue{Additionally, one must still make the replacement $[\hat{\mathbf{Z}}_{\rm A}(\mathbf{x})]_{\perp}\rightarrow \mu(t)[\hat{\mathbf{Z}}_{\rm A}(\mathbf{x})]_{\perp}$ to ensure $\hat{U}_{\rm mp} = 1$ even under the new time-dependent gauge condition.

In summary, we can implement time-dependent phenomenlogical interactions between the transverse field and the active material particles in any gauge by letting
\begin{subequations}
\begin{equation}
[\hat{\mathbf{Z}}_{\rm A}(\mathbf{x})]_{\perp}\rightarrow \mu(t)[\hat{\mathbf{Z}}_{\rm A}(\mathbf{x})]_{\perp}
\end{equation}
\begin{equation}
\mathbf{K}_{\perp}^g(\mathbf{x},\mathbf{x'}) \rightarrow \mathbf{K}_{\perp}^{g}(\mathbf{x},\mathbf{x'},t) = \mu(t)\mathbf{K}_{\perp}^g(\mathbf{x},\mathbf{x'}) 
\end{equation}
\begin{equation}
\hat{H} \rightarrow \hat{H} + \dot{\mu}(t)\int d^3x \hat{\mathbf{A}}_{\perp}(\mathbf{x}) \cdot \hat{\mathbf{P}}^g_{\perp}(\mathbf{x})
\end{equation}
\end{subequations}
} 

 \blue{To transform between gauges,} we can perform the time-dependent gauge transformation via $\hat{W}_{\rm C, mp}(t)$ (see Eq.~\eqref{eq:wdef}, following the rule in Eq.~\eqref{eq:td1}) to transform from one gauge to another, and the additional term is correctly accounted for by this transformation. Consequently, one can introduce time-dependent interactions in systems with ultrastrong coupling in either gauge without introducing any ambiguity or violating gauge invariance. Using our formalism of the time-dependent quantization function $\mathbf{K}(\mathbf{x},\mathbf{x'},t)$, this is done without treating any gauge as more fundamental than another, although as noted in Ref.~\cite{PhysRevResearch.3.023079}, this is most easily done in the Coulomb gauge, where the coupling strength can straightforwardly be made time-dependent.

In contrast to Ref.~\cite{Stokes2021Feb}, which claims that introducing time-dependent light-matter interaction strengths necessarily breaks the gauge-invariance of the fundamental light-matter Lagrangian, we note that this is circumvented by applying the time-dependent modulation of the interaction strength only to the transverse part of the interaction, since the transverse vector potential is gauge-invariant. This in fact produces equivalent results to the replacements made at the level of the Coulomb and multipolar gauge Hamiltonians described above, and allows one to introduce gauge-invariant time-dependent couplings in a completely unambiguous way.

Thus ultrastrong time-dependent
light-matter interactions are {\em not} gauge relative, as fully expected, if using a correct and unambiguous theory.
Furthermore, it should also be noted that modulating the longitudinal (quasistatic) interaction
is likely to lead to undesired and unphysical predictions beyond just the breaking of gauge invariance, as this would involve the modulation of the Coulomb forces which bind together the constituent atoms of the matter degrees of freedom. Clearly the desired phenomenological model in many cases should be that of a time-dependent transverse coupling only.

\section{The dipole approximation and two-level systems}\label{sec:dtls}
In this section, we apply the results of Sec.~\ref{sec:quantization} to some commonly used models and approximations in quantum optics. Specifically, we discuss the \emph{dipole approximation} (or long-wavelength approximation) in Sec.~\ref{ssec:dipole}, TLSs in Sec.~\ref{ssec:tls}, and how to go beyond the dipole approximation for effective single-particle models in Sec.~\ref{ssec:beyonddipole}. We focus on the case of the dispersive and absorbing dielectric, but analogous results for the case of a real dielectric, where the fields are expanded in terms of the generalized transverse normal modes of the system (see Appendix~\ref{sec:appB}) can easily be derived by applying the appropriate approximations to the generalized Coulomb and multipolar gauges in Eqs.~\eqref{eq:gc} and~\eqref{eq:gmp}.

\subsection{Dipole Approximation}\label{ssec:dipole}
In the Coulomb and multipolar gauges, the Hamiltonian is in a particularly convenient form to expand the field potential functions around $\mathbf{x} = \mathbf{r}_{\rm A} =0$, the center of the (e.g.) molecular charge distribution. In particular, taking this expansion to first order in $\mathbcallc{\hat r}_\alpha$ results in the \emph{dipole approximation}.
For example, under the dipole approximation, the operator $\hat{\mathcal{U}}_{\blue{g}}$ becomes (with no mode truncation)\blue{~\cite{DiStefano2019Aug,Taylor2020Sep, PhysRevResearch.3.023079}}
\begin{equation}
    \hat{\mathcal{U}}_{\blue{g}} = \exp{\left[\frac{i}{\hbar} \mathbcallc{\hat d} \cdot \hat{\mathbf{A}}_0^{\blue{g}} \right]},
\end{equation}
where we have defined the \emph{dipole operator} $\mathbcallc{\hat d} = \sum_{\alpha}q_{\alpha}\mathbcallc{\hat r}_{\alpha}$.
Similarly, $\blue{\hat{V}_{\rm mp}}$ becomes $\blue{\hat{V}_{\rm mp}} = \exp{\left[-\frac{i}{\hbar}\mathbcallc{\hat d} \cdot \hat{\mathbf{A}}_{\perp,0} \right]}$. Note one could also define a dipole operator for each particle, which is naturally more suited to multi-particle models such as the Dicke and Hopfield models~\cite{Garziano2020Aug}.
In this section we shall use fields with the 0 subscript to correspond to the evaluation at the origin (or, more generally at the center of a charge distribution $\mathbf{x}= \mathbf{r}_{\rm A}$); for example, here $\hat{\mathbf{A}}_0 \blue{\equiv} \hat{\mathbf{A}}(\mathbf{0})$. The Hamiltonian in the Coulomb gauge, after applying the dipole approximation to Eq.~\eqref{eq:Hc} (with no mode truncation), is:
\begin{align}\label{eq:dipoleH}
    \hat{\mathcal{H}}^{\blue{\rm C}}  = & \exp{\left[\frac{i}{\hbar} \mathbcallc{\hat d} \cdot \hat{\mathbf{A}}_{\perp,0} \right]}\hat{\mathcal{H}}_{0}\exp{\left[-\frac{i}{\hbar} \mathbcallc{\hat d} \cdot \hat{\mathbf{A}}_{\perp,0} \right]} \nonumber \\ & +\hat{H}_{\rm F} - \mathbcallc{\hat d}  \cdot \blue{\left[\hat{\mathbf{E}}_{\rm F}\right]_{\parallel,0}},
\end{align}
and in the multipolar (or dipole) gauge
\begin{align}\label{eq:dipolemp}
    \hat{\mathcal{H}}^{\blue{\rm mp}}  = &\hat{H}_{\rm F} + \hat{\mathcal{H}}_0  - \mathbcallc{\hat d}  \cdot \blue{\hat{\mathbf{E}}_{{\rm F},0}}   \nonumber \\ &+ \sum_{\alpha,\alpha'}\frac{q_{\alpha}q_{\alpha'}}{2\epsilon_0}   \mathbcallc{\hat r}_{\alpha} \cdot {\bm \delta}^{\perp}(\mathbf{0}) \cdot \mathbcallc{\hat r}_{\alpha'}.
\end{align}
The last term in Eq.~\eqref{eq:dipolemp} is not well-defined, which is a well-known problem~\cite{Woolley2020Feb,Taylor2022Mar}. However, under mode truncation, the divergence becomes finite:
\begin{align}\label{eq:hmpprime}
    \hat{\tilde{\mathcal{H}}}^{\blue{\rm mp}} =& \hat{\mathcal{H}}_{\rm F} + \hat{\mathcal{H}}_{0} +   \left(i\sum_{\mu \nu} \hbar\chi_{\mu\nu}^*\hat{a}_{\mu}\hat{\xi}^{\blue{g}\dagger}_{\nu} + \text{H.c.}\right) \nonumber \\ & + \sum_{\mu\nu}\hbar \chi_{\mu \nu} \hat{\xi}^{\blue{g}\dagger}_{\mu}\hat{\xi}^{\blue{g}}_{\nu} - \hat{\mathbcallc{d}} \cdot \blue{\left[\hat{\mathbf{E}}_{\rm F}\right]_{\parallel,0}},
\end{align}
where, within the dipole approximation,
\begin{equation}
\hat{\xi}^{\blue{g}}_{\mu} = -\frac{ \hat{\mathbcallc{d}}\cdot \mathbf{f}_{\mu}^*(\mathbf{0})}{\sqrt{2\epsilon_0 \hbar\chi_{\mu\mu}}},
\end{equation}
and the mode-truncated result for the Coulomb gauge simply replaces $\hat{\mathbf{A}}_{\perp,0}$ with $\ahatbt_{\perp,0}$. Note that, as mentioned in Sec.~\ref{ssec:modeT} and further justified in Sec.~\ref{sec:res}, the third term on the right-hand side  of Eq.~\eqref{eq:hmpprime} can also be written as $-\hat{\mathbcallc{d}} \cdot \blue{\left[\hat{\bm{\mathcal{E}}}_{\rm F}\right]_{\perp,0}}$, where $\blue{\left[\hat{\bm{\mathcal{E}}}_{\rm F}\right]_{\perp}(\mathbf{x})}$ is the \emph{correctly-truncated} transverse \blue{and} bosonic part of the electric field operator defined in Eq.~\eqref{eq:eTcorrect}.

\subsection{Two-level systems}\label{ssec:tls}
If we restrict ourselves to only two quantized material states $\ket{e}$ and $\ket{g}$ (for the active media), with energies $+\hbar\omega_0/2$ and $-\hbar\omega_0/2$, respectively, then we can take advantage of the SU(2) Pauli matrix algebra, and write $\hat{\mathcal{H}}_0 = \hbar \omega_0\hat{\sigma}_z/2$, where $\hat{\sigma}_z = \ket{e}\bra{e}-\ket{g}\bra{g}$. Furthermore, the only information about the TLS which is required, in addition to its energy level separation, is its dipole moment. We shall assume the two states to have parity symmetry, and take the off-diagonal matrix element to be real:
\begin{subequations}\label{eq: d_def}
\begin{equation}\label{eq:dia}
    \bra{e}\hat{\mathbf{d}}\ket{e} = \bra{g}\hat{\mathbf{d}}\ket{g}=0,
\end{equation}
\begin{equation}\label{eq:off}
    \bra{e}\hat{\mathbf{d}}\ket{g} = \bra{g}\hat{\mathbf{d}}\ket{e} = \mathbf{d}, 
\end{equation}
\end{subequations}
such that $\mathbcallc{\hat d} = \mathbf{d} \hat{\sigma}_x$, where $\hat{\sigma}_x = \ket{e}\bra{g} + \ket{g}\bra{e}$. We can then calculate the matrix exponentials in Eq.~\eqref{eq:dipoleH} to obtain the Coulomb gauge Hamiltonian under the dipole approximation for a TLS:
\begin{align}\label{eq:coulombtls}
&\hat{\mathcal{H}}_{\rm C} = \hat{H}_{\rm F} - \mathbf{d} \cdot \blue{\left[\hat{\mathbf{E}}_{\rm F}\right]_{\parallel,0}}\hat{\sigma}_x \nonumber \\ &  + \frac{\hbar \omega_0}{2}\left[\cos{\left(\frac{2\mathbf{d} \cdot \hat{\mathbf{A}}_{\perp,0}}{\hbar}\right)}\hat{\sigma}_z + \sin{\left(\frac{2\mathbf{d} \cdot \hat{\mathbf{A}}_{\perp,0}}{\hbar}\right)}\hat{\sigma}_y\right],
\end{align}
where $\hat{\sigma}_y = i\ket{g}\bra{e} - i\ket{e}\bra{g}$. The corresponding multipolar gauge Hamiltonian is also easily obtained:
\begin{equation}\label{eq:multipolartls}
    \hat{\mathcal{H}}_{\rm mp} = \hat{H}_{\rm F} + \frac{\hbar \omega_0}{2} \hat{\sigma}_z - \mathbf{d} \cdot \hat{\mathbf{E}}\blue{_{{\rm F},0}} \hat{\sigma}_x.
\end{equation}

In Eq.~\eqref{eq:multipolartls}, we have been able to drop the problematic divergent term $\mathbf{d} \cdot {\bm \delta}^{\perp}(\mathbf{0}) \cdot \mathbf{d}/(2\epsilon_0)$, which although not well-defined, is a c-number which does not contribute to the Hamiltonian dynamics in the two-level subspace. The mode-truncated Hamiltonian in the TLS approximation is easily obtained by expressing Eqs.~\eqref{eq:coulombtls} in terms of the truncated field $\ahatbt_{\perp,0}$ for the Coulomb gauge, and letting $\hat{\mathbcallc{d}} \rightarrow \mathbf{d}\hat{\sigma}_x$ in Eq.~\eqref{eq:hmpprime}. For example, in the single-mode limit where $\chi_{\mu \nu} = \chi$, we have, in alignment with previous works~\cite{DiStefano2019Aug,PhysRevResearch.3.023079,Salmon2022Mar}
\begin{align}\label{eq:coulombtls1}
&\hat{\tilde{\mathcal{H}}}\blue{^{\rm C}_{(1)}}= \hbar\chi\hat{a}^{\dagger}\hat{a} - \mathbf{d} \cdot \blue{\left[\hat{\mathbf{E}}_{\rm F}\right]_{\parallel,0}}\hat{\sigma}_x + \frac{\hbar \omega_0}{2} \nonumber \\ & \times \left[\cos{\left(2\left\{\eta \hat{a}^{\dagger} + \eta^*\hat{a}^{\dagger}\right\}\right)}\hat{\sigma}_z + \sin{\left(2\left\{\eta \hat{a}^{\dagger} + \eta^*\hat{a}^{\dagger}\right\}\right)}\hat{\sigma}_y\right],
\end{align}
and
\begin{align}\label{eq:mptls1}
\hat{\tilde{\mathcal{H}}}\blue{^{\rm mp}_{(1)}}=& \hbar\chi\hat{a}^{\dagger}\hat{a} - \mathbf{d} \cdot \blue{\left[\hat{\mathbf{E}}_{\rm F}\right]_{\parallel,0}}\hat{\sigma}_x \nonumber \\ &+ i\hbar\chi\left[\eta\hat{a}^{\dagger} \! -\!  \eta^*\hat{a}\right]\! \blue{\hat{\sigma}_x} +\frac{\hbar\omega_0}{2}\hat{\sigma}_z,
\end{align}
where $\eta = \mathbf{d} \cdot \mathbf{f}^*(\mathbf{0})/\sqrt{2\epsilon_0 \hbar\chi}$\blue{, and we have dropped a term proportional to the identity.}

\subsection{Beyond the dipole approximation for a two-level system}\label{ssec:beyonddipole}

Here, we discuss an effective single particle model for a TLS. The results in this section can be generalized to higher-level systems as well.

To derive general results beyond the dipole approximation, one can 
assume in some instances an effective single-particle model for the polarization consisting of a single charge $q$ with position operator $\mathbcallc{\hat r}$, and a charge ${-}q$ which remains in a position eigenstate of the Hamiltonian with eigenvalue $\mathbf{r}_{\rm A}=0$. This is just one toy model (reminiscent of the hydrogen atom), and other effective single-particle models exist; for example, we could also have a charge $+q$ with position operator $\mathbcallc{\hat r}$ and a charge $-q$ with position operator $-\mathbcallc{\hat r}$. Sticking with the former case, we have 
\begin{equation}
    \hat{\bm{\mathcal{P}}}^{\blue{g}}(\mathbf{x}) = -q\mathbf{K}^{\blue{g}}(\mathbf{x},\mathbcallc{\hat r}),
\end{equation}
where $\mathbcallc{\hat r} = \mathbf{d}\hat{\sigma}_x/q$. We can derive simpler expressions by noting that, since $\hat{\sigma}_x$ is proportional to the position operator, one can immediately evaluate any function of operators which depends only on the position operator by using the eigenbasis of $\hat{\sigma}_x$. By construction (to preserve gauge invariance upon truncation), we have formulated our theory such that the entire Hamiltonian can be simplified in this manner. For example, for a function $\hat{f}(\mathbcallc{\hat r})$, one can write
\begin{equation}
    \hat{f}(\mathbcallc{\hat r}) = \frac{f(\mathbf{r}_{\rm dip}) + f(-\mathbf{r}_{\rm dip})}{2} + \frac{f(\mathbf{r}_{\rm dip}) - f(-\mathbf{r}_{\rm dip})}{2}\hat{\sigma}_x,
\end{equation}
where $\mathbf{r}_{\rm dip} = \mathbf{d}/q$ is an effective position coordinate.
Such a representation could be of course generalized to include cases where the position operator is a generic Hermitian matrix in the two-level subspace.
We find, not yet considering mode truncation,
\begin{align}\label{eq:genH_beyonddipole}
\mathcal{\hat{H}}^{\blue{g}} &= \hat{H}_{\rm F} + \frac{\hbar \omega_0}{2}\left[\cos{(\hat{\Phi}^{\blue{g}}_{\mathbf{A}})}\hat{\sigma}_z + \sin{(\hat{\Phi}^{\blue{g}}_{\mathbf{A}})}\hat{\sigma}_y\right] \nonumber \\
& +\frac{q^2}{4\epsilon_0}\int d^3x\left[\blue{\left\{\mathbf{K}^{g}_{\perp}(\mathbf{x},\mathbf{r}_{\rm dip})\right\}^2} - \blue{\left\{\mathbf{K}^{g}_{\perp}(\mathbf{x},\mathbf{r}_{\rm dip})\right\}^2}\right]\hat{\sigma}_x \nonumber \\ 
& + \frac{q}{2}\int d^3x \hat{\mathbf{E}}\blue{_{\rm F}}(\mathbf{x}) \cdot \left[\mathbf{K}^{\blue{g}}(\mathbf{x},\mathbf{r}_{\rm dip}) + \mathbf{K}^{\blue{g}}(\mathbf{x},-\mathbf{r}_{\rm dip})\right] \nonumber \\ 
& +\frac{q}{2}\int d^3x \hat{\mathbf{E}}_\blue{\rm F}(\mathbf{x}) \cdot \left[\mathbf{K}^{\blue{g}}(\mathbf{x},\mathbf{r}_{\rm dip}) - \mathbf{K}^\blue{g}(\mathbf{x},-\mathbf{r}_{\rm dip})\right]\hat{\sigma}_x, 
\end{align}
where $\hat{\Phi}{^{\blue{g}}_{\mathbf{A}}} = \mathbf{d} \cdot \int_{-1}^1 ds \hat{\mathbf{A}}^\blue{g}(s\mathbf{r}_{\rm dip})/\hbar$, and we have dropped the c-number term,
\begin{equation}
    \frac{q^2}{4\epsilon_0} \int d^3x \left[\blue{\left\{\mathbf{K}^{g}_{\perp}(\mathbf{x},\mathbf{r}_{\rm dip})\right\}^2} + \blue{\left\{\mathbf{K}^{g}_{\perp}(\mathbf{x},\mathbf{r}_{\rm dip})\right\}^2}\right],
\end{equation}
which does not contribute dynamically. In the case of mode truncation, the Hamiltonian $\hat{\tilde{\mathcal{H}}}^{\blue{g}}$ is similar to Eq.~\eqref{eq:genH_beyonddipole}, but with replacements $\hat{H}_{\rm F} \rightarrow \hat{\mathcal{H}}_{\rm F}$, $\hat{\Phi}^{\blue{g}}_{\mathbf{A}} \rightarrow \hat{\Phi}^{\blue{g}}_{\mathbf{\mathcal{A}}}$, and $\hat{\mathbf{E}}_{\blue{{\rm F}}} \rightarrow \hat{\bm{\mathcal{E}}}_{\blue{{\rm F}}}=\blue{\left[\hat{\bm{\mathcal{E}}}_{\rm F}\right]_{\perp}} + \blue{\left[\hat{\mathbf{E}}_{\rm F}\right]_{\parallel}} $, where $\blue{\left[\hat{\bm{\mathcal{E}}}_{\rm F}\right]_{\perp}}$ is the bosonic portion of the correctly-truncated transverse electric field operator, defined in Eq.~\eqref{eq:eTcorrect}.
Additionally, the second line instead becomes
\begin{align}
    \frac{q^2}{4\epsilon_0}  \sum_{\mu}&\Bigg[ \left(\int d^3x \mathbf{K}^{\blue{g}}_{\perp}(\mathbf{x},\mathbf{r}_{\rm dip}) \cdot \mathbf{f}_{\mu}(\mathbf{x})\right)^2 \nonumber \\ & - \left(\int d^3x \mathbf{K}^{\blue{g}}_{\perp}(\mathbf{x},-\mathbf{r}_{\rm dip}) \cdot \mathbf{f}_{\mu}(\mathbf{x})\right)^2\Bigg]\hat{\sigma}_x,
\end{align}
and the c-number  \blue{term} that is dropped is modified accordingly to smooth out the divergence.

From this result, it is straightforward to obtain the Coulomb gauge Hamiltonian beyond the dipole approximation:
\begin{align}\label{eq:HC_bd}
\hat{\mathcal{H}}^{\blue{{\rm C}}} &= \hat{H}_{\rm F} + \frac{\hbar \omega_0}{2}\left[\cos{(\hat{\Phi}^{\blue{\rm C}}_{\mathbf{A}})}\hat{\sigma}_z + \sin{(\hat{\Phi}^{\blue{\rm C}}_{\mathbf{A}})}\hat{\sigma}_y\right] \nonumber \\ &  - \frac{\mathbf{d} }{2} \cdot \int_0^1 ds\left[ \blue{\left[\hat{\mathbf{E}}_{\rm F}(s\mathbf{r}_{\rm dip})\right]_{\parallel}} - \blue{\left[\hat{\mathbf{E}}_{\rm F}(-s\mathbf{r}_{\rm dip})\right]_{\parallel}}\right] \nonumber \\ & - \frac{\mathbf{d} }{2} \cdot \int_{-1}^{1}ds \blue{\left[\hat{\mathbf{E}}_{\rm F}(s\mathbf{r}_{\rm dip})\right]_{\parallel}} \hat{\sigma}_x.
\end{align}

Similarly, for the multipolar gauge,
\begin{align}\label{eq:HMP_bd}
\hat{\mathcal{H}}_{\rm mp} &= \hat{H}_{\rm F} +\frac{\hbar\omega_0}{2}\hat{\sigma}_z \nonumber \\ &  -\frac{\mathbf{d} }{2} \cdot \int_0^1 ds\left[ \hat{\mathbf{E}}_{\blue{\rm F}}(s\mathbf{r}_{\rm dip}) - \hat{\mathbf{E}}_{\blue{\rm F}}(-s\mathbf{r}_{\rm dip})\right] \nonumber \\ & - \frac{\mathbf{d} }{2} \cdot \int_{-1}^{1}ds \hat{\mathbf{E}}_{\blue{\rm F}}(s\mathbf{r}_{\rm dip}) \hat{\sigma}_x.
\end{align}

Note that the second line of Eqs.~\eqref{eq:HC_bd} and~\eqref{eq:HMP_bd} vanish for the common situation in which the electric field is an even function of position (e.g., in cavity-QED where a dipole is placed at a modal antinode). With mode truncation, Eqs.~\eqref{eq:HC_bd} and~\eqref{eq:HMP_bd} should be expressed in terms of the truncated field variables, with the additional term to the multipolar gauge Hamiltonian:
\begin{equation}
    \frac{1}{4\epsilon_0} \sum_{\mu} \left[\left( \mathbf{d} \!  \cdot \! \! \int_0^{1} \! \! ds \mathbf{f}_{\mu}(s\mathbf{r}_{\rm dip}
    )\right)^2 \! \!-\! \!\left( \mathbf{d}\!  \cdot \! \! \int_{-1}^{0} \! \! ds \mathbf{f}_{\mu}(s\mathbf{r}_{\rm dip}
    )\right)^2  \right]\hat{\sigma}_x.
\end{equation}

Considering a one-dimensional system and neglecting the longitudinal field terms, the Coulomb gauge result shown by Eq.~\eqref{eq:HC_bd} was also found in Ref.~\cite{Savasta2021May}, where it was noted that due to the spatial integral over the transverse vector potential, going beyond the dipole approximation introduces a natural cut-off for high frequency interactions, as $\hat{\Phi}^{\blue{\rm C}}_{\mathbf{A}}$ vanishes for mode wavelengths much shorter than $|\mathbf{r}_{\rm dip}|$. Here, we have extended these results to consider an arbitrary gauge Hamiltonian for the general three-dimensional case, and including the longitudinal terms.

\section{Resolution of gauge ambiguity associated with photon detection and field observables}\label{sec:res}

A potential ambiguity that can arise when determining \emph{observables} of the electromagnetic field is the gauge-dependent nature of the field operators. As an example, consider the transverse electric field operator in the Coulomb gauge $\hat{\mathbf{E}}^{\blue{\rm C}}_{\perp} = \blue{[\hat{\mathbf{E}}_{\rm F}]_{\perp}}$, and the multipolar gauge, $\hat{\mathbf{E}}^{\blue{\rm mp}}_{\perp} = \blue{[\hat{\mathbf{E}}_{\rm F}]_{\perp}} - \frac{1}{\epsilon_0}\hat{\mathbf{P}}^{\rm mp}_{\perp}$. In this section we focus on gauge ambiguities associated with (electromagnetic) mode truncation, so the polarization can be truncated or untruncated.

Suppose we want to calculate an observable which is a function of the electric field, $\langle \hat{O}(\hat{\mathbf{E}}^\blue{g})\rangle_g$, for gauges $g = {\rm C}, {\rm mp}$, and the subscript on the expectation value indicates it is to be calculated with respect to a state vector (or density operator) in the gauge $g$.
Without any truncation, we have, by construction of the manifestly gauge invariant quantum theory presented in previous sections, $\langle \hat{O}(\hat{\mathbf{E}}^{\blue{\rm C}})\rangle_{\rm C} = \langle \hat{O}(\hat{\mathbf{E}}^{\blue{{\rm mp}}})\rangle_{\rm mp}$. Since $\hat{\mathbf{P}}$ is only non-zero in the vicinity of the matter charged particles, one should be able to calculate any observable evaluated at locations far away from the position of the free matter particles  as $\langle \hat{O}(\hat{\mathbf{E}}^{\blue{{\rm C}}})\rangle_{\rm C} = \langle \hat{O}(\hat{\mathbf{E}}^{\blue{{\rm mp}}})\rangle_{\rm mp} = \langle \hat{O}(\blue{[\hat{\mathbf{E}}_{\rm F}]_{\perp}})\rangle_{\rm C}= \langle \hat{O}(\blue{[\hat{\mathbf{E}}_{\rm F}]_{\perp}})\rangle_{\rm mp}$. However, as the multipolar and Coulomb gauges have different Hamiltonians, $\langle \hat{O}(\blue{[\hat{\mathbf{E}}_{\rm F}]_{\perp}})\rangle_{\rm C}$ and $\langle \hat{O}(\blue{[\hat{\mathbf{E}}_{\rm F}]_{\perp}})\rangle_{\rm mp}$ will generally be different when mode truncation is considered.

To illustrate this issue, consider the following expansion of the bosonic part of the  electric field:
\begin{equation}\label{eq:efanoc}
    \blue{\left[\hat{\mathbf{E}}_{\rm F}(\mathbf{x})\right]_{\perp}} = i\sum_{\mu}\sqrt{\frac{\hbar\chi_{\mu\mu}}{2\epsilon_0}} \mathbf{f}^{{\bm E}_{\perp}}_{\mu}(\mathbf{x}) \hat{a}_{\mu} + \text{H.c.},
\end{equation}
where $\mathbf{f}_{\mu}^{\bm E}$ are proposed modal functions for the electric field expansion \blue{defined through naive projection},  \blue{given} in Appendix~\ref{sec:appA}. We will ultimately show that this is generally not the correct form of the electric field operator mode expansion when mode truncation is to be performed.

Suppose Eq.~\eqref{eq:efanoc} is expressed in the Coulomb gauge, such that $\blue{[\hat{\mathbf{E}}_{\rm F}]_{\perp}} = \hat{\mathbf{E}}_{\perp}$. Then, to obtain the multipolar gauge expression for the transverse electric field, we apply the PZW transformation (using the dipole approximation for simplicity):
\begin{equation}
    \hat{W}_{\rm mp, C}\hat{a}_{\mu}\hat{W}_{\rm mp, C}^{\dagger} = \hat{a}_{\mu} + i\frac{\hat{\mathbcallc{d}} \cdot \mathbf{f}_{\mu}^*(\mathbf{0})}{\sqrt{2\hbar\epsilon_0\chi_{\mu\mu}}},
\end{equation}
and so
\begin{align}\label{eq:efanoc2}
    &\hat{W}_{\rm mp, C} \blue{\left[\hat{\mathbf{E}}_{\rm F}(\mathbf{x})\right]_{\perp}}\hat{W}_{\rm mp, C}^{\dagger} = \nonumber \\ &  \blue{\left[\hat{\mathbf{E}}_{\rm F}(\mathbf{x})\right]_{\perp}} \blue{-} \frac{\hat{\mathbcallc{d}}}{\epsilon_0} \cdot \left[\frac{1}{2}\sum_{\mu} \mathbf{f}_{\mu}^{\mathbf{E}_{\perp}}(\mathbf{x})\mathbf{f}_{\mu}^*(\mathbf{0}) +\text{c.c.}\right].
\end{align}
The quantity in square brackets is equal to ${\bm \delta}^{\perp}(\mathbf{x})$ if the mode expansion is complete (see Appendix~\ref{sec:appA}), but for a finite mode expansion, it becomes nonzero even away from the dipole location at $\mathbf{x} = \mathbf{0}$. Thus, upon mode truncation we have a potential ambiguity; should we truncate with respect to the photonic Hilbert space, using $\hat{P}_{\rm M}$, or with respect to the sum over mode index $\mu$? 

In Ref.~\cite{PhysRevResearch.3.023079}, this potential ambiguity was identified, with the proposed solution to truncate with respect to the mode index. However, no definitive argument was given for why this should be the case, beyond the fact that it is unitarily equivalent to the seemingly less ambiguous situation in the Coulomb gauge. Here, we show that this can be justified by a proper restoration of the gauge invariance lost under naive mode truncation. 

To resolve this ambiguity, we take an approach similar in spirit to that of Ref.~\cite{PhysRevResearch.3.023079}, by explicitly modelling the detector degree of freedom, and using the mode truncation procedure outlined in this work to derive a gauge-invariant model of the detector constituent particles.

To do so, we can apply the theory in \blue{previously} developed sections \blue{(and the appendices)} by generalizing the quantum charge and current densities to become $\hat{\rho}_{\rm A}(\mathbf{x}) \rightarrow \hat{\rho}_{\rm A}(\mathbf{x}) + \hat{\rho}_{\rm d}(\mathbf{x})$, and $\hat{\mathbf{J}}_{\rm A}(\mathbf{x}) \rightarrow \hat{\mathbf{J}}_{\rm A}(\mathbf{x}) + \hat{\mathbf{J}}_{\rm d}(\mathbf{x})$, where 
\begin{equation}
    \hat{\rho}_{\rm d}(\mathbf{x}) = \sum_{\alpha_{\rm d}} q_{\alpha_{\rm d}} \delta(\mathbf{x}-\hat{\mathbf{r}}_{\alpha_{\rm d}})
\end{equation}
\begin{equation}
    \hat{\mathbf{J}}_{\rm d}(\mathbf{x}) = \sum_{\alpha_{\rm d}} q_{\alpha_{\rm d}} \dot{\hat{\mathbf{r}}}_{\alpha_{\rm d}}\delta(\mathbf{x}-\hat{\mathbf{r}}_{\alpha_{\rm d}}),
\end{equation}
corresponding to ``detector'' particles with position operators $\hat{\mathbf{r}}_{\alpha_{\rm d}}$ and charge $q_{\alpha_{\rm d}}$ where again we assume $\sum_{\alpha_{\rm d}}q_{\alpha_{\rm d}} = 0$. This change induces $\hat{\mathbf{P}}(\mathbf{x}) \rightarrow \hat{\mathbf{P}}(\mathbf{x}) + \hat{\mathbf{P}}_{\rm d}(\mathbf{x})$, to reflect the additional detector degrees of freedom in the polarization. We assume these particles to be localized around a position $\mathbf{r}_{\rm d}$, and as such, we take the following form for the multipolar gauge polarization (as well as the corresponding change in $\mathbf{Z}_{\rm A}(\mathbf{x})$):
\begin{equation}
    \hat{\mathbf{P}}^{\blue{\rm mp}}_{\rm d}(\mathbf{x}) = \sum_{\alpha_{\rm d}}q_{\alpha_{\rm d}}(\hat{\mathbf{r}}_{\alpha_{\rm d}} - \mathbf{r}_{\rm d})  \int_0^{1} ds \delta(\mathbf{x}-\mathbf{r}_{\rm d} - s(\hat{\mathbf{r}}_{\alpha_{\rm d}}- \mathbf{r}_{\rm d})).
\end{equation}
Obviously this is an oversimplified model of photodetection~\cite{Gardiner2004Aug}, but it is sufficient to capture the essential features with respect to gauge invariance and mode truncation. For the sake of simplicity (and because we assume weak coupling of the detector to the field), we will assume a truncation of the total material subspace such that the detector degree of freedom can be reduced to a TLS with energy separation $\hbar\omega_{\rm d}$. Similarly, we assume the dipole approximation at the location of the detector $\mathbf{r}_{\rm d}$.

The Hamiltonian for this system under mode truncation is that of Eq.~\eqref{eq:hmode}, generalized to incorporate the additional detector. Expanding to first order in the detector dipole moment operator $\hat{\mathbcallc{d}}_{\rm d} = \sum_{\alpha_{\rm d}}q_{\alpha_{\rm d}}(\hat{\mathbcallc{r}}_{\alpha_{\rm d}}-\mathbf{r}_{\rm d})$, we obtain the Coulomb and multipolar gauge Hamiltonians:
\begin{equation}
    \hat{\tilde{\mathcal{H}}}\blue{_{\rm d}^{\rm C}} = \hat{\tilde{\mathcal{H}}}^{\blue{{\rm C}}} + \frac{\hbar\omega_{\rm d}}{2}\hat{\sigma}^{\rm d}_{z} + \omega_{\rm d} \mathbf{d}_{\rm d} \cdot \ahatbt_{\perp}(\mathbf{r}_{\rm d}) \hat{\sigma}_y^{\rm d},
\end{equation}
\begin{equation}
    \hat{\tilde{\mathcal{H}}}\blue{_{\rm d}^{\rm mp}} = \hat{\tilde{\mathcal{H}}}^{\blue{{\rm mp}}} + \frac{\hbar\omega_{\rm d}}{2}\hat{\sigma}^{\rm d}_{z}  -\left(i\sum_{\mu\nu}\hbar\chi_{\mu\nu}^*\eta_{\nu}^{\rm d*}\hat{a}_{\mu}' + \text{H.c.}\right)\hat{\sigma}_x^{\rm d},
\end{equation}
where we have used $\hat{\mathbcallc{d}}_{\rm d} = \mathbf{d}_{\rm d} \hat{\sigma}_x^{\rm d}$, and we have defined $\eta_{\mu}^{\rm d} = \mathbf{d}_{\rm d} \cdot \mathbf{f}_{\mu}^*(\mathbf{r}_{\rm d})/(\sqrt{2\epsilon_0 \hbar \chi_{\mu\mu}})$
and $\hat{a}_{\mu}' = \hat{a}_{\mu} + i \eta_{\mu}\hat{\sigma}_{x}$, where $\eta_{\mu} = \mathbf{d} \cdot \mathbf{f}_{\mu}^*(\mathbf{0})/(\sqrt{2\epsilon_0 \hbar \chi_{\mu\mu}})$. Suppose, now, that the detector TLS frequency $\omega_{\rm d}$ is resonant with a transition of the Hamiltonian (neglecting the detector part) between eigenstates $\ket{j}$ and $\ket{i}$ with frequency $\omega_{j}-\omega_{i} = \omega_{\rm d}$~\cite{PhysRevResearch.3.023079}. By using perturbation theory (Fermi's golden rule), the photodetection rate should be proportional to $R_{\rm C}$ and $R_{\rm mp}$ in the Coulomb and multipolar gauges, respectively, where
\begin{equation}
    R_{\rm C} = |\bra{i_{\rm C}} \omega_{\rm d} \mathbf{d}_{\rm d} \cdot \ahatbt_{\perp}(\mathbf{r}_{\rm d})\ket{j_{\rm C}}|^2
\end{equation}
\begin{equation}\label{eq:R-mp}
    R_{\rm mp} = |\bra{i_{\rm mp}}i\sum_{\mu\nu}\hbar\chi_{\mu\nu}^*\eta_{\nu}^{\rm d*}\hat{a}_{\mu}' + \text{H.c.}\ket{j_{\rm mp}}|^2.
\end{equation}
The aim now is to show these are equivalent.
First, we note that in the Coulomb gauge
\begin{align}\label{eq:aperpdot}
    \frac{\partial}{\partial t}\ahatbt_{\perp}(\mathbf{x}) &= \frac{i}{\hbar}[\hat{\tilde{\mathcal{H}}}^{\blue{{\rm C}}},\ahatbt_{\perp}(\mathbf{x})] \nonumber \\ & = - i\sum_{\mu\nu}\frac{\hbar\chi_{\mu\nu}}{\sqrt{2\epsilon_0 \hbar \chi_{\mu\mu}}}\mathbf{f}_{\mu}(\mathbf{x})\hat{a}_{\nu} + \text{H.c.} \nonumber \\
    & = -\hat{\bm{\mathcal{E}}}\blue{^{\rm C}_{\perp}}(\mathbf{x}),
\end{align}
where $\hat{\bm{\mathcal{E}}}\blue{^{\rm C}_{\perp}}$ is the correctly-truncated transverse electric field operator for the Coulomb gauge defined in Eq.~\eqref{eq:eTcorrect}. It is generally \emph{not} equal to the transverse electric field operator which one would obtain by simply truncating the expansion in Eq.~\eqref{eq:efanoc}, and this suggests that truncating Eq.~\eqref{eq:efanoc} directly is the incorrect approach in a dispersive and absorbing medium.

Using Eq.~\eqref{eq:aperpdot}, if we take matrix elements with respect to the eigenstates of the Hamiltonian $\hat{\tilde{\mathcal{H}}}^{\blue{{\rm C}}}$, we can derive the relationship:
\begin{equation}
\bra{i_{\rm C}} \ahatbt_{\perp} \ket{j_{\rm C}} = \sum_{\mu\nu}\frac{1}{\sqrt{2\epsilon_0 \hbar \chi_{\mu\mu}}}\frac{\bra{i_{\rm C}}\hbar \chi_{\mu \nu}\mathbf{f}_{\mu} \hat{a}_{\nu} - \text{H.c.}\ket{j_{\rm C}}}{\omega_{j}-\omega_i}.
\end{equation}
Using the fact that $\omega_{\rm d} = \omega_{j} - \omega_i$, and multiplying by a factor of $|i|^2 =1$, we find
\begin{equation}\label{eq:Rc2}
    R_{\rm C} = \left|\bra{i_{\rm C}}\left(i\sum_{\mu\nu}\hbar\chi_{\mu\nu}^*\eta_{\nu}^{\rm d*}\hat{a}_{\mu} + \text{H.c.}\right)\ket{j_{\rm C}}\right |^2.
\end{equation}

Now, we note that the PZW transformation (excluding the detector subspace), applied to Eq.~\eqref{eq:Rc2}, simply takes $\hat{a}_{\mu}$ to $\hat{a}_{\mu}'$, and the matrix elements to be evaluated in the eigenstates of the multipolar gauge. Thus, $R_{\rm C} = R_{\rm mp}$, and we can confirm the final result: the operator which should be used to model photodetection at \blue{transition} frequency $\omega$ is $\omega\ahatbt_{\perp}(\mathbf{x})$, and this quantity is gauge invariant. Equivalently, the correctly-truncated transverse electric field operator $\hat{\bm{\mathcal{E}}}^{\blue{g}}_{\perp}(\mathbf{x})$ can be used, which is $\hat{\bm{\mathcal{E}}}^{\blue{g}}_{\perp} = \hat{\mathcal{V}}_{\blue{g}}\blue{[\hat{\bm{\mathcal{E}}}_{\rm F}]_{\perp}}\hat{\mathcal{V}}^{\dagger}_{\blue{g}}$, and $\blue{[\hat{\bm{\mathcal{E}}}_{\rm F}]_{\perp}}$ is defined in Eq.~\eqref{eq:eTcorrect}. This fact can be understood by noting  that the matrix element of Eq.~\eqref{eq:R-mp} is that of the operator $\mathbf{d}_{\rm d} \cdot \hat{\bm{\mathcal{E}}}^{\blue{\rm mp}}_{\perp}(\mathbf{r}_{\rm d})$.

We also comment that, as has been pointed out previously, the notion of what the ``correct'' truncated electric field is can be fundamentally ambiguous, and depends on what is ultimately being measured in experiment~\cite{PhysRevResearch.3.023079,Stokes2022Nov}. In this case, this definition is consistent with photodetection experiments. We also note that this definition of the truncated field violates causality~\cite{PhysRevResearch.3.023079}, however this is perfectly consistent with a truncated mode approximation~\cite{SanchezMunoz2018May}, at least assuming an anharmonic mode spectrum.

\section{Conclusions}\label{sec:conc}
In conclusion, we have presented a general theory of gauge-invariant light-matter interactions under material and photonic subspace truncation in an arbitrary medium, focusing on the realistic case of an inhomogeneous dispersive and absorbing dielectric. Our theory is thus applicable to a wide range of optical and photonic systems.

In Sec.~\ref{sec:action}, we gave the fundamental minimal-coupling action which gave Maxwell's equations and the Lorentz force law for a system with medium reservoir oscillator degrees of freedom (ultimately characterized by the dielectric function $\epsilon(\mathbf{x},\omega)$), as well as free material particles. We then quantized the system using a quantization function approach in Sec.~\ref{sec:quantization}, where the gauge of the system was left arbitrary, allowing us to show manifestly that the quantum theory was gauge invariant, and easily recover known results for the common Coulomb and multipolar gauges. In Sec.~\ref{sec:GI}, we showed how gauge invariance manifests in the quantum theory, and how material and photonic space truncation can be included without sacrificing gauge invariance in the reduced subspace, allowing one to derive models which can be used to model ultrastrong light-matter interactions for common models in quantum optics, including TLSs and single-mode models. We drew contrast with a discrete mode expansion in lossy systems, where the mode expansion Hamiltonian $\hat{\mathcal{H}}_{\rm F}$ is non-diagonal, with the more familiar normal mode expansion, which can be implemented in free space, or a lossless and dispersionless dielectric. In particular, we argued that this difference requires rigorous open quantum system models of loss in  \blue{resonant} mode systems (e.g., a cavity-qubit system with photon loss) to be derived in the Coulomb gauge, as this gauge is the unique gauge in which information regarding the part of the ``reservoir'' electromagnetic field (to be traced out) can be encoded in a purely photonic operator subspace.

Next, in Sec.~\ref{sec:time-dep}, we extended the arbitrary-gauge quantization function approach to canonical quantization by allowing for gauge conditions with explicit time-dependence, and showed how this could be used to introduce unambiguous phenomenological time-dependent modulation of the transverse light-matter coupling strength. We applied our theory to common models and approximations used in the literature in Sec.~\ref{sec:dtls}, including the dipole approximation, the TLS, and how to go beyond the dipole approximation for effective single-particle models. 
Finally, in Sec.~\ref{sec:res}, we discussed how a particular gauge ambiguity could arise with respect to observables of a mode-truncated field, and how to resolve this ambiguity by means of an explicit ``detector particle'' model, where gauge invariance is preserved using the techniques of Sec.~\ref{sec:GI}. This allowed us to show that the vector potential is the fundamental field coordinate to be truncated directly, and subsequently, we showed how to obtain the correctly-truncated electric field from this result.

Finally, we note that our results can be used as a starting point for more fundamental models of loss and dissipation in quantum optics. For example, the techniques of material and mode truncation can be used to derive rigorous and accurate models of dipole interactions with a few discrete modes, modelled as quantized QNMs (as shown in Appendix~\ref{sec:appA}). Separating the entire electromagnetic field into QNM and reservoir components~\cite{frankequantization,franke2020quantized}, and using standard methods of open quantum systems, a master equation could then be derived to govern the lossy dipole-QNM system. This would solve the long-standing problem of how to go beyond a phenomenological formulation of the system-reservoir coupling in quantum optics and input-output theory~\cite{Gardiner1985Jun} in the USC regime for general three-dimensional resonators, which has recently been shown to be essential to predicting emission spectra and other observables from cavity-QED systems in the USC regime~\cite{Salmon2022Mar}.

\begin{acknowledgements}
We acknowledge funding from Queen's University, the Canadian Foundation for Innovation (CFI), the Natural Sciences and Engineering Research Council of Canada (NSERC), and the Alexander von Humboldt Foundation through a Humboldt Research Award.
\end{acknowledgements}

\appendix
\blue{
\section{Action and Equations of Motion}\label{app:action}

The total action for the system described in Sec.~\ref{sec:action} is~\cite{Suttorp2004Sep,Philbin2010Dec}
\begin{equation}
    \mathcal{S} = \mathcal{S}_{\rm em}[A_\mu] + \mathcal{S}_{\mathbf{X}}[\mathbf{X}_{\omega}] + \mathcal{S}_{\rm A}[\mathbf{r}_\alpha] + \mathcal{S}_{\rm int}[A_\mu;  \mathbf{r}_\alpha; \mathbf{X}_{\omega}],
\end{equation}
where 
\begin{subequations}
\begin{equation}\label{eq:A1}
    \mathcal{S}_{\rm em}[A_\mu] = \int d^4x \left[\frac{\epsilon_0}{2}\mathbf{E}^2 - \frac{1}{2\mu_0}\mathbf{B}^2\right],
\end{equation}
\begin{equation}\label{eq:A2}
    \mathcal{S}_{\rm A}[\mathbf{r}_{\alpha}] =\frac{1}{2}\sum_{\alpha}m_{\alpha}\dot{\mathbf{r}}_{\alpha}^2,
\end{equation}
\begin{equation}\label{eq:A3}
    \mathcal{S}_{\mathbf{X}}[\mathbf{X}_\omega] = \frac{1}{2}\int d^4x \int_0^{\infty}d\omega \left[ \dot{\mathbf{X}}^2_{\omega} - \omega^2 \mathbf{X}_{\omega}^2\right],
\end{equation}
\begin{align}\label{eq:A4}
    \mathcal{S}_{\rm int}[A_{\mu}; \mathbf{r}_\alpha; \mathbf{X}_\omega] = \int d^4x &\int_0^{\infty} d\omega \alpha(\mathbf{x},\omega) \mathbf{X}_{\omega} \cdot \mathbf{E}\nonumber \\ & - \int d^4x J_{\rm A}^\mu A_\mu,
\end{align}
\end{subequations}
with
\begin{equation}\label{eq:alpha}
    \alpha(\mathbf{x},\omega)= \sqrt{\frac{2\epsilon_0 \omega}{\pi}\epsilon_I(\mathbf{x},\omega)}\, 
\end{equation}
and the four-current for the free particles is
$J_{\rm A}^{\mu}(x) = (c\rho_{\rm A}(\mathbf{x},t),\mathbf{J}_{\rm A}(\mathbf{x},t))$.
This current satisfies a continuity equation ${\bm \nabla} \cdot \mathbf{J}_{\rm A} + \dot{\rho}_{\rm A}=0$, where
\begin{subequations}
\begin{equation}\label{eq:charged}
    \rho_{\rm A}(\mathbf{x},t) = \sum_{\alpha} q_{\alpha}  \delta(\mathbf{x}-\mathbf{r}_{\alpha})
\end{equation}
\begin{equation}\label{eq:currentd}
    \mathbf{J}_{\rm A}(\mathbf{x},t) = \sum_{\alpha} q_{\alpha} \dot{\mathbf{r}}_{\alpha} \delta(\mathbf{x}-\mathbf{r}_{\alpha}).
\end{equation}
\end{subequations}
The electric and magnetic fields are expressed in terms of the potentials as $\mathbf{E} = -\dot{\mathbf{A}} - {\bm \nabla} \phi$ and $\mathbf{B} = {\bm \nabla} \times \mathbf{A}$, respectively.

Requiring that the action be stationary, we obtain the following set of equations:
\begin{subequations}\label{eq:eom}
\begin{equation}\label{eq:eom1}
    \epsilon_0 \nabla \cdot \mathbf{E} + \int_0^{\infty} d\omega \nabla \cdot [\alpha(\mathbf{x},\omega) \mathbf{X}_{\omega}] = \rho_{\rm A},
\end{equation}
\begin{equation}\label{eq:eom2}
    \frac{1}{\mu_0}\nabla \times \mathbf{B} - \epsilon_0 \dot{\mathbf{E}} - \int_0^{\infty}d\omega\alpha(\mathbf{x},\omega)\mathbf{\dot{X}}_{\omega} = \mathbf{J}_{\rm A},
\end{equation}
\begin{equation}\label{eq:eom3}
    m_{\alpha}\ddot{\mathbf{r}}_{\alpha} = q_{\alpha}\mathbf{E}(\mathbf{r}_{\alpha}) + q_{\alpha} \dot{\mathbf{r}}_{\alpha} \times \mathbf{B}(\mathbf{r}_{\alpha}),
\end{equation}
\begin{equation}\label{eq:eom4}
    \ddot{\mathbf{X}}_{\omega} + \omega^2\mathbf{X}_{\omega} = \alpha(\mathbf{x},\omega)\mathbf{E}.
\end{equation}
\end{subequations}
Clearly, Eqs.~\eqref{eq:eom1} and~\eqref{eq:eom2} are the Gauss and Amp{\`e}re-Maxwell laws with active material sources, as well as a polarization field $\mathbf{P}_{\rm M} = \int d\omega \alpha(\mathbf{x},\omega) \mathbf{X}_{\omega}$ arising from the passive medium reservoir which also acts as a source---the ``passivity'' of the medium is demonstrated by the ``Fano diagonalization'' of Sec.~\ref{sec:Fano_diagnonalization}, which allows the fields to be expressed in terms of bosonic polariton operators and the photonic Green's function without explicit reference to the medium interaction. Equation~\eqref{eq:eom3} is the Lorentz force equation of motion for the active free particles, and Eq.~\eqref{eq:eom4} is the equation of motion for the passive medium reservoir field. 

\section{Quantization via Dirac's prescription with constraints}\label{app:dirac}

In this appendix, we quantize the system consisting of the electromagnetic fields, passive medium, and active material particles using Dirac's prescription.

We note that we have three constraints for this system: $\chi_0 = \chi_1 = \chi_2 = 0$, as given by Eqs.~\eqref{eq:c0},~\eqref{eq:c1}, and~\eqref{eq:c2}.
These constraints are to be understood as only applying on a subset of phase space in which all the constraints are applied---in the parlance of Dirac, these are \emph{weak equalities}~\cite{Dirac1958Aug}.
To proceed with the quantization procedure, we must evaluate Poisson brackets between these constraints. In particular, these should be evaluated \emph{before} applying the constraints.
It is easy to verify that the Poisson brackets of $\chi_0$ with the other two constraints vanish. This means $\chi_0$ is a so-called first-class constraint, and the remaining Poisson brackets (over the ``second-class'' constraints) thus form an orthogonal matrix $C_{ij}(\mathbf{x},\mathbf{x'}) = \{\chi_i(\mathbf{x}),\chi_{j}(\mathbf{x'})\}$\footnote{\blue{Where, for example, the Poisson bracket for fields $f(\mathbf{x})$ and $g(\mathbf{x'})$ which depend on fields $\theta_{i}$ and their conjugates $\Pi_{\theta_i}$ takes the form  $\{f(\mathbf{x}),g(\mathbf{x'})\} = \sum_i\int \! d^3y\left[ \frac{\delta f(\mathbf{x})}{\delta \theta_i(\mathbf{y})}\frac{\delta g(\mathbf{x'})}{\delta \Pi_{\theta_i}(\mathbf{y})} - \frac{\delta g(\mathbf{x'})}{\delta \theta_i(\mathbf{y})}\frac{\delta f(\mathbf{x})}{\delta \Pi_{\theta_i}(\mathbf{y})}\right]$, with all fields evaluated at equal times.}}, which can be evaluated as
\begin{equation}
    \mathbf{C}(\mathbf{x},\mathbf{x'}) = \delta(\mathbf{x}-\mathbf{x'})\begin{bmatrix}
0 & -1 \\
1 & 0 
\end{bmatrix},
\end{equation}
with inverse (in this case) $\mathbf{C}^{-1} = \mathbf{C}^{T}$,
which satisfies 
\begin{equation}\sum_k\int d^3x'' 
C_{ik}(\mathbf{x},\mathbf{x''})C^{-1}_{kj}(\mathbf{x''},\mathbf{x'}) =\delta_{ij} \delta(\mathbf{x}-\mathbf{x'}).
\end{equation}
From this, one then defines the \emph{Dirac bracket}~\cite{Woolley1999Jan,Weinberg1995Jun,sundermeyer1982constrained}:
\begin{align}
    \{&A,  B\}_{\rm D} \equiv \{A, B\} -  \nonumber \\ & \sum_{ij} \!\int \! d^3x \! \int \! d^3x' \! \{A,\chi_{i}(\mathbf{x})\} C^{-1}_{ij}(\mathbf{x},\mathbf{x'}) \{\chi_j(\mathbf{x'}),B\},
\end{align}
and quantizes by imposing $[\hat{A},\hat{B}] = i\hbar \{A,B\}_{\rm D}.$

Evaluating in this manner, we find the nonzero commutators to be (in the Schr{\"o}dinger picture)
\begin{equation}
    [\mathbf{\hat{X}}_{\omega}(\mathbf{x}),\mathbf{\hat{\Pi}}_{\omega'}(\mathbf{x'})] = i\hbar\mathbf{I}\delta(\mathbf{x}-\mathbf{x'})\delta(\omega-\omega'),
\end{equation}
\begin{equation}\label{eq:commr}
    [\mathbf{\hat{r}}_{\alpha},\mathbf{\hat{p}}_{\alpha'}] = i\hbar \mathbf{I}\delta_{\alpha \alpha'},
\end{equation}
\begin{align}
    [\mathbf{\hat{A}}(\mathbf{x}),\mathbf{\hat{\Pi}}_{\mathbf{A}}(\mathbf{x'})] &= i\hbar\left[\mathbf{I}\delta(\mathbf{x}-\mathbf{x'}) + {\bm \nabla}^{\mathbf{x}}\mathbf{K}(\mathbf{x'},\mathbf{x})\right] \nonumber \\ & = i\hbar\left[{\bm \delta}^{\perp}(\mathbf{x}-\mathbf{x'}) + {\bm \nabla}^{\mathbf{x}}\mathbf{K}_{\perp}(\mathbf{x'},\mathbf{x})\right],
\end{align}
and
\begin{equation}
    [\mathbf{\hat{p}}_{\alpha},\mathbf{\hat{\Pi}}_{\mathbf{A}}(\mathbf{x})] = i\hbar q_{\alpha} {\bm \nabla}^{\hat{\mathbf{r}}_{\alpha}}\mathbf{K}(\mathbf{x},\hat{\mathbf{r}}_{\alpha}).
\end{equation}
We can also construct the quantum Hamiltonian:
\begin{align}
    \hat{H} =& \sum_{\alpha}\dot{\hat{\mathbf{r}}}_{\alpha} \cdot \hat{\mathbf{p}}_{\alpha}  + \int d^3x \dot{\hat{\mathbf{A}}} \cdot \hat{\mathbf{\Pi}}_{\mathbf{A}} \nonumber \\ & + \int d^3x \int_0^{\infty} d\omega \dot{\hat{\mathbf{X}}}_{\omega} \cdot \hat{\mathbf{\Pi}}_{\mathbf{X}_{\omega}} - \hat{\mathcal{L}}.
\end{align}
While these commutators define, in principle, a working quantum theory of fields, they do not separate photonic and material degrees of freedom in the form of canonical commutation relations. That is, the operators $\hat{\mathbf{A}}$ and $\hat{\mathbf{\Pi}}_{\mathbf{A}}$ cannot be expanded as a superposition of bosonic creation and annihilation operators. Nonetheless, a canonical transformation can be easily found to describe the system in terms of new coordinates which do satisfy the usual canonical commutation relations.

After quantization, the ``weak equalities'' of the constraints become strong equalities, and can be applied to the quantum operators. The constraint $\chi_1(\mathbf{x}) = 0$ means that the longitudinal part of $\hat{\mathbf{\Pi}}_{\mathbf{A}}(\mathbf{x})$ can be expressed analytically in terms of $\hat{\rho}_{\rm A}(\mathbf{x})$:
\begin{align}
    &\hat{\mathbf{\Pi}}_{\mathbf{A},\parallel}(\mathbf{x}) = \hat{\mathbf{P}}_{\parallel}(\mathbf{x}) \nonumber \\ &= \sum_{\alpha}q_\alpha(\hat{\mathbf{r}}_{\alpha} - \mathbf{r}_{\rm A})\cdot \int_0^{1} ds {\bm \delta}^{\parallel}(\mathbf{x}-\mathbf{r}_{\rm A} - s(\hat{\mathbf{r}}_{\alpha} - \mathbf{r}_{\rm A})),
    \end{align}
    where the second equality can be shown to be consistent with Eq.~\eqref{eq:PL}. For this work, we take $\mathbf{r}_{\rm A} = \mathbf{0}$ hereafter without loss of generality. Note that, more generally, one can also consider situations where $\mathbf{r}_{\rm A} \rightarrow \hat{\mathbf{r}}_{\rm A}$ corresponds to a dynamical degree of freedom (e.g., a center-of-mass coordinate)~\cite{Buhmann2004Nov, Scheel2009Feb}.
    
    By means of a canonical transformation, we can define new \emph{unconstrained} canonical coordinates $\hat{\mathbf{A}}_{\perp}$ and $\hat{\mathbf{\Pi}} = \hat{\mathbf{\Pi}}_{\mathbf{A}} - \hat{\mathbf{P}}$~\cite{Weinberg1995Jun}. In this new coordinate system, the field variables are manifestly transverse, and the Gauss's law constraint simply becomes ${\bm \nabla} \cdot \hat{\mathbf{\Pi}} = 0$. One can then verify that we have the modified commutators from Sec.~\ref{sec:can_quantization}, which now take their usual canonical form.}

\section{Construction of discrete modes from the bosonic continuum and relationship to quantized quasinormal mode theory}\label{sec:appA}
In this appendix, we first give detail in Sec.~\ref{ssec:discrete} on how discrete modes can be constructed out of the continuous mode expansion expressed in terms of the macroscopic QED operators $\hat{\mathbf{b}}(\mathbf{r},\omega)$, $\hat{\mathbf{b}}^{\dagger}(\mathbf{r},\omega)$, as was used to discuss mode truncation in Sec.~\ref{ssec:modeT}, and then in Sec.~\ref{ssec:qnm} connect to the important example of quantized QNMs.

\subsection{Construction of discrete modes in system with dispersion and absorption}\label{ssec:discrete}
In general, we can construct a discrete ``modal'' operator as $\hat{a}_{\mu} = \int d^3x \int_0^{\infty} \mathbf{L}_{\mu}(\mathbf{x},\omega) \cdot \hat{\mathbf{b}}(\mathbf{x},\omega)$, where $\mathbf{L}_{\mu}(\mathbf{x},\omega)$ is a function which projects the full bosonic subspace of the Fano diagonalization onto a discrete Fock subspace. For this to be true, we must require $[\hat{a}_{\mu},\hat{a}^{\dagger}_{\eta}] = \delta_{\mu\eta}$, and so
\begin{equation}
    \int d^3x \int_0^{\infty} d\omega \mathbf{L}_{\mu}(\mathbf{x},\omega) \cdot \mathbf{L}^*_{\eta}(\mathbf{x},\omega) = \delta_{\mu\eta}.
\end{equation}

Under this definition, we have $\hat{\mathcal{H}}_{\rm F} = \sum_{\mu \eta} \hbar \chi_{\mu \eta} \hat{a}_{\mu}^{\dagger} \hat{a}_{\eta}$, where 
\begin{equation}\label{eq:chimatrix}
\chi_{\mu\eta} = \int d^3x \int_0^{\infty} d\omega \omega \mathbf{L}_{\mu}(\mathbf{x},\omega) \cdot \mathbf{L}^*_{\eta}(\mathbf{x},\omega),
\end{equation}
\begin{equation}
\hat{P}_{\rm M}\hat{\mathbf{b}}(\mathbf{x},\omega)\hat{P}_{\rm M} = \sum_{\mu} \mathbf{L}_{\mu}^*(\mathbf{x},\omega) \hat{a}_{\mu},
\end{equation}
and
\begin{align}\label{eq:fdef}
   & \mathbf{f}_{\mu}(\mathbf{x}) = \nonumber \\ & \sqrt{\frac{2\chi_{\mu \mu}}{\pi}}\! \! \int \! \! d^3x\blue{'} \! \! \int_0^{\infty} \! \!  \frac{d\omega}{\omega}\sqrt{\epsilon_I(\mathbf{x'},\omega)} \mathbf{G}^{\perp}(\mathbf{x},\mathbf{x'},\omega) \cdot \mathbf{L}_{\mu}^{*}(\mathbf{x'},\omega),
\end{align}
are the mode profiles for the expansion of the vector potential in Eq.~\eqref{eq:aexp}. Alternatively, one could define mode profiles for an expansion of the transverse electric field, as in Eq.~\eqref{eq:efanoc}:
\begin{align}\label{eq:fdefe}
   & \mathbf{f}_{\mu}^{\mathbf{E}_{\perp}}(\mathbf{x}) = \nonumber \\ & \sqrt{\frac{2}{\chi_{\mu \mu}\pi}}\! \! \int \! \! d^3x' \! \! \int_0^{\infty} \! \!  \blue{d\omega}\sqrt{\epsilon_I(\mathbf{x'},\omega)} \mathbf{G}^{\perp}(\mathbf{x},\mathbf{x'},\omega) \cdot \mathbf{L}_{\mu}^{*}(\mathbf{x'},\omega).
\end{align}
\blue{However, to preserve gauge invariance under truncation, the} properly-truncated electric field operator should \blue{instead} be expanded in terms of the functions $\mathbf{f}'_{\mu} = \sum_{\eta} \frac{\chi^*_{\mu \eta}}{\sqrt{\chi_{\mu\mu}\chi_{\eta\eta}}}\mathbf{f}_{\eta}$.

This construction is general, but can be used to describe, e.g., quantized QNMs~\cite{frankequantization}. Additionally, these ``modes'' need not be true modes in the sense that they satisfy Maxwell's equations with the appropriate boundary conditions---rather, they constitute, in the most general case, a truncation of the spatial and frequency-dependent degrees of freedom of the ``Fano-diagonalized'' polariton operators in which the electromagnetic and passive medium fields are expressed as a linear functional thereof.

By appropriate choice of projection functions, the untruncated results can be recovered by imposing an appropriate completeness relation.
Specifically, if the projection functions satisfy: 
\begin{equation}\label{eq:A6}
    \sum_{\mu}\mathbf{L}_{\mu}(\mathbf{x},\omega)\mathbf{L}^{*}_{\mu}(\mathbf{x'},\omega') = \mathbf{I}\delta(\mathbf{x}-\mathbf{x'})\delta(\omega-\omega'),
\end{equation}
then,
\begin{align}
& \sum_{\mu\eta} \frac{\chi_{\mu \eta}}{\sqrt{\chi_{\mu \mu} \chi_{\eta \eta}}} \mathbf{f}_{\mu}(\mathbf{x})\mathbf{f}^*_{\eta}(\mathbf{x'})  \nonumber \\ 
&= \frac{2}{\pi}\int d^3y \int_0^{\infty}\frac{d\omega}{\omega}\epsilon_I(\mathbf{y},\omega) \mathbf{G}^{\perp}(\mathbf{x},\mathbf{y},\omega) \! \cdot \! \mathbf{G}^{\perp,*}(\mathbf{y},\mathbf{x'},\omega) \nonumber \\ & = \frac{2}{\pi} \int_0^{\infty} \frac{d\omega}{\omega} \text{Im}\{\mathbf{G}^{\perp}(\mathbf{x},\mathbf{x'},\omega)\} \nonumber \\ & = {\bm \delta}^{\perp}(\mathbf{x}-\mathbf{x'}),
\end{align}
where in the second line we have used Eqs.~\eqref{eq:chimatrix} and~\eqref{eq:A6}, in the third the relations~\eqref{eq:gf1},~\eqref{eq:gf2}, and~\eqref{eq:chimatrix}, and in the fourth an identity related to causality proven in Ref.~\cite{scheel1998qed}. 

Similarly one can show $\sum_{\mu}\mathbf{f}^{\mathbf{E}_{\perp}}_{\mu}(\mathbf{x})\mathbf{f}_{\mu}(\mathbf{x'}) = {\bm \delta}^{\perp}(\mathbf{x}-\mathbf{x'})$. Note that the completeness relation can be put in the more usual form by introducing symmetrized mode functions $\mathbf{f}_{\mu}^{\rm s}(\mathbf{x}) =\sum_{\eta} \left[\chi^{\frac{1}{2}}\right]_{\eta\mu} \mathbf{f}_{\eta}(\mathbf{x})/\sqrt{\chi_{\eta\eta}}$. That the matrix $\boldsymbol{\chi}^{\frac{1}{2}}$ necessarily exists relies on $\boldsymbol{\chi}$ being invertible. From the form of Eq.~\eqref{eq:chimatrix}, and the orthogonality and completeness relations imposed on the projection functions $\mathbf{L}_{\mu}(\mathbf{x},\omega)$, $\boldsymbol{\chi}$ takes the form of an inner product with positive weight function $\omega$, and thus is invertible. The symmetrized mode functions then satisfy $\sum_{\mu} \mathbf{f}_{\mu}^{\rm s}(\mathbf{x})\mathbf{f}_{\mu}^{\rm s, *}(\mathbf{x'}) = {\bm \delta}^{\perp}(\mathbf{x}-\mathbf{x'})$.

For an example, a mode expansion that is complete in the spatial part can be found by using plane waves as the basis set for $\mathbf{L}$. In this case, the spatial part could take the form $\left(\frac{k_0}{2\pi}\right)^{\frac{3}{2}}e^{i\mathbf{k}_n \cdot \mathbf{x}}$ where $\mathbf{k}_{n} = (n_x,n_y,n_z)k_0$, with $n_x,n_y,n_z$ integers, and $k_0$ being a momentum cut-off which would go to zero as the quantization volume $\rightarrow \infty$.

It should be noted that this expansion is not the same as the common \emph{normal mode} approach often taken in dielectric media without dispersion or absorption, where the mode functions are solutions to the Helmholtz equation. In the absence of loss or dispersion, and with closed or periodic boundary conditions, these are the generalized eigenfunctions which satisfy $\int d^3x \epsilon(\mathbf{x}) \mathbf{h}_{i}(\mathbf{x}) \cdot \mathbf{h}_{j}(\mathbf{x}) = \delta_{ij}$.
These functions are \emph{generalized transverse}, in that they satisfy ${\bm \nabla} \cdot (\epsilon(\mathbf{x})\mathbf{h}_i(\mathbf{x})) =0$. Only in the limit $\epsilon_R(\mathbf{x},\omega) \rightarrow \epsilon(\mathbf{x})$ and $\epsilon_I(\mathbf{x},\omega) \rightarrow 0$, and for closed or periodic boundary conditions, can one expand the fields as  linear combinations of these generalized transverse eigenfunctions. In fact, our approach does not even require the modal functions $\mathbf{f}_{\mu}(\mathbf{r})$ to satisfy the Helmholtz equation.

In the case that the projection functions $\mathbf{L}_{\mu}(\mathbf{x},\omega)$ are well-isolated in frequency, one can approximate the matrix $\chi_{\mu \eta}$ in  Eq.~\eqref{eq:chimatrix} as diagonal, with $\omega_{\mu} \equiv \chi_{\mu \mu}$, and in this case the generalized completeness relation, as well as the mode expansions take their more usual forms $\ahatbt_{\perp}(\mathbf{x}) \approx \sum_{\mu}\sqrt{\frac{\hbar}{2\epsilon_0\omega_{\mu}}}\mathbf{f}(\mathbf{x})\hat{a}_{\mu} + \text{H.c.}$, and $\blue{[\hat{\bm{\mathcal{E}}}_{\rm F}(\mathbf{x})]_{\perp}}\approx i\sum_{\mu}\sqrt{\frac{\hbar\omega_{\mu}}{2\epsilon_0}}\mathbf{f}(\mathbf{x})\hat{a}_{\mu} + \text{H.c.}$

As shown in Appendix~\ref{sec:appB}, the development of the arbitrary-gauge quantization of the electromagnetic field is implemented more naturally for lossless and dispersionless systems with closed or periodic boundary conditions by utilizing a constraint equation (analogous to $\chi_2$---Eq.~\eqref{eq:c2}) more suitable for the implementation of the generalized Coulomb and multipolar gauges, as well as an appropriate modified Lagrangian. Such an approach yields mode expansions directly in terms of the generalized eigenfunctions to the Helmholtz equations.
%
%\sh{Perhaps this paragraph should come before the previous one, to avoid mixing up with those modes, with have no overlap}
%good point, fixed!

\subsection{Relationship to quantized quasinormal modes}\label{ssec:qnm}
A particularly noteworthy example of the construction of discrete modes out of the bosonic continuum is in the quantized quasinormal mode (QNM) theory~\cite{frankequantization,franke2020fluctuation,PhysRevA.105.023702}; QNMs, with mode profiles $\tilde{\mathbf{f}}_{\mu}(\mathbf{x})$, are eigenfunctions of the Helmholtz equation, which for open boundary conditions is a non-Hermitian eigenproblem:
\begin{equation}
    {\bm \nabla} \times {\bm \nabla} \times \tilde{\mathbf{f}}_{\mu}(\mathbf{x}) - \frac{\tilde{\omega}_{\mu}^2}{c^2}\epsilon(\mathbf{x},\tilde{\omega}_{\mu})\tilde{\mathbf{f}}_{\mu}(\mathbf{x})=0,
\end{equation}
along with suitable boundary conditions (i.e., the Silver-M{\"u}ller conditions of Eq.~\eqref{eq: SM_Cond_GF}. The eigenfrequency $\tilde{\omega}_{\mu} =\omega_{\mu} -i \gamma_{\mu}$ is the complex QNM frequency, with real part $\omega_{\mu}$ and imaginary (dissipative) part $\gamma_{\mu}$; QNMs allow one to calculate relevant quantum optics quantities such as cavity $Q$ factors, complex effective mode volumes, Purcell factors, and radiative $\beta$ factors~\cite{ 2ndquant2,Kristensen2011,Lalanne_review,Kristensen:20}.
Notably, QNMs are the open-cavity modes for both dielectric resonators and systems with material loss (e.g., also describing localized plasmon modes of metallic resonators, which are also QNMs, with radiative and nonradiative loss). In the case of lossless dielectrics (i.e., with no material loss), the dissipation captured in
$\gamma_\mu$ is entirely through radiative loss.
%\sh{This may have been said earlier, so check I am not repeating the same thing (and also below) - just to make this clear again}
%looks good

For quantized QNMs, the mode projection functions $\mathbf{L}_{\mu}(\mathbf{x},\omega)$ take the form
%\sh{not using $\omega_{\rm m}$ now, though maybe ok to use here, but then change also the function below?}
\begin{equation}\label{eq:A8}
    \mathbf{L}_{\mu}(\mathbf{x},\omega) = \sum_{\eta}\left[S^{-\frac{1}{2}}\right]_{\mu\eta}\sqrt{\frac{\omega_{\eta}}{2\pi}}\frac{\sqrt{\epsilon_I(\mathbf{x},\omega)}}{\tilde{\omega}_{\eta}-\omega}\tilde{\mathbf{F}}'_{\eta}(\mathbf{x},\omega).
\end{equation}
While these are orthonormal, they are not complete, and so approximating the field expansion in terms of purely QNM operators is necessarily a form of mode truncation, which requires the procedures outlined in Sec.~\ref{ssec:modeT} to restore gauge invariance.
The symmetrizing matrix $S_{\mu\eta}$ is given by the frequency integral
\begin{equation}
    S_{\mu\eta}=\frac{\sqrt{\omega_{\mu}\omega_{\eta}}}{2\pi}\int_0^\infty{\rm d}\omega\frac{S_{\mu\eta}^{\rm nrad}(\omega)+S_{\mu\eta}^{\rm rad}(\omega)}{(\omega-\tilde{\omega}_\mu)(\omega-\tilde{\omega}_\eta^*)},
\end{equation}
where
\begin{equation}
   S_{\mu\eta}^{\rm nrad}(\omega)= \int_V{\rm d}^3x \epsilon_I(\mathbf{x},\omega)\tilde{\mathbf{f}}_\mu(\mathbf{x})\cdot\tilde{\mathbf{f}}_\eta^*(\mathbf{x}),
\end{equation}
gives the non-radiative loss contribution through a QNM overlap integral over a region $V$ containing the absorptive material, which is assumed to be separated from the rest of space by a discontinuity in the dielectric function, and
\begin{align}
S^{\rm rad}_{\mu\eta}(\omega)&=
\int_{\mathbb{R}^{3}-V}\! \! \! \!{\rm d}  ^3x \epsilon_I(\mathbf{x},\omega)\tilde{\mathbf{F}}_\mu(\mathbf{x},\omega)\cdot\tilde{\mathbf{F}}_\eta^{*}(\mathbf{x},\omega)
\nonumber \\ 
&=\frac{n_{\rm B}c}{\omega}\oint_{\mathcal{S}_\infty} {\rm d}A_\mathbf{s}\tilde{\mathbf{F}}_{\mu}(\mathbf{s},\omega)\cdot\tilde{\mathbf{F}}_{\eta}^*(\mathbf{s},\omega),
\end{align}
gives the radiative loss through a far-field surface $S_{\infty}$.

The function $\tilde{\mathbf{F}}_{\mu}'(\mathbf{x},\omega)$, in Eq.~\eqref{eq:A8}, is defined piece-wise, where within the absorptive region $V$ it is equal to $\tilde{\mathbf{f}}_{\mu}(\mathbf{x})$, and outside of this region it is defined as $\tilde{\mathbf{F}}_{\mu}(\mathbf{x},\omega)$, which is a regularized form to avoid the divergence of $\tilde{\mathbf{f}}_{\mu}(\mathbf{x})$ in the far-field, and can be obtained from a Dyson equation approach~\cite{franke2020quantized}, or a near-field to far-field transformation~\cite{ren_near-field_2020}.

The transverse Green's function for locations $\mathbf{x}$, %located 
within the volume $V$, can be expanded as~\cite{ge_quasinormal_2014}
\begin{equation}
    \mathbf{G}^{\perp}(\mathbf{x},\mathbf{x'},\omega) = \sum_{\mu}\frac{\omega}{2(\tilde{\omega}_{\mu}-\omega)}\tilde{\mathbf{f}}(\mathbf{x})\tilde{\mathbf{F}}'_{\mu}(\mathbf{x'},\omega).
\end{equation}

Substituting these results into Eq.~\eqref{eq:fdef}, we find, for $\mathbf{x}$ within
%\sh{or near, since for plasmonics, the emitter will always be outside, and we want it to work at those emitter positions}
the volume $V$,
\begin{equation}
\mathbf{f}_{\mu}(\mathbf{x}) = \sum_{\eta} \tilde{\mathbf{f}}_{\eta}(\mathbf{x}) \left[S^{\frac{1}{2}}\right]_{\eta\mu}\sqrt{\frac{\chi_{\mu\mu}}{\omega_{\eta}}}.
\end{equation}
This expression also holds approximately for locations near the volume $V$, e.g., in plasmonic systems where the field is dominated by the QNMs in the near-vicinity of the scattering structure.~\cite{ge_quasinormal_2014,Ge2015Mar}.

The functions $\mathbf{f}_{\mu}(\mathbf{x})$ are similar to symmetrized QNM mode functions which have appeared in previous works. However, previous works have performed the mode expansion in terms of the electric field, as opposed to the vector potential (which plays a more fundamental role in \blue{ultrastrong} light-matter interactions), so the expressions we show here are slightly different.

As a consistency check, note that due to the pole in Eq.~\eqref{eq:A8}, for $Q_{\mu} \equiv \omega_{\mu}/(2\gamma_{\mu}) \gg 1$, the frequency integral in the definition of $\chi_{\mu \mu}$ is sharply peaked about $\omega = \omega_{\mu}$ and thus we can approximate $\chi_{\mu \mu} \approx \omega_{\mu}$. In this common case, we can also expand $\blue{
    [\hat{\bm{\mathcal{E}}}_{\rm F}(\mathbf{x})]_{\perp} \approx i\sum_{\mu}\sqrt{\frac{\hbar\omega_{\mu}}{2\epsilon_0}}\mathbf{f}_{\mu}(\mathbf{x})\hat{a}_{\mu} + \text{H.c.}}$,
which is the same result reached in previous works on quantized QNMs~\cite{frankequantization}.

\section{Quantization and truncation in a dielectric medium without dispersion or dissipation}\label{sec:appB}
For a medium with no dispersion or absorption, the dielectric function is real and independent of frequency such that $\epsilon = \epsilon(\mathbf{x})$. In this case, Maxwell's equations yield eigenmodes of the medium, $\mathbf{h}_{\mu}(\mathbf{x})$,
which satisfy the Helmholtz equation, 
\begin{equation}
    {\bm \nabla} \times {\bm \nabla} \times \mathbf{h}_{\mu}(\mathbf{x}) -\frac{\omega_{\mu}^2}{c^2}\epsilon(\mathbf{x})\mathbf{h}_{\mu}(\mathbf{x}) =0,
\end{equation}
and when solved with 
closed or periodic boundary conditions, yields
 corresponding (real) eigenvalues $\omega_{\mu}$.
The eigenfunctions are generalized transverse, in that they satisfy ${\bm \nabla} \cdot [\epsilon(\mathbf{x})\mathbf{h}_{\mu}(\mathbf{x})]=0$, and can be chosen to be real~\cite{Wubs2003Jul}. 
These are \emph{normal modes}, which are orthogonal in the sense that $\int d^3x \epsilon(\mathbf{x})\mathbf{h}_{\mu}(\mathbf{x})\cdot \mathbf{h}_{\nu}(\mathbf{x}) = \delta_{\mu\nu}$, and complete in the sense that any generalized transverse function can be expanded as a linear combination of them~\cite{Glauber1991Jan,motu3,Wubs2003Jul}. In Sec.~\ref{ssec:c_q_gc}, we perform canonical quantization using Dirac's constrained quantization in the generalized Coulomb gauge, and show how the fields can be expanded in terms of the generalized transverse normal mode expansion; then in Sec.~\ref{ssec:pzw_gen}, we perform a PZW transformation to obtain the result for the generalized multipolar gauge, recovering previously known results~\cite{Glauber1991Jan,motu3,Dalton1997Jul,Wubs2003Jul}. In Sec.~\ref{ssec:gen_trunc}, we truncate the material and mode degrees of freedom for this system in a manner consistent with gauge invariance in the truncated space.

Note for open resonators and cavities, even with no material loss (i.e., a real dielectric),
one should formally use QNMs if performing few mode quantization (as desired in cavity-QED), though  mode quantization using normal modes with heuristic dissipation is often a very good approximation for single mode high-Q resonators outside of the USC regime. However, with several cavity modes, it can be essential to capture the effects of the QNM phase~\cite{deLasson,frankequantization,2108.10194}, even when not considering USC. Performing quantization with material loss and carefully taking the limit of a lossless dielectric naturally recovers a description that is valid for the QNMs of dielectric cavity systems, which is also consistent with the fluctuation-dissipation theorem~\cite{franke2020fluctuation}. As noted in Appendix~\ref{sec:appA}, in that case, the dissipation is fully contained through the {\it radiative} loss, which is an inherent part of the QNM.

\subsection{Canonical quantization in the generalized Coulomb gauge}\label{ssec:c_q_gc}
While results for a nonabsorbing and nondispersive dielectric can be obtained as a limiting case of the theory presented in the main text~\cite{franke2020fluctuation}, it is more convenient to start with a different effective Lagrangian which does not explicitly contain the medium reservoir degrees of freedom, and gives the macroscopic Maxwell's equations upon extremization of the action:
\begin{align}
    \mathcal{L}_{\rm r} &= \frac{1}{2}\sum_{\alpha}m_{\alpha}\dot{\mathbf{r}}_{\alpha}^2 + \frac{1}{2}\int d^3x \left[\epsilon_0\epsilon(\mathbf{x})\mathbf{E}^2 - \frac{1}{\mu_0}\mathbf{B}^2\right] \nonumber \\ & - \int d^3x J_{\rm A}^{\mu} A_{\mu}.
\end{align}

This Lagrangian gives the same canonical momenta as the one in the main text for the scalar potential and particle coordinates: $\Pi_{\phi}=0$, $\mathbf{p}_{\alpha} = m_{\alpha} \dot{\mathbf{r}}_{\alpha} + q_{\alpha} \mathbf{A}(\mathbf{r}_{\alpha})$, and gives for the vector potential the canonical momentum $    \mathbf{\Pi}_{\mathbf{A}} = -\epsilon_0 \epsilon(\mathbf{x})\mathbf{E}$. The constraint $\chi_0 = \Pi_{\phi}=0$ is thus unchanged, as well as Gauss's law $\chi_1 = {\bm \nabla} \cdot \mathbf{\Pi}_{\mathbf{A}} - \rho_{\rm A} =0$, provided it is expressed in terms of the new canonical momentum.

To quantize this system in the generalized Coulomb gauge, one approach is to use the quantization function $\mathbf{K}(\mathbf{x},\mathbf{x'})$ as in Sec.~\ref{sec:can_quantization}, where the transverse part of this function determines the gauge. We briefly mention how this can be done later, with appropriate constraints on the form of $\mathbf{K}$. Alternatively, we can take a more direct approach, and explicitly let the gauge constraint be $\chi_2 = {\bm \nabla} \cdot (\epsilon \mathbf{A})=0$, and calculate the Dirac brackets from this. Choosing the latter approach, the nonzero elements of the constraint matrix are $C_{12}(\mathbf{x},\mathbf{x'})$ and $C_{21}(\mathbf{x},\mathbf{x'}) = -C_{12}(\mathbf{x'},\mathbf{x})$, where
\begin{equation}
    C_{12}(\mathbf{x},\mathbf{x'}) ={\bm \nabla}^{\mathbf{x}} \cdot\left[\epsilon(\mathbf{x}){\bm \nabla}^{\mathbf{x}} \delta(\mathbf{x}-\mathbf{x'})\right].
\end{equation}
This implies 
\begin{equation}\label{eq:b3}
    -{\bm \nabla}^{\mathbf{x}} \cdot \left[\epsilon(\mathbf{x}) {\bm \nabla}^{\mathbf{x}}C^{-1}_{12}(\mathbf{x},\mathbf{x'})\right] = \delta(\mathbf{x}-\mathbf{x'}).
\end{equation}
While we do not solve for an explicit form for the inverse constraint matrix, that $\mathbf{C}_{12}^{-1}$ satisfies Eq.~\eqref{eq:b3} is in fact the only information needed to quantize the theory.

Constructing the Dirac bracket between the field coordinate and momentum, we obtain,
\begin{align}
    Q_{ij}(\mathbf{x},\mathbf{x'}) &\equiv \{A_i(\mathbf{x}),\Pi_{\mathbf{A} j}(\mathbf{x'})\}_{\rm D}  \nonumber \\ & =\delta_{ij}\delta (\mathbf{x}-\mathbf{x'}) - \epsilon(\mathbf{x'})\nabla_j^{\mathbf{x'}}\nabla_i^{\mathbf{x}}C_{12}^{-1}(\mathbf{x},\mathbf{x'}).
\end{align}
We can show that this function is generalized transverse in its first argument, in that it satisfies:
\begin{equation}
\nabla_i^{\mathbf{x}}\left[\epsilon(\mathbf{x}) Q_{ij}(\mathbf{x},\mathbf{x'})\right] = 0.
\end{equation}
To prove this, we use Eq.~\eqref{eq:b3}, obtaining
\begin{align}
     & \nabla_i^{\mathbf{x}}\left[\epsilon(\mathbf{x}) Q_{ij}(\mathbf{x},\mathbf{x'})\right] \nonumber \\ & = \nabla_j^{\mathbf{x}}\left[\epsilon(\mathbf{x})\delta(\mathbf{x}-\mathbf{x'})\right] + \epsilon(\mathbf{x'})\nabla_{j}^{\mathbf{x'}}\delta(\mathbf{x}-\mathbf{x'}),
\end{align}
and to show this is zero, we can integrate against a test function:
\begin{equation}
    \int d^3x' \nabla_i^{\mathbf{x}}\left[\epsilon(\mathbf{x}) Q_{ij}(\mathbf{x},\mathbf{x'})\right]f_{j}(\mathbf{x'}) =0.
\end{equation}

Similarly, one can show that $\nabla_j^{\mathbf{x'}}Q_{ij}(\mathbf{x},\mathbf{x'}) =0$, i.e., it is transverse with respect to its second argument). 
Since $\blue{Q}_{ij}(\mathbf{x},\mathbf{x'})$ is generalized transverse with respect to $\mathbf{x}$ and $i$, it must have an expansion within the generalized eigenfunctions of the Helmholtz equation: $Q_{ij}(\mathbf{x},\mathbf{x'}) = \sum_{\mu} Q_{\mu,j}(\mathbf{x'})h_{\mu,i}(\mathbf{x})$. The expansion coefficients $Q_{\mu,j}(\mathbf{x'})$ can be found as $\int d^3x \mathbf{h}_{\mu}(\mathbf{x}) \cdot \mathbf{Q}(\mathbf{x},\mathbf{x'})$, and so
\begin{align}\label{eq:Qexp}
    Q_{\mu,j}(\mathbf{x'}) & = \epsilon(\mathbf{x'})h_{\mu,j}(\mathbf{x'}) \nonumber \\ & - \epsilon(\mathbf{x'}){\nabla}_j^{\mathbf{x'}}\int d^3x \epsilon(\mathbf{x})h_{\mu,i}(\mathbf{x})\nabla_i^{\mathbf{x}}C_{12}^{-1}(\mathbf{x},\mathbf{x'}).
\end{align}
Since the integral in the second line of Eq.~\eqref{eq:Qexp} consists of a transverse function $\epsilon \mathbf{h}_{\mu}$ integrated against a longitudinal function ${\bm \nabla}C_{12}^{-1}$, it vanishes. As such, we can write
\begin{equation}
    Q_{ij}(\mathbf{x},\mathbf{x'}) = \sum_{\mu} \epsilon(\mathbf{x'})h_{\mu,i}(\mathbf{x})h_{\mu,j}(\mathbf{x'}).
\end{equation}
In this form, it can be seen that $\mathbf{Q}$ is the generalized transverse delta function, introduced in other works~\cite{Dalton1997Jul,Wubs2003Jul}. For any vector function $\mathbf{f}(\mathbf{x})$, one can define
\begin{align}\label{eq:genT_def}
    \left[\hat{\mathbf{f}}(\mathbf{x})\right]_{\perp}^{(\epsilon)} &= \int d^3x \mathbf{Q}(\mathbf{x},\mathbf{x'}) \cdot \mathbf{f}(\mathbf{x'})\nonumber \\ &
    = \sum_{\mu} \mathbf{h}_{\mu}(\mathbf{x})\int d^3x' \epsilon(\mathbf{x'})\mathbf{h}_{\mu}(\mathbf{x'}) \cdot \mathbf{f}(\mathbf{x'}),
\end{align}
where $ \left[\mathbf{f}(\mathbf{x})\right]_{\perp}^{(\epsilon)} = \left[\mathbf{f}_{\perp}(\mathbf{x})\right]_{\perp}^{(\epsilon)} $ denotes the generalized transverse component of $\mathbf{f}$, which satisfies
\begin{equation}
    {\bm \nabla} \cdot \left(\epsilon(\mathbf{x}) \left[\hat{\mathbf{f}}(\mathbf{x})\right]_{\perp}^{(\epsilon)}\right) = 0.
\end{equation}

Now, the only other nonzero Dirac brackets are $\{r_{\alpha,i},p_{\alpha',j}\}_{\rm D} =\delta_{\alpha\alpha'}\delta_{ij}$, and 
\begin{equation}
    \{p_{\alpha,i},{\Pi_{\mathbf{A}}}_j(\mathbf{x})\}_{\rm D} = 
    -q_{\alpha}\epsilon(\mathbf{x})\nabla_i^{\mathbf{r}_{\alpha}}\nabla_j^{\mathbf{x}}C_{12}^{-1}(\mathbf{r}_{\alpha},\mathbf{x}).
\end{equation}
Thus, we can construct separable photonic and matter Hilbert spaces by, similarly as in Sec.~\ref{sec:can_quantization}, defining a new canonical variable: $\mathbf{\Pi} = \mathbf{\Pi}_{\mathbf{A}} - \mathbf{P}^{\rm gC}$, where
\begin{equation}
\mathbf{P}^{\rm gC}(\mathbf{x}) = -\int d^3x' \mathbf{F}(\mathbf{x},\mathbf{x'})\rho_{\rm A}(\mathbf{x}),
\end{equation}
and $\mathbf{F}(\mathbf{x},\mathbf{x'}) = -\epsilon(\mathbf{x}){\bm \nabla}^{\mathbf{x}}C_{12}^{-1}(\mathbf{x'},\mathbf{x})$. This function is defined by two important properties: $\mathbf{F}(\mathbf{x},\mathbf{x'})/\epsilon(\mathbf{x})$ is longitudinal, and its longitudinal part is the Green's function for the divergence operator; i.e., $\mathbf{F}_{\parallel}(\mathbf{x},\mathbf{x'}) = \mathbf{K}_{\parallel}(\mathbf{x},\mathbf{x'})$. This latter property ensures that $\mathbf{\Pi}$ is transverse, as can seen from the constraint $\chi_1$ (because ${\bm \nabla} \cdot \mathbf{P}^{\rm gC} = -\rho_{\rm A}$). Note that one could alternatively arrive at the quantized theory in the generalized Coulomb gauge by using a quantization function $\mathbf{K}$ which satisfies these same properties as $\mathbf{F}$ as the quantization function in the approach from Sec.~\ref{sec:can_quantization}.

We can now promote Dirac brackets (multiplied by $i\hbar$) to commutators in the usual fashion, and promote canonical variables to operators. We thus have
\begin{equation}\label{eq:comm_d}
    [\hat{\mathbf{A}}(\mathbf{x}),\hat{\mathbf{\Pi}}(\mathbf{x'})] = i\hbar \mathbf{Q}(\mathbf{x},\mathbf{x'}), 
\end{equation}
which implies that the fields have modal expansions~\cite{motu3}
\begin{subequations}\label{eq:exp}
\begin{equation}
\hat{\mathbf{A}}(\mathbf{x}) = \sum_{\mu}\sqrt{\frac{\hbar}{2\epsilon_0\omega_{\mu}}}\mathbf{h}_{\mu}(\mathbf{x})\hat{a}_{\mu} + \text{H.c.},
\end{equation}
\begin{equation}\label{eq:piexp}
\hat{\mathbf{\Pi}}(\mathbf{x}) = -i\epsilon_0\epsilon(\mathbf{x})\sum_{\mu}\sqrt{\frac{\hbar\omega_{\mu}}{2\epsilon_0}}\mathbf{h}_{\mu}(\mathbf{x})\hat{a}_{\mu} + \text{H.c.}
\end{equation}
\end{subequations}
We then construct the Hamiltonian in the usual manner, finding
\begin{align}
    \hat{H} = &  \sum_{\alpha}\frac{\left[\hat{\mathbf{p}}_{\alpha} - q_{\alpha} \hat{\mathbf{A}}(\hat{\mathbf{r}}_{\alpha})\right]^2}{2m_{\alpha}}  \nonumber \\ &+ \frac{1}{2}\left[\int d^3x \frac{(\hat{\mathbf{\Pi}}+\hat{\mathbf{P}}^{\rm gC})^2}{\epsilon_0 \epsilon(\mathbf{x})} + \frac{\hat{\mathbf{B}}^2}{\mu_0}\right] - \int \! d^3x \hat{\mathbf{P}}^{\rm gC} \cdot {\bm \nabla}\hat{\phi}.
\end{align}

We can now make several simplifications. First, we note that the term $\int \hat{\mathbf{\Pi}} \cdot \hat{\mathbf{P}}^{\rm gC}/\epsilon(\mathbf{x})$ vanishes, because $\hat{\mathbf{\Pi}}$ is transverse, and $\hat{\mathbf{P}}^{\rm gC}/\epsilon(\mathbf{x})$ is longitudinal. Second, the last term can be written as
\begin{equation}
     \int d^3x \hat{\mathbf{P}^{\rm gC}} \cdot {\bm \nabla} \hat{\phi} = \int d^3x \frac{\hat{\mathbf{P}}^{\rm gC}}{\epsilon(\mathbf{x})}\left(\epsilon(\mathbf{x}){\bm \nabla}\hat{\phi}\right)_{\parallel},
\end{equation}
but the longitudinal part of $\left(\epsilon(\mathbf{x}){\bm \nabla}\hat{\phi}\right)$ can be found from the constraint $\chi_1$, by noting that since $\epsilon(\mathbf{x})\dot{\mathbf{A}}$ is transverse; then the constraint simply becomes \begin{equation}
    {\bm \nabla} \cdot (\epsilon_0\epsilon(\mathbf{x}){\bm \nabla}\hat{\phi}) = -\hat{\rho}_{\rm A} = {\bm \nabla} \cdot \hat{\mathbf{P}^{\rm gC}}.
\end{equation}

Thus, the total Hamiltonian in the generalized Coulomb gauge $\hat{H}^{\blue{\rm gC}}$ can be written
\begin{align}
    \hat{H}^{\blue{\rm gC}} = &  \sum_{\alpha}\frac{\left[\hat{\mathbf{p}}_{\alpha} - q_{\alpha} \hat{\mathbf{A}}(\hat{\mathbf{r}}_{\alpha})\right]^2}{2m_{\alpha}} +\hat{H}_{\rm F} + \hat{V}_{\rm Coul},
\end{align}
with
\begin{align}
    \hat{H}_{\rm F} &=\frac{1}{2}\int d^3x\left[ \frac{\hat{\mathbf{\Pi}}^2}{\epsilon_0 \epsilon(\mathbf{x})} + \frac{\hat{\mathbf{B}}^2}{\mu_0}\right] \nonumber \\ &=
    \sum_{\mu}\hbar\omega_{\mu}\hat{a}_{\mu}^{\dagger}\hat{a}_{\mu},
\end{align}
where in the second line we have used the expansions in Eq.~\eqref{eq:exp} and dropped the zero point energy, 
and
\begin{equation} \hat{V}_{\rm Coul} = -\int d^3x \frac{\hat{\mathbf{P}}^{\rm gC}\cdot \hat{\mathbf{P}}^{\rm gC}_{\parallel}}{2\epsilon_0\epsilon(\mathbf{x})}.
\end{equation}
Explicit expressions can be derived for $\hat{\mathbf{P}}^{\rm gC}$, and thus $\hat{V}_{\rm Coul}$ using the following method. First, note that the longitudinal part is already fixed by Gauss's law:
\begin{align}
    \hat{\mathbf{P}}^{\rm gC}_{\parallel}(\mathbf{x}) & 
    = - \int d^3x' \mathbf{K}_{\parallel}(\mathbf{x},\mathbf{x'})\hat{\rho}_{\rm A}(\mathbf{x'}) \nonumber \\ &
    = {\bm \nabla}^{\mathbf{x}} \int d^3x' \frac{\hat{\rho}_{\rm A}(\mathbf{x'})}{4\pi|\mathbf{x}-\mathbf{x'}|}.
    \end{align}
    
    Now we can introduce the auxiliary function $\hat{\mathbf{N}}(\mathbf{x}) = (\hat{\mathbf{P}}^{\rm gC} - \hat{\mathbf{P}}^{\rm gC}_{\parallel})/\epsilon(\mathbf{x})$, which is generalized transverse. Thus, this function can be expanded in terms of the modal functions $\hat{\mathbf{N}}(\mathbf{x}) = \sum_{\mu} \hat{n}_{\mu} \mathbf{h}_{\mu}(\mathbf{x})$, where $\hat{n}_{\mu}$ are operator-valued expansion coefficients. These can be found as
    \begin{align}
        \hat{n}_{\mu} &= \int d^3x \mathbf{h}_{\mu}(\mathbf{x}) \cdot \left[ \hat{\mathbf{P}^{\rm gC}}(\mathbf{x})-\hat{\mathbf{P}^{\rm gC}}_{\parallel}(\mathbf{x}) \right] \nonumber \\ & =-\int d^3x \mathbf{h}_{\mu}(\mathbf{x}) \cdot \hat{\mathbf{P}}^{\rm gC}_{\parallel}(\mathbf{x}),
    \end{align}
    and thus we can find an explicit expression:
    \begin{align}\label{eq:pprime}
        \hat{\mathbf{P}}^{\rm gC}(\mathbf{x}) &= \hat{\mathbf{P}}^{\rm gC}_{\parallel}(\mathbf{x}) + \epsilon(\mathbf{x})\hat{\mathbf{N}}(\mathbf{x}) \nonumber \\ & = \hat{\mathbf{P}}^{\rm gC}_{\parallel}(\mathbf{x}) -\sum_{\mu} \epsilon(\mathbf{x})\mathbf{h}_{\mu}(\mathbf{x})\int d^3x \hat{\mathbf{P}}^{\rm gC}_{\parallel}(\mathbf{x'}) \cdot \mathbf{h}_{\mu}(\mathbf{x}) \nonumber \\ & = \hat{\mathbf{P}}^{\rm gC}_{\parallel}(\mathbf{x})  - \epsilon(\mathbf{x})\left[\frac{\hat{\mathbf{P}}^{\rm gC}_{\parallel}(\mathbf{x})}{\epsilon(\mathbf{x})}\right]_{\perp}^{(\epsilon)}.
    \end{align}
    
These results fully recover previous findings on quantization in the generalized Coulomb gauge~\cite{Glauber1991Jan,motu3,Dalton1997Jul,Wubs2003Jul}, using a constrained quantization approach. Note that the electric field operator is, in the generalized Coulomb gauge,
\begin{equation}\label{eq:egc}
    \hat{\mathbf{E}}^{\blue{{\rm gC}}}(\mathbf{x}) = -\frac{1}{\epsilon_0\epsilon(\mathbf{x})}\left[\hat{\mathbf{\Pi}}(\mathbf{x}) + \hat{\mathbf{P}}^{\rm gC}(\mathbf{x})\right].
\end{equation}

\subsection{PZW transformation to generalized multipolar gauge}\label{ssec:pzw_gen}
As in Sec.~\ref{sec:quantization}, we can transform from the generalized Coulomb gauge Hamiltonian $\hat{H}^{\blue{\rm gC}}$ to the generalized multipolar Hamiltonian $\hat{H}^{\blue{\rm gmp}}$ by the PZW operator:
\begin{equation}
    \hat{H}^{\blue{\rm gmp}} = \hat{U}^{\dagger}_{\rm PZW}\hat{H}^{\blue{\rm gC}}\hat{U}_{\rm PZW},
\end{equation}
and the unitary operator takes the same form as the usual PZW transformation:
\begin{equation}
    \hat{U}_{\rm PZW} = \exp{\left[\frac{i}{\hbar}\int d^3x \hat{\mathbf{A}}(\mathbf{x}) \cdot \hat{\mathbf{Z}}_{\rm A}(\mathbf{x})\right]}.
\end{equation}
Note  we have used the variable $\mathbf{Z}_{\rm A}(\mathbf{x}) = \sum_{\alpha} q_{\alpha}\hat{\mathbf{r}}_{\alpha} \int_0^1 ds\delta(\mathbf{x}-s\hat{\mathbf{r}}_{\alpha})$ from the main text to represent a quantity usually referred to as the polarization in this context, rather than $\mathbf{P}$, to avoid confusion with the gauge-dependent polarization defined in the main text and previous subsection, which is used in the canonical transformation from $\hat{\mathbf{\Pi}}_{\mathbf{A}}$ to $\hat{\mathbf{\Pi}}$.

Using the commutation relation~\eqref{eq:comm_d}, we can find the transformation of the canonical momentum coordinate for the field as
\begin{equation}\label{eq:tran_pi}
    \hat{U}^{\dagger}_{\rm PZW}\mathbf{\hat{\Pi}}(\mathbf{x})\hat{U}_{\rm PZW} = \mathbf{\hat{\Pi}}(\mathbf{x}) + \int d^3x' \hat{\mathbf{Z}}_{\rm A}(\mathbf{x'}) \cdot \mathbf{Q}(\mathbf{x'},\mathbf{x}).
\end{equation}
Note that the second term in Eq.~\eqref{eq:tran_pi} is not the generalized transverse component of the polarization $\hat{\mathbf{Z}}_{\rm A}$; rather, the second term, when integrated against a vector function $\mathbf{f}(\mathbf{x})$, is an integral over the dot product of $\hat{\mathbf{Z}}_{\rm A}$ and the generalized transverse component of $\mathbf{f}(\mathbf{x})$.

Applying the PZW \blue{transformation} to the Hamiltonian term involving $\hat{\mathbf{\Pi}}$, we thus obtain:
\begin{align}\label{eq:Ht1}
    &\hat{U}^{\dagger}_{\rm PZW}\int d^3x \frac{\hat{\mathbf{\Pi}}^2}{2\epsilon_0 \epsilon(\mathbf{x})}\hat{U}_{\rm PZW}  \nonumber \\ &=\int d^3x \frac{\hat{\mathbf{\Pi}}^2}{2\epsilon_0 \epsilon(\mathbf{x})}  + \int d^3x \frac{\hat{\mathbf{\Pi}} \cdot \hat{\mathbf{Z}}_{\rm A}}{\epsilon_0\epsilon(\mathbf{x})} + \int d^3x \frac{(\hat{\mathbf{Z}}_{\rm A}-\hat{\mathbf{P}}^{\rm gC})^2}{2\epsilon_0 \epsilon(\mathbf{x})}.
\end{align}
To arrive at the second term in Eq.~\eqref{eq:Ht1}, we have used the fact that $\hat{\mathbf{\Pi}}/\epsilon(\mathbf{x})$ is already generalized transverse. For the third term, we first note that we can use the identity 
\begin{equation}\label{eq:qidentity}
\epsilon(\mathbf{x'})Q_{ji}(\mathbf{x'},\mathbf{x}) = \epsilon(\mathbf{x})Q_{ij}(\mathbf{x},\mathbf{x'})
\end{equation} 
to write it as
\begin{equation}
    \int d^3x\frac{\epsilon(\mathbf{x})}{2\epsilon_0}\left(\left[\frac{\hat{\mathbf{Z}}_{\rm A}(\mathbf{x})}{\epsilon(\mathbf{x})}\right]_{\perp}^{(\epsilon)}\right)^2.
\end{equation}
Then, we use the decomposition $\hat{\mathbf{Z}}_{\rm A} = \hat{\mathbf{Z}}_{\rm A,\perp} + \hat{\mathbf{Z}}_{\rm A,\parallel}$, and subsequently,
\begin{equation}\label{eq:z_identity}
   \left[\frac{\hat{\mathbf{Z}}_{\rm A}(\mathbf{x})}{\epsilon(\mathbf{x})}\right]_{\perp}^{(\epsilon)} = \frac{\hat{\mathbf{Z}}_{\rm A, \perp}(\mathbf{x})}{\epsilon(\mathbf{x})} + \left[\frac{\hat{\mathbf{Z}}_{\rm A, \parallel}(\mathbf{x})}{\epsilon(\mathbf{x})}\right]_{\perp}^{(\epsilon)}.
\end{equation}
Finally, we use Eq.~\eqref{eq:pprime}, and note that $\hat{\mathbf{Z}}_{\rm A,\parallel} = \hat{\mathbf{P}}^{\rm gC}_{\parallel}$. The entire generalized multipolar gauge Hamiltonian is then, neglecting the magnetic terms as in the main text,
\begin{equation}\label{eq:Hgmp}
    \hat{H}^{\blue{\rm gmp}} = \sum_{\alpha}\frac{\hat{\mathbf{p}}_{\alpha}^2}{2m_{\alpha}} + \hat{H}_{\rm F} + \int d^3x \frac{\hat{\mathbf{\Pi}} \cdot \hat{\mathbf{Z}}_{\rm A}}{\epsilon_0\epsilon(\mathbf{x})} + \int d^3x \frac{\hat{\mathbf{Z}}_{\rm A}^2}{2\epsilon_0 \epsilon(\mathbf{x})},
\end{equation}
again in agreement with Refs.~\cite{motu3,Dalton1997Jul,Wubs2003Jul}.

The electric field operator in the generalized multipolar gauge is, using Eqs.~\eqref{eq:egc} and~\eqref{eq:tran_pi},
\begin{align}
      \hat{\mathbf{E}}^{\blue{\rm gmp}}(\mathbf{x}) &=  \hat{\mathbf{E}}^{\blue{\rm gC}}(\mathbf{x}) -\int d^3x \frac{\hat{\mathbf{Z}}_{\rm A}(\mathbf{x'}) \cdot \mathbf{Q}(\mathbf{x'},\mathbf{x})}{\epsilon_0\epsilon(\mathbf{x})} \nonumber \\ 
      & = \hat{\mathbf{E}}^{\blue{\rm gC}}(\mathbf{x}) - 
      \left[\frac{\hat{\mathbf{Z}}_{\rm A}(\mathbf{x})}{\epsilon_0\epsilon(\mathbf{x})}\right]_{\perp}^{(\epsilon)} \nonumber \\ 
      & =\hat{\mathbf{E}}^{\blue{\rm gC}}(\mathbf{x}) - 
     \frac{1}{\epsilon_0\epsilon(\mathbf{x})} \left[\hat{\mathbf{Z}}_{\rm A,\perp}(\mathbf{x}) - \hat{\mathbf{P}}^{\rm gC}_{\perp}(\mathbf{x})\right] \nonumber \\ 
     & = \blue{\left[\hat{\mathbf{E}}_{\rm F}(\mathbf{x}) \right]_{\perp}}- \frac{\hat{\mathbf{Z}}_{\rm A}(\mathbf{x})}{\epsilon_0\epsilon(\mathbf{x})},
\end{align}
where in the second line we have used Eqs.~\eqref{eq:genT_def} and~\eqref{eq:qidentity}, in the third line we have used Eqs.~\eqref{eq:pprime} and~\eqref{eq:z_identity}, and in the final line we have used defined $\blue{[\hat{\mathbf{E}}_{\rm F}(\mathbf{x}) ]_{\perp}}= -\hat{\mathbf{\Pi}}(\mathbf{x})/[\epsilon_0\epsilon(\mathbf{x})]$ as the part of the transverse electric field operator that can be expanded in terms of bosonic operators, and Eq.~\eqref{eq:egc}.

\subsection{Material and Mode Truncation}\label{ssec:gen_trunc}

In this subsection, we show how to introduce material and mode truncation in a manner which preserves gauge invariance. The procedure is essentially the same as in Sec.~\ref{sec:GI}, but considering only two gauges (generalized Coulomb and generalized multipolar).

First note that we can write the generalized Coulomb gauge Hamiltonian in the following form:
\begin{equation}
    \hat{H}^{\blue{\rm gC}} = \hat{H}_{\rm F} + \hat{U}_{\rm PZW}\hat{H}_0\hat{U}^{\dagger}_{\rm PZW},
\end{equation}
where $\hat{H}_0 = \sum_{\alpha} \frac{\hat{\mathbf{p}}_{\alpha}^2}{2m_{\alpha}} + \hat{V}_{\rm Coul}$, and again we have neglected the magnetic terms. From this form, it is easy to see that we can write the generalized multipolar Hamiltonian as 
\begin{equation}
    \hat{H}^{\blue{\rm gmp}} = \hat{U}^{\dagger}_{\rm PZW}\hat{H}_{\rm F}\hat{U}_{\rm PZW} + \hat{H}_0.
\end{equation}

As in Sec.~\ref{ssec:matT}, we can introduce gauge-invariant interactions within the truncated material space by assuming $\hat{H}_0$ to be represented by a discrete energy basis of a few material eigenstates, found in the absence of coupling with the medium. We then truncate the material particle position degrees of freedom in the operator $\hat{U}_{\rm PZW}$ to obtain the correct materially-truncated Hamiltonian in either gauge.
Note that this procedure, as pointed out in Ref.~\cite{Wubs2003Jul}, neglects the influence of local variations in $\epsilon(\mathbf{x})$ on the energy structure of the truncated material system. Assuming a material system of molecular or atomic scales, this variation can sometimes reduce the free-space Coulomb interaction between material particles by approximately a factor of $1/\epsilon(\mathbf{x}_0)$, at the location of the system of particles $\mathbf{x}_0$, but a full treatment of this requires a model of dispersion, and is beyond the scope of this formalism~\cite{Wubs2003Jul}. We can partially circumvent this problem by instead assuming the truncated $\hat{\mathcal{H}}_0$ to consist of the medium-modified energy levels, and focus on the interactions with the transverse field $\hat{\mathbf{\Pi}}$.

To truncate the mode expansion, we can use the mode truncation operator defined in Eq.~\eqref{eq:pmdef}; this is equivalent to letting the sum in the field expansions~\eqref{eq:exp} run over a finite number of modes, as well as in $\hat{H}_{\rm F} = \sum_{\mu}\hbar\omega_{\mu}\hat{a}^{\dagger}_{\mu}\hat{a}_{\mu}$. As with material truncation, we should apply this directly to the unitary operators which generate the minimal coupling Hamiltonian.

Applying both material and mode truncation, then, we find $\hat{U}_{\rm PZW} \rightarrow \hat{\tilde{\mathcal{U}}}_{\rm PZW}$, such that
\begin{equation}
    \hat{\tilde{\mathcal{U}}}_{\rm PZW} = \exp{\left[\frac{i}{\hbar}\int d^3x \ahatbt(\mathbf{x}) \cdot \hat{\bm{\mathcal{Z}}}_{\rm A}(\mathbf{x})\right]}.
\end{equation}
Thus, we find the correctly-truncated generalized Coulomb gauge Hamiltonian:
\begin{equation}\label{eq:gc}
    \hat{\tilde{\mathcal{H}}}^{\blue{\rm gC}} = \hat{\mathcal{H}}_{\rm F} + \hat{\tilde{\mathcal{U}}}_{\rm PZW}\hat{\mathcal{H}}_0\hat{\tilde{\mathcal{U}}}_{\rm PZW}^{\dagger},
\end{equation}
and the correctly-truncated multipolar gauge Hamiltonian
\begin{equation}\label{eq:gmp}
    \hat{\tilde{\mathcal{H}}}^{\blue{\rm gmp}} = \hat{\tilde{\mathcal{U}}}_{\rm PZW}^{\dagger}\hat{\mathcal{H}}_{\rm gC}\hat{\tilde{\mathcal{U}}}_{\rm PZW} = \hat{\tilde{\mathcal{U}}}_{\rm PZW}^{\dagger}\hat{H}_{\rm F}\hat{\tilde{\mathcal{U}}}_{\rm PZW} + \hat{\mathcal{H}}_0,
\end{equation}
and, explicitly
\begin{align}
   & \hat{\tilde{\mathcal{U}}}_{\rm PZW}^{\dagger}\hat{H}_{\rm F}\hat{\tilde{\mathcal{U}}}_{\rm PZW} = \hat{\mathcal{H}}_{\rm F}\nonumber \\ 
    &   - \int \! \!  d^3x \blue{\left[\hat{\bm{\mathcal{E}}}_{\rm F}(\mathbf{x})\right]_{\perp}} \cdot \hat{\bm{\mathcal{Z}}}_{\rm A}(\mathbf{x}) + \sum_{\mu}\frac{\left[\int \! d^3x \hat{\bm{\mathcal{Z}}}_{\rm A}(\mathbf{x}) \! \cdot \! \mathbf{h}_{\mu}(\mathbf{x})\right]^2}{2\epsilon_0},
\end{align}
where $\blue{[\hat{\bm{\mathcal{E}}}_{\rm F}(\mathbf{x})]_{\perp}} = \hat{P}_{\rm M}\blue{[\hat{\mathbf{E}}_{\rm F}(\mathbf{x})]_{\perp}}\hat{P}_{\rm M}$. 
The difference with naive, direct mode truncation of the total Hamiltonian is 
in the third term (the polarization-squared term), as is clear from comparison with Eq.~\eqref{eq:Hgmp}.

\bibliography{main_new.bib}

\begin{thebibliography}{87}%
\makeatletter
\providecommand \@ifxundefined [1]{%
 \@ifx{#1\undefined}
}%
\providecommand \@ifnum [1]{%
 \ifnum #1\expandafter \@firstoftwo
 \else \expandafter \@secondoftwo
 \fi
}%
\providecommand \@ifx [1]{%
 \ifx #1\expandafter \@firstoftwo
 \else \expandafter \@secondoftwo
 \fi
}%
\providecommand \natexlab [1]{#1}%
\providecommand \enquote  [1]{``#1''}%
\providecommand \bibnamefont  [1]{#1}%
\providecommand \bibfnamefont [1]{#1}%
\providecommand \citenamefont [1]{#1}%
\providecommand \href@noop [0]{\@secondoftwo}%
\providecommand \href [0]{\begingroup \@sanitize@url \@href}%
\providecommand \@href[1]{\@@startlink{#1}\@@href}%
\providecommand \@@href[1]{\endgroup#1\@@endlink}%
\providecommand \@sanitize@url [0]{\catcode `\\12\catcode `\$12\catcode
  `\&12\catcode `\#12\catcode `\^12\catcode `\_12\catcode `\%12\relax}%
\providecommand \@@startlink[1]{}%
\providecommand \@@endlink[0]{}%
\providecommand \url  [0]{\begingroup\@sanitize@url \@url }%
\providecommand \@url [1]{\endgroup\@href {#1}{\urlprefix }}%
\providecommand \urlprefix  [0]{URL }%
\providecommand \Eprint [0]{\href }%
\providecommand \doibase [0]{https://doi.org/}%
\providecommand \selectlanguage [0]{\@gobble}%
\providecommand \bibinfo  [0]{\@secondoftwo}%
\providecommand \bibfield  [0]{\@secondoftwo}%
\providecommand \translation [1]{[#1]}%
\providecommand \BibitemOpen [0]{}%
\providecommand \bibitemStop [0]{}%
\providecommand \bibitemNoStop [0]{.\EOS\space}%
\providecommand \EOS [0]{\spacefactor3000\relax}%
\providecommand \BibitemShut  [1]{\csname bibitem#1\endcsname}%
\let\auto@bib@innerbib\@empty
%</preamble>
\bibitem [{\citenamefont {Gruner}\ and\ \citenamefont
  {Welsch}(1996)}]{grunwel}%
  \BibitemOpen
  \bibfield  {author} {\bibinfo {author} {\bibfnamefont {T.}~\bibnamefont
  {Gruner}}\ and\ \bibinfo {author} {\bibfnamefont {D.-G.}\ \bibnamefont
  {Welsch}},\ }\bibfield  {title} {\bibinfo {title} {Green-function approach to
  the radiation-field quantization for homogeneous and inhomogeneous
  {Kramers-Kronig} dielectrics},\ }\href
  {https://doi.org/10.1103/PhysRevA.53.1818} {\bibfield  {journal} {\bibinfo
  {journal} {Phys. Rev. A}\ }\textbf {\bibinfo {volume} {53}},\ \bibinfo
  {pages} {1818} (\bibinfo {year} {1996})}\BibitemShut {NoStop}%
\bibitem [{\citenamefont {Agarwal}(1975{\natexlab{a}})}]{PhysRevA.11.230}%
  \BibitemOpen
  \bibfield  {author} {\bibinfo {author} {\bibfnamefont {G.~S.}\ \bibnamefont
  {Agarwal}},\ }\bibfield  {title} {\bibinfo {title} {Quantum electrodynamics
  in the presence of dielectrics and conductors. i. electromagnetic-field
  response functions and black-body fluctuations in finite geometries},\ }\href
  {https://doi.org/10.1103/PhysRevA.11.230} {\bibfield  {journal} {\bibinfo
  {journal} {Phys. Rev. A}\ }\textbf {\bibinfo {volume} {11}},\ \bibinfo
  {pages} {230} (\bibinfo {year} {1975}{\natexlab{a}})}\BibitemShut {NoStop}%
\bibitem [{\citenamefont {Huttner}\ and\ \citenamefont
  {Barnett}(1992)}]{Huttner}%
  \BibitemOpen
  \bibfield  {author} {\bibinfo {author} {\bibfnamefont {B.}~\bibnamefont
  {Huttner}}\ and\ \bibinfo {author} {\bibfnamefont {S.~M.}\ \bibnamefont
  {Barnett}},\ }\bibfield  {title} {\bibinfo {title} {Quantization of the
  electromagnetic field in dielectrics},\ }\href
  {https://doi.org/10.1103/PhysRevA.46.4306} {\bibfield  {journal} {\bibinfo
  {journal} {Phys. Rev. A}\ }\textbf {\bibinfo {volume} {46}},\ \bibinfo
  {pages} {4306} (\bibinfo {year} {1992})}\BibitemShut {NoStop}%
\bibitem [{\citenamefont {Dung}\ \emph {et~al.}(1998)\citenamefont {Dung},
  \citenamefont {Kn\"oll},\ and\ \citenamefont {Welsch}}]{Dung}%
  \BibitemOpen
  \bibfield  {author} {\bibinfo {author} {\bibfnamefont {H.~T.}\ \bibnamefont
  {Dung}}, \bibinfo {author} {\bibfnamefont {L.}~\bibnamefont {Kn\"oll}},\ and\
  \bibinfo {author} {\bibfnamefont {D.-G.}\ \bibnamefont {Welsch}},\ }\bibfield
   {title} {\bibinfo {title} {Three-dimensional quantization of the
  electromagnetic field in dispersive and absorbing inhomogeneous
  dielectrics},\ }\href {https://doi.org/10.1103/PhysRevA.57.3931} {\bibfield
  {journal} {\bibinfo  {journal} {Phys. Rev. A}\ }\textbf {\bibinfo {volume}
  {57}},\ \bibinfo {pages} {3931} (\bibinfo {year} {1998})}\BibitemShut
  {NoStop}%
\bibitem [{\citenamefont {Scheel}\ \emph {et~al.}(1998)\citenamefont {Scheel},
  \citenamefont {Kn{\"o}ll},\ and\ \citenamefont {Welsch}}]{scheel1998qed}%
  \BibitemOpen
  \bibfield  {author} {\bibinfo {author} {\bibfnamefont {S.}~\bibnamefont
  {Scheel}}, \bibinfo {author} {\bibfnamefont {L.}~\bibnamefont {Kn{\"o}ll}},\
  and\ \bibinfo {author} {\bibfnamefont {D.-G.}\ \bibnamefont {Welsch}},\
  }\bibfield  {title} {\bibinfo {title} {{QED} commutation relations for
  inhomogeneous kramers-kronig dielectrics},\ }\href@noop {} {\bibfield
  {journal} {\bibinfo  {journal} {Physical Review A}\ }\textbf {\bibinfo
  {volume} {58}},\ \bibinfo {pages} {700} (\bibinfo {year} {1998})}\BibitemShut
  {NoStop}%
\bibitem [{\citenamefont {Suttorp}\ and\ \citenamefont
  {Wubs}(2004)}]{suttorp2004field}%
  \BibitemOpen
  \bibfield  {author} {\bibinfo {author} {\bibfnamefont {L.}~\bibnamefont
  {Suttorp}}\ and\ \bibinfo {author} {\bibfnamefont {M.}~\bibnamefont {Wubs}},\
  }\bibfield  {title} {\bibinfo {title} {Field quantization in inhomogeneous
  absorptive dielectrics},\ }\href@noop {} {\bibfield  {journal} {\bibinfo
  {journal} {Physical Review A}\ }\textbf {\bibinfo {volume} {70}},\ \bibinfo
  {pages} {013816} (\bibinfo {year} {2004})}\BibitemShut {NoStop}%
\bibitem [{\citenamefont {Philbin}(2010)}]{Philbin2010Dec}%
  \BibitemOpen
  \bibfield  {author} {\bibinfo {author} {\bibfnamefont {T.~G.}\ \bibnamefont
  {Philbin}},\ }\bibfield  {title} {\bibinfo {title} {{Canonical quantization
  of macroscopic electromagnetism}},\ }\href
  {https://doi.org/10.1088/1367-2630/12/12/123008} {\bibfield  {journal}
  {\bibinfo  {journal} {New J. Phys.}\ }\textbf {\bibinfo {volume} {12}},\
  \bibinfo {pages} {123008} (\bibinfo {year} {2010})}\BibitemShut {NoStop}%
\bibitem [{\citenamefont {Agarwal}(1975{\natexlab{b}})}]{PhysRevA.12.1475}%
  \BibitemOpen
  \bibfield  {author} {\bibinfo {author} {\bibfnamefont {G.~S.}\ \bibnamefont
  {Agarwal}},\ }\bibfield  {title} {\bibinfo {title} {Quantum electrodynamics
  in the presence of dielectrics and conductors. iv. general theory for
  spontaneous emission in finite geometries},\ }\href
  {https://doi.org/10.1103/PhysRevA.12.1475} {\bibfield  {journal} {\bibinfo
  {journal} {Phys. Rev. A}\ }\textbf {\bibinfo {volume} {12}},\ \bibinfo
  {pages} {1475} (\bibinfo {year} {1975}{\natexlab{b}})}\BibitemShut {NoStop}%
\bibitem [{\citenamefont {Kristensen}\ and\ \citenamefont
  {Hughes}(2014)}]{NormKristHughes}%
  \BibitemOpen
  \bibfield  {author} {\bibinfo {author} {\bibfnamefont {P.~T.}\ \bibnamefont
  {Kristensen}}\ and\ \bibinfo {author} {\bibfnamefont {S.}~\bibnamefont
  {Hughes}},\ }\bibfield  {title} {\bibinfo {title} {Modes and mode volumes of
  leaky optical cavities and plasmonic nanoresonators},\ }\href@noop {}
  {\bibfield  {journal} {\bibinfo  {journal} {ACS Photonics}\ }\textbf
  {\bibinfo {volume} {1}},\ \bibinfo {pages} {2} (\bibinfo {year}
  {2014})}\BibitemShut {NoStop}%
\bibitem [{\citenamefont {Kristensen}\ \emph {et~al.}(2020)\citenamefont
  {Kristensen}, \citenamefont {Herrmann}, \citenamefont {Intravaia},\ and\
  \citenamefont {Busch}}]{Kristensen:20}%
  \BibitemOpen
  \bibfield  {author} {\bibinfo {author} {\bibfnamefont {P.~T.}\ \bibnamefont
  {Kristensen}}, \bibinfo {author} {\bibfnamefont {K.}~\bibnamefont
  {Herrmann}}, \bibinfo {author} {\bibfnamefont {F.}~\bibnamefont
  {Intravaia}},\ and\ \bibinfo {author} {\bibfnamefont {K.}~\bibnamefont
  {Busch}},\ }\bibfield  {title} {\bibinfo {title} {Modeling electromagnetic
  resonators using quasinormal modes},\ }\href
  {https://doi.org/10.1364/AOP.377940} {\bibfield  {journal} {\bibinfo
  {journal} {Adv. Opt. Photon.}\ }\textbf {\bibinfo {volume} {12}},\ \bibinfo
  {pages} {612} (\bibinfo {year} {2020})}\BibitemShut {NoStop}%
\bibitem [{\citenamefont {Suttorp}\ and\ \citenamefont {van
  Wonderen}(2004)}]{Suttorp2004Sep}%
  \BibitemOpen
  \bibfield  {author} {\bibinfo {author} {\bibfnamefont {L.~G.}\ \bibnamefont
  {Suttorp}}\ and\ \bibinfo {author} {\bibfnamefont {A.~J.}\ \bibnamefont {van
  Wonderen}},\ }\bibfield  {title} {\bibinfo {title} {{Fano diagonalization of
  a polariton model}},\ }\href {https://doi.org/10.1209/epl/i2004-10131-8}
  {\bibfield  {journal} {\bibinfo  {journal} {Europhys. Lett.}\ }\textbf
  {\bibinfo {volume} {67}},\ \bibinfo {pages} {766} (\bibinfo {year}
  {2004})}\BibitemShut {NoStop}%
\bibitem [{\citenamefont {Weinberg}\ and\ \citenamefont
  {Weinberg}(1995)}]{Weinberg1995Jun}%
  \BibitemOpen
  \bibfield  {author} {\bibinfo {author} {\bibfnamefont {S.}~\bibnamefont
  {Weinberg}}\ and\ \bibinfo {author} {\bibfnamefont {S.}~\bibnamefont
  {Weinberg}},\ }\href
  {https://books.google.com/books/about/The_Quantum_Theory_of_Fields.html?id=doeDB3_WLvwC}
  {\emph {\bibinfo {title} {{The Quantum Theory of Fields, Volume 1}}}}\
  (\bibinfo  {publisher} {Cambridge University Press},\ \bibinfo {address}
  {Cambridge, England, UK},\ \bibinfo {year} {1995})\BibitemShut {NoStop}%
\bibitem [{\citenamefont {Dung}\ \emph {et~al.}(2000)\citenamefont {Dung},
  \citenamefont {Kn\"oll},\ and\ \citenamefont {Welsch}}]{PhysRevA.62.053804}%
  \BibitemOpen
  \bibfield  {author} {\bibinfo {author} {\bibfnamefont {H.~T.}\ \bibnamefont
  {Dung}}, \bibinfo {author} {\bibfnamefont {L.}~\bibnamefont {Kn\"oll}},\ and\
  \bibinfo {author} {\bibfnamefont {D.-G.}\ \bibnamefont {Welsch}},\ }\bibfield
   {title} {\bibinfo {title} {Spontaneous decay in the presence of dispersing
  and absorbing bodies: General theory and application to a spherical cavity},\
  }\href {https://doi.org/10.1103/PhysRevA.62.053804} {\bibfield  {journal}
  {\bibinfo  {journal} {Phys. Rev. A}\ }\textbf {\bibinfo {volume} {62}},\
  \bibinfo {pages} {053804} (\bibinfo {year} {2000})}\BibitemShut {NoStop}%
\bibitem [{\citenamefont {Fleischhauer}(1999)}]{PhysRevA.60.2534}%
  \BibitemOpen
  \bibfield  {author} {\bibinfo {author} {\bibfnamefont {M.}~\bibnamefont
  {Fleischhauer}},\ }\bibfield  {title} {\bibinfo {title} {Spontaneous emission
  and level shifts in absorbing disordered dielectrics and dense atomic gases:
  A {Green's-function} approach},\ }\href
  {https://doi.org/10.1103/PhysRevA.60.2534} {\bibfield  {journal} {\bibinfo
  {journal} {Phys. Rev. A}\ }\textbf {\bibinfo {volume} {60}},\ \bibinfo
  {pages} {2534} (\bibinfo {year} {1999})}\BibitemShut {NoStop}%
\bibitem [{\citenamefont {Yao}\ \emph {et~al.}(2009)\citenamefont {Yao},
  \citenamefont {Van~Vlack}, \citenamefont {Reza}, \citenamefont {Patterson},
  \citenamefont {Dignam},\ and\ \citenamefont {Hughes}}]{PhysRevB.80.195106}%
  \BibitemOpen
  \bibfield  {author} {\bibinfo {author} {\bibfnamefont {P.}~\bibnamefont
  {Yao}}, \bibinfo {author} {\bibfnamefont {C.}~\bibnamefont {Van~Vlack}},
  \bibinfo {author} {\bibfnamefont {A.}~\bibnamefont {Reza}}, \bibinfo {author}
  {\bibfnamefont {M.}~\bibnamefont {Patterson}}, \bibinfo {author}
  {\bibfnamefont {M.~M.}\ \bibnamefont {Dignam}},\ and\ \bibinfo {author}
  {\bibfnamefont {S.}~\bibnamefont {Hughes}},\ }\bibfield  {title} {\bibinfo
  {title} {Ultrahigh purcell factors and lamb shifts in slow-light metamaterial
  waveguides},\ }\href {https://doi.org/10.1103/PhysRevB.80.195106} {\bibfield
  {journal} {\bibinfo  {journal} {Phys. Rev. B}\ }\textbf {\bibinfo {volume}
  {80}},\ \bibinfo {pages} {195106} (\bibinfo {year} {2009})}\BibitemShut
  {NoStop}%
\bibitem [{\citenamefont {Dung}\ \emph {et~al.}(2002)\citenamefont {Dung},
  \citenamefont {Kn\"{o}ll},\ and\ \citenamefont {Welsch}}]{Dung2002}%
  \BibitemOpen
  \bibfield  {author} {\bibinfo {author} {\bibfnamefont {H.~T.}\ \bibnamefont
  {Dung}}, \bibinfo {author} {\bibfnamefont {L.}~\bibnamefont {Kn\"{o}ll}},\
  and\ \bibinfo {author} {\bibfnamefont {D.-G.}\ \bibnamefont {Welsch}},\
  }\bibfield  {title} {\bibinfo {title} {{Resonant dipole-dipole interaction in
  the presence of dispersing and absorbing surroundings}},\ }\href
  {https://doi.org/10.1103/PhysRevA.66.063810} {\bibfield  {journal} {\bibinfo
  {journal} {Phys. Rev. A}\ }\textbf {\bibinfo {volume} {66}},\ \bibinfo
  {pages} {063810} (\bibinfo {year} {2002})}\BibitemShut {NoStop}%
\bibitem [{\citenamefont {Kristensen}\ \emph {et~al.}(2011)\citenamefont
  {Kristensen}, \citenamefont {M{\o}rk}, \citenamefont {Lodahl},\ and\
  \citenamefont {Hughes}}]{Kristensen2011}%
  \BibitemOpen
  \bibfield  {author} {\bibinfo {author} {\bibfnamefont {P.~T.}\ \bibnamefont
  {Kristensen}}, \bibinfo {author} {\bibfnamefont {J.}~\bibnamefont {M{\o}rk}},
  \bibinfo {author} {\bibfnamefont {P.}~\bibnamefont {Lodahl}},\ and\ \bibinfo
  {author} {\bibfnamefont {S.}~\bibnamefont {Hughes}},\ }\bibfield  {title}
  {\bibinfo {title} {{Decay dynamics of radiatively coupled quantum dots in
  photonic crystal slabs}},\ }\href
  {https://doi.org/10.1103/PhysRevB.83.075305} {\bibfield  {journal} {\bibinfo
  {journal} {Phys. Rev. B}\ }\textbf {\bibinfo {volume} {83}},\ \bibinfo
  {pages} {075305} (\bibinfo {year} {2011})}\BibitemShut {NoStop}%
\bibitem [{\citenamefont {Buhmann}\ \emph {et~al.}(2004)\citenamefont
  {Buhmann}, \citenamefont {Kn{\ifmmode\ddot{o}\else\"{o}\fi}ll}, \citenamefont
  {Welsch},\ and\ \citenamefont {Dung}}]{Buhmann2004Nov}%
  \BibitemOpen
  \bibfield  {author} {\bibinfo {author} {\bibfnamefont {S.~Y.}\ \bibnamefont
  {Buhmann}}, \bibinfo {author} {\bibfnamefont {L.}~\bibnamefont
  {Kn{\ifmmode\ddot{o}\else\"{o}\fi}ll}}, \bibinfo {author} {\bibfnamefont
  {D.-G.}\ \bibnamefont {Welsch}},\ and\ \bibinfo {author} {\bibfnamefont
  {H.~T.}\ \bibnamefont {Dung}},\ }\bibfield  {title} {\bibinfo {title}
  {{Casimir-Polder forces: A nonperturbative approach}},\ }\href
  {https://doi.org/10.1103/PhysRevA.70.052117} {\bibfield  {journal} {\bibinfo
  {journal} {Phys. Rev. A}\ }\textbf {\bibinfo {volume} {70}},\ \bibinfo
  {pages} {052117} (\bibinfo {year} {2004})}\BibitemShut {NoStop}%
\bibitem [{\citenamefont {Kamandar~Dezfouli}\ and\ \citenamefont
  {Hughes}(2017)}]{KamandarDezfouli2017May}%
  \BibitemOpen
  \bibfield  {author} {\bibinfo {author} {\bibfnamefont {M.}~\bibnamefont
  {Kamandar~Dezfouli}}\ and\ \bibinfo {author} {\bibfnamefont {S.}~\bibnamefont
  {Hughes}},\ }\bibfield  {title} {\bibinfo {title} {{Quantum Optics Model of
  Surface-Enhanced Raman Spectroscopy for Arbitrarily Shaped Plasmonic
  Resonators}},\ }\href {https://doi.org/10.1021/acsphotonics.7b00157}
  {\bibfield  {journal} {\bibinfo  {journal} {ACS Photonics}\ }\textbf
  {\bibinfo {volume} {4}},\ \bibinfo {pages} {1245} (\bibinfo {year}
  {2017})}\BibitemShut {NoStop}%
\bibitem [{\citenamefont {Frisk~Kockum}\ \emph {et~al.}(2019)\citenamefont
  {Frisk~Kockum}, \citenamefont {Miranowicz}, \citenamefont {De~Liberato},
  \citenamefont {Savasta},\ and\ \citenamefont
  {Nori}}]{frisk_kockum_ultrastrong_2019}%
  \BibitemOpen
  \bibfield  {author} {\bibinfo {author} {\bibfnamefont {A.}~\bibnamefont
  {Frisk~Kockum}}, \bibinfo {author} {\bibfnamefont {A.}~\bibnamefont
  {Miranowicz}}, \bibinfo {author} {\bibfnamefont {S.}~\bibnamefont
  {De~Liberato}}, \bibinfo {author} {\bibfnamefont {S.}~\bibnamefont
  {Savasta}},\ and\ \bibinfo {author} {\bibfnamefont {F.}~\bibnamefont
  {Nori}},\ }\bibfield  {title} {\bibinfo {title} {Ultrastrong coupling between
  light and matter},\ }\href {https://doi.org/10.1038/s42254-018-0006-2}
  {\bibfield  {journal} {\bibinfo  {journal} {Nature Reviews Physics}\ }\textbf
  {\bibinfo {volume} {1}},\ \bibinfo {pages} {19} (\bibinfo {year}
  {2019})}\BibitemShut {NoStop}%
\bibitem [{\citenamefont {Forn-D\'{\i}az}\ \emph {et~al.}(2019)\citenamefont
  {Forn-D\'{\i}az}, \citenamefont {Lamata}, \citenamefont {Rico}, \citenamefont
  {Kono},\ and\ \citenamefont {Solano}}]{RevModPhys.91.025005}%
  \BibitemOpen
  \bibfield  {author} {\bibinfo {author} {\bibfnamefont {P.}~\bibnamefont
  {Forn-D\'{\i}az}}, \bibinfo {author} {\bibfnamefont {L.}~\bibnamefont
  {Lamata}}, \bibinfo {author} {\bibfnamefont {E.}~\bibnamefont {Rico}},
  \bibinfo {author} {\bibfnamefont {J.}~\bibnamefont {Kono}},\ and\ \bibinfo
  {author} {\bibfnamefont {E.}~\bibnamefont {Solano}},\ }\bibfield  {title}
  {\bibinfo {title} {Ultrastrong coupling regimes of light-matter
  interaction},\ }\href {https://doi.org/10.1103/RevModPhys.91.025005}
  {\bibfield  {journal} {\bibinfo  {journal} {Rev. Mod. Phys.}\ }\textbf
  {\bibinfo {volume} {91}},\ \bibinfo {pages} {025005} (\bibinfo {year}
  {2019})}\BibitemShut {NoStop}%
\bibitem [{\citenamefont {De~Bernardis}\ \emph
  {et~al.}(2018{\natexlab{a}})\citenamefont {De~Bernardis}, \citenamefont
  {Jaako},\ and\ \citenamefont {Rabl}}]{DeBernardis2018Apr}%
  \BibitemOpen
  \bibfield  {author} {\bibinfo {author} {\bibfnamefont {D.}~\bibnamefont
  {De~Bernardis}}, \bibinfo {author} {\bibfnamefont {T.}~\bibnamefont
  {Jaako}},\ and\ \bibinfo {author} {\bibfnamefont {P.}~\bibnamefont {Rabl}},\
  }\bibfield  {title} {\bibinfo {title} {{Cavity quantum electrodynamics in the
  nonperturbative regime}},\ }\href
  {https://doi.org/10.1103/PhysRevA.97.043820} {\bibfield  {journal} {\bibinfo
  {journal} {Phys. Rev. A}\ }\textbf {\bibinfo {volume} {97}},\ \bibinfo
  {pages} {043820} (\bibinfo {year} {2018}{\natexlab{a}})}\BibitemShut
  {NoStop}%
\bibitem [{\citenamefont {De~Bernardis}\ \emph
  {et~al.}(2018{\natexlab{b}})\citenamefont {De~Bernardis}, \citenamefont
  {Pilar}, \citenamefont {Jaako}, \citenamefont {De~Liberato},\ and\
  \citenamefont {Rabl}}]{DeBernardis2018Nov}%
  \BibitemOpen
  \bibfield  {author} {\bibinfo {author} {\bibfnamefont {D.}~\bibnamefont
  {De~Bernardis}}, \bibinfo {author} {\bibfnamefont {P.}~\bibnamefont {Pilar}},
  \bibinfo {author} {\bibfnamefont {T.}~\bibnamefont {Jaako}}, \bibinfo
  {author} {\bibfnamefont {S.}~\bibnamefont {De~Liberato}},\ and\ \bibinfo
  {author} {\bibfnamefont {P.}~\bibnamefont {Rabl}},\ }\bibfield  {title}
  {\bibinfo {title} {{Breakdown of gauge invariance in ultrastrong-coupling
  cavity QED}},\ }\href {https://doi.org/10.1103/PhysRevA.98.053819} {\bibfield
   {journal} {\bibinfo  {journal} {Phys. Rev. A}\ }\textbf {\bibinfo {volume}
  {98}},\ \bibinfo {pages} {053819} (\bibinfo {year}
  {2018}{\natexlab{b}})}\BibitemShut {NoStop}%
\bibitem [{\citenamefont {Taylor}\ \emph {et~al.}(2022)\citenamefont {Taylor},
  \citenamefont {Mandal}, \citenamefont {Huo},\ and\ \citenamefont
  {Huo}}]{Taylor2022Mar}%
  \BibitemOpen
  \bibfield  {author} {\bibinfo {author} {\bibfnamefont {M.~A.~D.}\
  \bibnamefont {Taylor}}, \bibinfo {author} {\bibfnamefont {A.}~\bibnamefont
  {Mandal}}, \bibinfo {author} {\bibfnamefont {P.}~\bibnamefont {Huo}},\ and\
  \bibinfo {author} {\bibfnamefont {P.}~\bibnamefont {Huo}},\ }\bibfield
  {title} {\bibinfo {title} {{Resolving ambiguities of the mode truncation in
  cavity quantum electrodynamics}},\ }\href {https://doi.org/10.1364/OL.450228}
  {\bibfield  {journal} {\bibinfo  {journal} {Opt. Lett.}\ }\textbf {\bibinfo
  {volume} {47}},\ \bibinfo {pages} {1446} (\bibinfo {year}
  {2022})}\BibitemShut {NoStop}%
\bibitem [{\citenamefont {Starace}(1971)}]{Starace1971Apr}%
  \BibitemOpen
  \bibfield  {author} {\bibinfo {author} {\bibfnamefont {A.~F.}\ \bibnamefont
  {Starace}},\ }\bibfield  {title} {\bibinfo {title} {{Length and Velocity
  Formulas in Approximate Oscillator-Strength Calculations}},\ }\href
  {https://doi.org/10.1103/PhysRevA.3.1242} {\bibfield  {journal} {\bibinfo
  {journal} {Phys. Rev. A}\ }\textbf {\bibinfo {volume} {3}},\ \bibinfo {pages}
  {1242} (\bibinfo {year} {1971})}\BibitemShut {NoStop}%
\bibitem [{\citenamefont {Keeling}(2007)}]{Keeling2007Jun}%
  \BibitemOpen
  \bibfield  {author} {\bibinfo {author} {\bibfnamefont {J.}~\bibnamefont
  {Keeling}},\ }\bibfield  {title} {\bibinfo {title} {{Coulomb interactions,
  gauge invariance, and phase transitions of the Dicke model}},\ }\href
  {https://doi.org/10.1088/0953-8984/19/29/295213} {\bibfield  {journal}
  {\bibinfo  {journal} {J. Phys.: Condens. Matter}\ }\textbf {\bibinfo {volume}
  {19}},\ \bibinfo {pages} {295213} (\bibinfo {year} {2007})}\BibitemShut
  {NoStop}%
\bibitem [{\citenamefont {Di~Stefano}\ \emph {et~al.}(2019)\citenamefont
  {Di~Stefano}, \citenamefont {Settineri}, \citenamefont
  {Macr{\ifmmode\grave{\imath}\else\`{\i}\fi}}, \citenamefont {Garziano},
  \citenamefont {Stassi}, \citenamefont {Savasta},\ and\ \citenamefont
  {Nori}}]{DiStefano2019Aug}%
  \BibitemOpen
  \bibfield  {author} {\bibinfo {author} {\bibfnamefont {O.}~\bibnamefont
  {Di~Stefano}}, \bibinfo {author} {\bibfnamefont {A.}~\bibnamefont
  {Settineri}}, \bibinfo {author} {\bibfnamefont {V.}~\bibnamefont
  {Macr{\ifmmode\grave{\imath}\else\`{\i}\fi}}}, \bibinfo {author}
  {\bibfnamefont {L.}~\bibnamefont {Garziano}}, \bibinfo {author}
  {\bibfnamefont {R.}~\bibnamefont {Stassi}}, \bibinfo {author} {\bibfnamefont
  {S.}~\bibnamefont {Savasta}},\ and\ \bibinfo {author} {\bibfnamefont
  {F.}~\bibnamefont {Nori}},\ }\bibfield  {title} {\bibinfo {title}
  {{Resolution of gauge ambiguities in ultrastrong-coupling cavity quantum
  electrodynamics}},\ }\href {https://doi.org/10.1038/s41567-019-0534-4}
  {\bibfield  {journal} {\bibinfo  {journal} {Nat. Phys.}\ }\textbf {\bibinfo
  {volume} {15}},\ \bibinfo {pages} {803} (\bibinfo {year} {2019})}\BibitemShut
  {NoStop}%
\bibitem [{\citenamefont {Stokes}\ and\ \citenamefont
  {Nazir}(2019)}]{Stokes2019Jan}%
  \BibitemOpen
  \bibfield  {author} {\bibinfo {author} {\bibfnamefont {A.}~\bibnamefont
  {Stokes}}\ and\ \bibinfo {author} {\bibfnamefont {A.}~\bibnamefont {Nazir}},\
  }\bibfield  {title} {\bibinfo {title} {{Gauge ambiguities imply
  Jaynes-Cummings physics remains valid in ultrastrong coupling QED}},\ }\href
  {https://doi.org/10.1038/s41467-018-08101-0} {\bibfield  {journal} {\bibinfo
  {journal} {Nat. Commun.}\ }\textbf {\bibinfo {volume} {10}},\ \bibinfo
  {pages} {1} (\bibinfo {year} {2019})}\BibitemShut {NoStop}%
\bibitem [{\citenamefont {Taylor}\ \emph {et~al.}(2020)\citenamefont {Taylor},
  \citenamefont {Mandal}, \citenamefont {Zhou},\ and\ \citenamefont
  {Huo}}]{Taylor2020Sep}%
  \BibitemOpen
  \bibfield  {author} {\bibinfo {author} {\bibfnamefont {M.~A.~D.}\
  \bibnamefont {Taylor}}, \bibinfo {author} {\bibfnamefont {A.}~\bibnamefont
  {Mandal}}, \bibinfo {author} {\bibfnamefont {W.}~\bibnamefont {Zhou}},\ and\
  \bibinfo {author} {\bibfnamefont {P.}~\bibnamefont {Huo}},\ }\bibfield
  {title} {\bibinfo {title} {{Resolution of Gauge Ambiguities in Molecular
  Cavity Quantum Electrodynamics}},\ }\href
  {https://doi.org/10.1103/PhysRevLett.125.123602} {\bibfield  {journal}
  {\bibinfo  {journal} {Phys. Rev. Lett.}\ }\textbf {\bibinfo {volume} {125}},\
  \bibinfo {pages} {123602} (\bibinfo {year} {2020})}\BibitemShut {NoStop}%
\bibitem [{\citenamefont {Wilson}(1974)}]{PhysRevD.10.2445}%
  \BibitemOpen
  \bibfield  {author} {\bibinfo {author} {\bibfnamefont {K.~G.}\ \bibnamefont
  {Wilson}},\ }\bibfield  {title} {\bibinfo {title} {Confinement of quarks},\
  }\href {https://doi.org/10.1103/PhysRevD.10.2445} {\bibfield  {journal}
  {\bibinfo  {journal} {Phys. Rev. D}\ }\textbf {\bibinfo {volume} {10}},\
  \bibinfo {pages} {2445} (\bibinfo {year} {1974})}\BibitemShut {NoStop}%
\bibitem [{\citenamefont {Savasta}\ \emph
  {et~al.}(2021{\natexlab{a}})\citenamefont {Savasta}, \citenamefont
  {Di~Stefano}, \citenamefont {Settineri}, \citenamefont {Zueco}, \citenamefont
  {Hughes},\ and\ \citenamefont {Nori}}]{Savasta2021May}%
  \BibitemOpen
  \bibfield  {author} {\bibinfo {author} {\bibfnamefont {S.}~\bibnamefont
  {Savasta}}, \bibinfo {author} {\bibfnamefont {O.}~\bibnamefont {Di~Stefano}},
  \bibinfo {author} {\bibfnamefont {A.}~\bibnamefont {Settineri}}, \bibinfo
  {author} {\bibfnamefont {D.}~\bibnamefont {Zueco}}, \bibinfo {author}
  {\bibfnamefont {S.}~\bibnamefont {Hughes}},\ and\ \bibinfo {author}
  {\bibfnamefont {F.}~\bibnamefont {Nori}},\ }\bibfield  {title} {\bibinfo
  {title} {{Gauge principle and gauge invariance in two-level systems}},\
  }\href {https://doi.org/10.1103/PhysRevA.103.053703} {\bibfield  {journal}
  {\bibinfo  {journal} {Phys. Rev. A}\ }\textbf {\bibinfo {volume} {103}},\
  \bibinfo {pages} {053703} (\bibinfo {year} {2021}{\natexlab{a}})}\BibitemShut
  {NoStop}%
\bibitem [{\citenamefont {Han}\ and\ \citenamefont
  {Madsen}(2010)}]{Han2010Jun}%
  \BibitemOpen
  \bibfield  {author} {\bibinfo {author} {\bibfnamefont {Y.-C.}\ \bibnamefont
  {Han}}\ and\ \bibinfo {author} {\bibfnamefont {L.~B.}\ \bibnamefont
  {Madsen}},\ }\bibfield  {title} {\bibinfo {title} {{Comparison between length
  and velocity gauges in quantum simulations of high-order harmonic
  generation}},\ }\href {https://doi.org/10.1103/PhysRevA.81.063430} {\bibfield
   {journal} {\bibinfo  {journal} {Phys. Rev. A}\ }\textbf {\bibinfo {volume}
  {81}},\ \bibinfo {pages} {063430} (\bibinfo {year} {2010})}\BibitemShut
  {NoStop}%
\bibitem [{\citenamefont {Li}\ \emph {et~al.}(2020)\citenamefont {Li},
  \citenamefont {Golez}, \citenamefont {Mazza}, \citenamefont {Millis},
  \citenamefont {Georges},\ and\ \citenamefont {Eckstein}}]{Li2020May}%
  \BibitemOpen
  \bibfield  {author} {\bibinfo {author} {\bibfnamefont {J.}~\bibnamefont
  {Li}}, \bibinfo {author} {\bibfnamefont {D.}~\bibnamefont {Golez}}, \bibinfo
  {author} {\bibfnamefont {G.}~\bibnamefont {Mazza}}, \bibinfo {author}
  {\bibfnamefont {A.~J.}\ \bibnamefont {Millis}}, \bibinfo {author}
  {\bibfnamefont {A.}~\bibnamefont {Georges}},\ and\ \bibinfo {author}
  {\bibfnamefont {M.}~\bibnamefont {Eckstein}},\ }\bibfield  {title} {\bibinfo
  {title} {{Electromagnetic coupling in tight-binding models for strongly
  correlated light and matter}},\ }\href
  {https://doi.org/10.1103/PhysRevB.101.205140} {\bibfield  {journal} {\bibinfo
   {journal} {Phys. Rev. B}\ }\textbf {\bibinfo {volume} {101}},\ \bibinfo
  {pages} {205140} (\bibinfo {year} {2020})}\BibitemShut {NoStop}%
\bibitem [{\citenamefont {Dalton}\ \emph {et~al.}(1999)\citenamefont {Dalton},
  \citenamefont {Barnett},\ and\ \citenamefont {Knight}}]{Dalton1999Jul}%
  \BibitemOpen
  \bibfield  {author} {\bibinfo {author} {\bibfnamefont {B.~J.}\ \bibnamefont
  {Dalton}}, \bibinfo {author} {\bibfnamefont {S.~M.}\ \bibnamefont
  {Barnett}},\ and\ \bibinfo {author} {\bibfnamefont {P.~L.}\ \bibnamefont
  {Knight}},\ }\bibfield  {title} {\bibinfo {title} {{Quasi mode theory of
  macroscopic canonical quantization in quantum optics and cavity quantum
  electrodynamics}},\ }\href {https://doi.org/10.1080/09500349908231338}
  {\bibfield  {journal} {\bibinfo  {journal} {J. Mod. Opt.}\ }\textbf {\bibinfo
  {volume} {46}},\ \bibinfo {pages} {1315} (\bibinfo {year}
  {1999})}\BibitemShut {NoStop}%
\bibitem [{\citenamefont {Fussell}\ and\ \citenamefont
  {Dignam}(2008)}]{Fussell2008May}%
  \BibitemOpen
  \bibfield  {author} {\bibinfo {author} {\bibfnamefont {D.~P.}\ \bibnamefont
  {Fussell}}\ and\ \bibinfo {author} {\bibfnamefont {M.~M.}\ \bibnamefont
  {Dignam}},\ }\bibfield  {title} {\bibinfo {title} {{Quasimode-projection
  approach to quantum-dot{\textendash}photon interactions in
  photonic-crystal-slab coupled-cavity systems}},\ }\href
  {https://doi.org/10.1103/PhysRevA.77.053805} {\bibfield  {journal} {\bibinfo
  {journal} {Phys. Rev. A}\ }\textbf {\bibinfo {volume} {77}},\ \bibinfo
  {pages} {053805} (\bibinfo {year} {2008})}\BibitemShut {NoStop}%
\bibitem [{\citenamefont {Vendromin}\ and\ \citenamefont
  {Dignam}(2022)}]{Vendromin2022Oct}%
  \BibitemOpen
  \bibfield  {author} {\bibinfo {author} {\bibfnamefont {C.}~\bibnamefont
  {Vendromin}}\ and\ \bibinfo {author} {\bibfnamefont {M.~M.}\ \bibnamefont
  {Dignam}},\ }\bibfield  {title} {\bibinfo {title} {{Nonlinear optical
  generation of entangled squeezed states in lossy nonorthogonal quasimodes: an
  analytic solution}},\ }\bibfield  {journal} {\bibinfo  {journal} {arXiv}\
  }\href {https://doi.org/10.48550/arXiv.2210.06521}
  {10.48550/arXiv.2210.06521} (\bibinfo {year} {2022}),\ \Eprint
  {https://arxiv.org/abs/2210.06521} {2210.06521} \BibitemShut {NoStop}%
\bibitem [{\citenamefont {Ching}\ \emph {et~al.}(1998)\citenamefont {Ching},
  \citenamefont {Leung}, \citenamefont {Maassen van~den Brink}, \citenamefont
  {Suen}, \citenamefont {Tong},\ and\ \citenamefont {Young}}]{2ndquant2}%
  \BibitemOpen
  \bibfield  {author} {\bibinfo {author} {\bibfnamefont {E.~S.~C.}\
  \bibnamefont {Ching}}, \bibinfo {author} {\bibfnamefont {P.~T.}\ \bibnamefont
  {Leung}}, \bibinfo {author} {\bibfnamefont {A.}~\bibnamefont {Maassen van~den
  Brink}}, \bibinfo {author} {\bibfnamefont {W.~M.}\ \bibnamefont {Suen}},
  \bibinfo {author} {\bibfnamefont {S.~S.}\ \bibnamefont {Tong}},\ and\
  \bibinfo {author} {\bibfnamefont {K.}~\bibnamefont {Young}},\ }\bibfield
  {title} {\bibinfo {title} {Quasinormal-mode expansion for waves in open
  systems},\ }\href {https://doi.org/10.1103/RevModPhys.70.1545} {\bibfield
  {journal} {\bibinfo  {journal} {Rev. Mod. Phys.}\ }\textbf {\bibinfo {volume}
  {70}},\ \bibinfo {pages} {1545} (\bibinfo {year} {1998})}\BibitemShut
  {NoStop}%
\bibitem [{\citenamefont {Lalanne}\ \emph {et~al.}(2018)\citenamefont
  {Lalanne}, \citenamefont {Yan}, \citenamefont {Vynck}, \citenamefont
  {Sauvan},\ and\ \citenamefont {Hugonin}}]{Lalanne_review}%
  \BibitemOpen
  \bibfield  {author} {\bibinfo {author} {\bibfnamefont {P.}~\bibnamefont
  {Lalanne}}, \bibinfo {author} {\bibfnamefont {W.}~\bibnamefont {Yan}},
  \bibinfo {author} {\bibfnamefont {K.}~\bibnamefont {Vynck}}, \bibinfo
  {author} {\bibfnamefont {C.}~\bibnamefont {Sauvan}},\ and\ \bibinfo {author}
  {\bibfnamefont {J.-P.}\ \bibnamefont {Hugonin}},\ }\bibfield  {title}
  {\bibinfo {title} {Light interaction with photonic and plasmonic
  resonances},\ }\href@noop {} {\bibfield  {journal} {\bibinfo  {journal}
  {Laser \& Photonics Reviews}\ }\textbf {\bibinfo {volume} {12}},\ \bibinfo
  {pages} {1700113} (\bibinfo {year} {2018})}\BibitemShut {NoStop}%
\bibitem [{\citenamefont {Franke}\ \emph {et~al.}(2019)\citenamefont {Franke},
  \citenamefont {Hughes}, \citenamefont {Dezfouli}, \citenamefont {Kristensen},
  \citenamefont {Busch}, \citenamefont {Knorr},\ and\ \citenamefont
  {Richter}}]{frankequantization}%
  \BibitemOpen
  \bibfield  {author} {\bibinfo {author} {\bibfnamefont {S.}~\bibnamefont
  {Franke}}, \bibinfo {author} {\bibfnamefont {S.}~\bibnamefont {Hughes}},
  \bibinfo {author} {\bibfnamefont {M.~K.}\ \bibnamefont {Dezfouli}}, \bibinfo
  {author} {\bibfnamefont {P.~T.}\ \bibnamefont {Kristensen}}, \bibinfo
  {author} {\bibfnamefont {K.}~\bibnamefont {Busch}}, \bibinfo {author}
  {\bibfnamefont {A.}~\bibnamefont {Knorr}},\ and\ \bibinfo {author}
  {\bibfnamefont {M.}~\bibnamefont {Richter}},\ }\bibfield  {title} {\bibinfo
  {title} {Quantization of quasinormal modes for open cavities and plasmonic
  cavity quantum electrodynamics},\ }\href
  {https://doi.org/10.1103/PhysRevLett.122.213901} {\bibfield  {journal}
  {\bibinfo  {journal} {Phys. Rev. Lett.}\ }\textbf {\bibinfo {volume} {122}},\
  \bibinfo {pages} {213901} (\bibinfo {year} {2019})}\BibitemShut {NoStop}%
\bibitem [{\citenamefont {Hughes}\ \emph {et~al.}(2019)\citenamefont {Hughes},
  \citenamefont {Franke}, \citenamefont {Gustin}, \citenamefont
  {Kamandar~Dezfouli}, \citenamefont {Knorr},\ and\ \citenamefont
  {Richter}}]{Hughes_SPS_2019}%
  \BibitemOpen
  \bibfield  {author} {\bibinfo {author} {\bibfnamefont {S.}~\bibnamefont
  {Hughes}}, \bibinfo {author} {\bibfnamefont {S.}~\bibnamefont {Franke}},
  \bibinfo {author} {\bibfnamefont {C.}~\bibnamefont {Gustin}}, \bibinfo
  {author} {\bibfnamefont {M.}~\bibnamefont {Kamandar~Dezfouli}}, \bibinfo
  {author} {\bibfnamefont {A.}~\bibnamefont {Knorr}},\ and\ \bibinfo {author}
  {\bibfnamefont {M.}~\bibnamefont {Richter}},\ }\bibfield  {title} {\bibinfo
  {title} {Theory and limits of on-demand single-photon sources using plasmonic
  resonators: A quantized quasinormal mode approach},\ }\href
  {https://doi.org/10.1021/acsphotonics.9b00849} {\bibfield  {journal}
  {\bibinfo  {journal} {{ACS} Photonics}\ }\textbf {\bibinfo {volume} {6}},\
  \bibinfo {pages} {2168} (\bibinfo {year} {2019})}\BibitemShut {NoStop}%
\bibitem [{\citenamefont {Franke}\ \emph
  {et~al.}(2020{\natexlab{a}})\citenamefont {Franke}, \citenamefont {Richter},
  \citenamefont {Ren}, \citenamefont {Knorr},\ and\ \citenamefont
  {Hughes}}]{franke2020quantized}%
  \BibitemOpen
  \bibfield  {author} {\bibinfo {author} {\bibfnamefont {S.}~\bibnamefont
  {Franke}}, \bibinfo {author} {\bibfnamefont {M.}~\bibnamefont {Richter}},
  \bibinfo {author} {\bibfnamefont {J.}~\bibnamefont {Ren}}, \bibinfo {author}
  {\bibfnamefont {A.}~\bibnamefont {Knorr}},\ and\ \bibinfo {author}
  {\bibfnamefont {S.}~\bibnamefont {Hughes}},\ }\bibfield  {title} {\bibinfo
  {title} {Quantized quasinormal-mode description of nonlinear cavity-qed
  effects from coupled resonators with a {Fano-like} resonance},\ }\href
  {https://doi.org/10.1103/PhysRevResearch.2.033456} {\bibfield  {journal}
  {\bibinfo  {journal} {Phys. Rev. Research}\ }\textbf {\bibinfo {volume}
  {2}},\ \bibinfo {pages} {033456} (\bibinfo {year}
  {2020}{\natexlab{a}})}\BibitemShut {NoStop}%
\bibitem [{\citenamefont {Franke}\ \emph {et~al.}(2022)\citenamefont {Franke},
  \citenamefont {Ren},\ and\ \citenamefont {Hughes}}]{PhysRevA.105.023702}%
  \BibitemOpen
  \bibfield  {author} {\bibinfo {author} {\bibfnamefont {S.}~\bibnamefont
  {Franke}}, \bibinfo {author} {\bibfnamefont {J.}~\bibnamefont {Ren}},\ and\
  \bibinfo {author} {\bibfnamefont {S.}~\bibnamefont {Hughes}},\ }\bibfield
  {title} {\bibinfo {title} {Quantized quasinormal-mode theory of coupled lossy
  and amplifying resonators},\ }\href
  {https://doi.org/10.1103/PhysRevA.105.023702} {\bibfield  {journal} {\bibinfo
   {journal} {Phys. Rev. A}\ }\textbf {\bibinfo {volume} {105}},\ \bibinfo
  {pages} {023702} (\bibinfo {year} {2022})}\BibitemShut {NoStop}%
\bibitem [{\citenamefont {Ren}\ \emph {et~al.}(2021)\citenamefont {Ren},
  \citenamefont {Franke},\ and\ \citenamefont {Hughes}}]{PhysRevX.11.041020}%
  \BibitemOpen
  \bibfield  {author} {\bibinfo {author} {\bibfnamefont {J.}~\bibnamefont
  {Ren}}, \bibinfo {author} {\bibfnamefont {S.}~\bibnamefont {Franke}},\ and\
  \bibinfo {author} {\bibfnamefont {S.}~\bibnamefont {Hughes}},\ }\bibfield
  {title} {\bibinfo {title} {Quasinormal modes, local density of states, and
  classical purcell factors for coupled loss-gain resonators},\ }\href
  {https://doi.org/10.1103/PhysRevX.11.041020} {\bibfield  {journal} {\bibinfo
  {journal} {Phys. Rev. X}\ }\textbf {\bibinfo {volume} {11}},\ \bibinfo
  {pages} {041020} (\bibinfo {year} {2021})}\BibitemShut {NoStop}%
\bibitem [{\citenamefont {Lentrodt}\ and\ \citenamefont
  {Evers}(2020)}]{Lentrodt2020Jan}%
  \BibitemOpen
  \bibfield  {author} {\bibinfo {author} {\bibfnamefont {D.}~\bibnamefont
  {Lentrodt}}\ and\ \bibinfo {author} {\bibfnamefont {J.}~\bibnamefont
  {Evers}},\ }\bibfield  {title} {\bibinfo {title} {{Ab Initio Few-Mode Theory
  for Quantum Potential Scattering Problems}},\ }\href
  {https://doi.org/10.1103/PhysRevX.10.011008} {\bibfield  {journal} {\bibinfo
  {journal} {Phys. Rev. X}\ }\textbf {\bibinfo {volume} {10}},\ \bibinfo
  {pages} {011008} (\bibinfo {year} {2020})}\BibitemShut {NoStop}%
\bibitem [{\citenamefont {Babiker}\ and\ \citenamefont
  {Rodney}(1983)}]{Babiker1983Feb}%
  \BibitemOpen
  \bibfield  {author} {\bibinfo {author} {\bibfnamefont {M.}~\bibnamefont
  {Babiker}}\ and\ \bibinfo {author} {\bibfnamefont {L.}~\bibnamefont
  {Rodney}},\ }\bibfield  {title} {\bibinfo {title} {{Derivation of the
  Power-Zienau-Woolley Hamiltonian in quantum electrodynamics by gauge
  transformation}},\ }\href {https://doi.org/10.1098/rspa.1983.0022} {\bibfield
   {journal} {\bibinfo  {journal} {Proc. R. Soc. Lond. A.}\ }\textbf {\bibinfo
  {volume} {385}},\ \bibinfo {pages} {439} (\bibinfo {year}
  {1983})}\BibitemShut {NoStop}%
\bibitem [{\citenamefont {Woolley}(2020)}]{Woolley2020Feb}%
  \BibitemOpen
  \bibfield  {author} {\bibinfo {author} {\bibfnamefont {R.~G.}\ \bibnamefont
  {Woolley}},\ }\bibfield  {title} {\bibinfo {title} {{Power-Zienau-Woolley
  representations of nonrelativistic QED for atoms and molecules}},\ }\href
  {https://doi.org/10.1103/PhysRevResearch.2.013206} {\bibfield  {journal}
  {\bibinfo  {journal} {Phys. Rev. Res.}\ }\textbf {\bibinfo {volume} {2}},\
  \bibinfo {pages} {013206} (\bibinfo {year} {2020})}\BibitemShut {NoStop}%
\bibitem [{\citenamefont {Woolley}(1999)}]{Woolley1999Jan}%
  \BibitemOpen
  \bibfield  {author} {\bibinfo {author} {\bibfnamefont {R.~G.}\ \bibnamefont
  {Woolley}},\ }\bibfield  {title} {\bibinfo {title} {{Charged particles, gauge
  invariance, and molecular electrodynamics}},\ }\href
  {https://doi.org/10.1002/(SICI)1097-461X(1999)74:5<531::AID-QUA9>3.0.CO;2-H}
  {\bibfield  {journal} {\bibinfo  {journal} {Int. J. Quantum Chem.}\ }\textbf
  {\bibinfo {volume} {74}},\ \bibinfo {pages} {531} (\bibinfo {year}
  {1999})}\BibitemShut {NoStop}%
\bibitem [{\citenamefont {Rousseau}\ and\ \citenamefont
  {Felbacq}(2017)}]{Rousseau2017Sep}%
  \BibitemOpen
  \bibfield  {author} {\bibinfo {author} {\bibfnamefont {E.}~\bibnamefont
  {Rousseau}}\ and\ \bibinfo {author} {\bibfnamefont {D.}~\bibnamefont
  {Felbacq}},\ }\bibfield  {title} {\bibinfo {title} {{The quantum-optics
  Hamiltonian in the Multipolar gauge}},\ }\href
  {https://doi.org/10.1038/s41598-017-11076-5} {\bibfield  {journal} {\bibinfo
  {journal} {Sci. Rep.}\ }\textbf {\bibinfo {volume} {7}},\ \bibinfo {pages}
  {1} (\bibinfo {year} {2017})}\BibitemShut {NoStop}%
\bibitem [{\citenamefont {Vukics}\ \emph {et~al.}(2021)\citenamefont {Vukics},
  \citenamefont {K{\ifmmode\acute{o}\else\'{o}\fi}nya},\ and\ \citenamefont
  {Domokos}}]{Vukics2021Aug}%
  \BibitemOpen
  \bibfield  {author} {\bibinfo {author} {\bibfnamefont {A.}~\bibnamefont
  {Vukics}}, \bibinfo {author} {\bibfnamefont {G.}~\bibnamefont
  {K{\ifmmode\acute{o}\else\'{o}\fi}nya}},\ and\ \bibinfo {author}
  {\bibfnamefont {P.}~\bibnamefont {Domokos}},\ }\bibfield  {title} {\bibinfo
  {title} {{The gauge-invariant Lagrangian, the
  Power{\textendash}Zienau{\textendash}Woolley picture, and the choices of
  field momenta in nonrelativistic quantum electrodynamics}},\ }\href
  {https://doi.org/10.1038/s41598-021-94405-z} {\bibfield  {journal} {\bibinfo
  {journal} {Sci. Rep.}\ }\textbf {\bibinfo {volume} {11}},\ \bibinfo {pages}
  {1} (\bibinfo {year} {2021})}\BibitemShut {NoStop}%
\bibitem [{\citenamefont {Andrews}\ \emph {et~al.}(2018)\citenamefont
  {Andrews}, \citenamefont {Jones}, \citenamefont {Salam},\ and\ \citenamefont
  {Woolley}}]{Andrews2018Jan}%
  \BibitemOpen
  \bibfield  {author} {\bibinfo {author} {\bibfnamefont {D.~L.}\ \bibnamefont
  {Andrews}}, \bibinfo {author} {\bibfnamefont {G.~A.}\ \bibnamefont {Jones}},
  \bibinfo {author} {\bibfnamefont {A.}~\bibnamefont {Salam}},\ and\ \bibinfo
  {author} {\bibfnamefont {R.~G.}\ \bibnamefont {Woolley}},\ }\bibfield
  {title} {\bibinfo {title} {{Perspective: Quantum Hamiltonians for optical
  interactions}},\ }\href {https://doi.org/10.1063/1.5018399} {\bibfield
  {journal} {\bibinfo  {journal} {J. Chem. Phys.}\ }\textbf {\bibinfo {volume}
  {148}},\ \bibinfo {pages} {040901} (\bibinfo {year} {2018})}\BibitemShut
  {NoStop}%
\bibitem [{\citenamefont {Stokes}\ and\ \citenamefont
  {Nazir}(2021{\natexlab{a}})}]{Stokes2021Sep}%
  \BibitemOpen
  \bibfield  {author} {\bibinfo {author} {\bibfnamefont {A.}~\bibnamefont
  {Stokes}}\ and\ \bibinfo {author} {\bibfnamefont {A.}~\bibnamefont {Nazir}},\
  }\bibfield  {title} {\bibinfo {title} {{Identification of
  Poincar{\ifmmode\backslash\else\textbackslash\fi}'e-gauge and multipolar
  nonrelativistic theories of QED}},\ }\href
  {https://doi.org/10.1103/PhysRevA.104.032227} {\bibfield  {journal} {\bibinfo
   {journal} {Phys. Rev. A}\ }\textbf {\bibinfo {volume} {104}},\ \bibinfo
  {pages} {032227} (\bibinfo {year} {2021}{\natexlab{a}})}\BibitemShut
  {NoStop}%
\bibitem [{\citenamefont {Gardiner}\ and\ \citenamefont
  {Collett}(1985)}]{Gardiner1985Jun}%
  \BibitemOpen
  \bibfield  {author} {\bibinfo {author} {\bibfnamefont {C.~W.}\ \bibnamefont
  {Gardiner}}\ and\ \bibinfo {author} {\bibfnamefont {M.~J.}\ \bibnamefont
  {Collett}},\ }\bibfield  {title} {\bibinfo {title} {{Input and output in
  damped quantum systems: Quantum stochastic differential equations and the
  master equation}},\ }\href {https://doi.org/10.1103/PhysRevA.31.3761}
  {\bibfield  {journal} {\bibinfo  {journal} {Phys. Rev. A}\ }\textbf {\bibinfo
  {volume} {31}},\ \bibinfo {pages} {3761} (\bibinfo {year}
  {1985})}\BibitemShut {NoStop}%
\bibitem [{\citenamefont {Salmon}\ \emph {et~al.}(2022)\citenamefont {Salmon},
  \citenamefont {Gustin}, \citenamefont {Settineri}, \citenamefont
  {Di~Stefano}, \citenamefont {Zueco}, \citenamefont {Savasta}, \citenamefont
  {Nori},\ and\ \citenamefont {Hughes}}]{Salmon2022Mar}%
  \BibitemOpen
  \bibfield  {author} {\bibinfo {author} {\bibfnamefont {W.}~\bibnamefont
  {Salmon}}, \bibinfo {author} {\bibfnamefont {C.}~\bibnamefont {Gustin}},
  \bibinfo {author} {\bibfnamefont {A.}~\bibnamefont {Settineri}}, \bibinfo
  {author} {\bibfnamefont {O.}~\bibnamefont {Di~Stefano}}, \bibinfo {author}
  {\bibfnamefont {D.}~\bibnamefont {Zueco}}, \bibinfo {author} {\bibfnamefont
  {S.}~\bibnamefont {Savasta}}, \bibinfo {author} {\bibfnamefont
  {F.}~\bibnamefont {Nori}},\ and\ \bibinfo {author} {\bibfnamefont
  {S.}~\bibnamefont {Hughes}},\ }\bibfield  {title} {\bibinfo {title}
  {{Gauge-independent emission spectra and quantum correlations in the
  ultrastrong coupling regime of open system cavity-QED}},\ }\href
  {https://doi.org/10.1515/nanoph-2021-0718} {\bibfield  {journal} {\bibinfo
  {journal} {Nanophotonics}\ }\textbf {\bibinfo {volume} {11}},\ \bibinfo
  {pages} {1573} (\bibinfo {year} {2022})}\BibitemShut {NoStop}%
\bibitem [{\citenamefont {Medina}\ \emph {et~al.}(2021)\citenamefont {Medina},
  \citenamefont {Garc{\ifmmode\acute{\imath}\else\'{\i}\fi}a-Vidal},
  \citenamefont
  {Fern{\ifmmode\acute{a}\else\'{a}\fi}ndez-Dom{\ifmmode\acute{\imath}\else\'{\i}\fi}nguez},\
  and\ \citenamefont {Feist}}]{Medina2021Mar}%
  \BibitemOpen
  \bibfield  {author} {\bibinfo {author} {\bibfnamefont {I.}~\bibnamefont
  {Medina}}, \bibinfo {author} {\bibfnamefont {F.~J.}\ \bibnamefont
  {Garc{\ifmmode\acute{\imath}\else\'{\i}\fi}a-Vidal}}, \bibinfo {author}
  {\bibfnamefont {A.~I.}\ \bibnamefont
  {Fern{\ifmmode\acute{a}\else\'{a}\fi}ndez-Dom{\ifmmode\acute{\imath}\else\'{\i}\fi}nguez}},\
  and\ \bibinfo {author} {\bibfnamefont {J.}~\bibnamefont {Feist}},\ }\bibfield
   {title} {\bibinfo {title} {{Few-Mode Field Quantization of Arbitrary
  Electromagnetic Spectral Densities}},\ }\href
  {https://doi.org/10.1103/PhysRevLett.126.093601} {\bibfield  {journal}
  {\bibinfo  {journal} {Phys. Rev. Lett.}\ }\textbf {\bibinfo {volume} {126}},\
  \bibinfo {pages} {093601} (\bibinfo {year} {2021})}\BibitemShut {NoStop}%
\bibitem [{\citenamefont {S{\ifmmode\acute{a}\else\'{a}\fi}nchez-Barquilla}\
  \emph {et~al.}(2022)\citenamefont
  {S{\ifmmode\acute{a}\else\'{a}\fi}nchez-Barquilla}, \citenamefont
  {Garc{\ifmmode\acute{\imath}\else\'{\i}\fi}a-Vidal}, \citenamefont
  {Fern{\ifmmode\acute{a}\else\'{a}\fi}ndez-Dom{\ifmmode\acute{\imath}\else\'{\i}\fi}nguez},\
  and\ \citenamefont {Feist}}]{Sanchez-Barquilla2022Aug}%
  \BibitemOpen
  \bibfield  {author} {\bibinfo {author} {\bibfnamefont {M.}~\bibnamefont
  {S{\ifmmode\acute{a}\else\'{a}\fi}nchez-Barquilla}}, \bibinfo {author}
  {\bibfnamefont {F.~J.}\ \bibnamefont
  {Garc{\ifmmode\acute{\imath}\else\'{\i}\fi}a-Vidal}}, \bibinfo {author}
  {\bibfnamefont {A.~I.}\ \bibnamefont
  {Fern{\ifmmode\acute{a}\else\'{a}\fi}ndez-Dom{\ifmmode\acute{\imath}\else\'{\i}\fi}nguez}},\
  and\ \bibinfo {author} {\bibfnamefont {J.}~\bibnamefont {Feist}},\ }\bibfield
   {title} {\bibinfo {title} {{Few-mode field quantization for multiple
  emitters}},\ }\bibfield  {journal} {\bibinfo  {journal} {Nanophotonics}\
  }\href {https://doi.org/10.1515/nanoph-2021-0795} {10.1515/nanoph-2021-0795}
  (\bibinfo {year} {2022})\BibitemShut {NoStop}%
\bibitem [{\citenamefont {del Pino}\ \emph {et~al.}(2018)\citenamefont {del
  Pino}, \citenamefont {Schr{\ifmmode\ddot{o}\else\"{o}\fi}der}, \citenamefont
  {Chin}, \citenamefont {Feist},\ and\ \citenamefont
  {Garcia-Vidal}}]{delPino2018Nov}%
  \BibitemOpen
  \bibfield  {author} {\bibinfo {author} {\bibfnamefont {J.}~\bibnamefont {del
  Pino}}, \bibinfo {author} {\bibfnamefont {F.~A. Y.~N.}\ \bibnamefont
  {Schr{\ifmmode\ddot{o}\else\"{o}\fi}der}}, \bibinfo {author} {\bibfnamefont
  {A.~W.}\ \bibnamefont {Chin}}, \bibinfo {author} {\bibfnamefont
  {J.}~\bibnamefont {Feist}},\ and\ \bibinfo {author} {\bibfnamefont {F.~J.}\
  \bibnamefont {Garcia-Vidal}},\ }\bibfield  {title} {\bibinfo {title} {{Tensor
  Network Simulation of Non-Markovian Dynamics in Organic Polaritons}},\ }\href
  {https://doi.org/10.1103/PhysRevLett.121.227401} {\bibfield  {journal}
  {\bibinfo  {journal} {Phys. Rev. Lett.}\ }\textbf {\bibinfo {volume} {121}},\
  \bibinfo {pages} {227401} (\bibinfo {year} {2018})}\BibitemShut {NoStop}%
\bibitem [{\citenamefont {Regidor}\ \emph {et~al.}(2021)\citenamefont
  {Regidor}, \citenamefont {Crowder}, \citenamefont {Carmichael},\ and\
  \citenamefont {Hughes}}]{Regidor2021Apr}%
  \BibitemOpen
  \bibfield  {author} {\bibinfo {author} {\bibfnamefont {S.~A.}\ \bibnamefont
  {Regidor}}, \bibinfo {author} {\bibfnamefont {G.}~\bibnamefont {Crowder}},
  \bibinfo {author} {\bibfnamefont {H.}~\bibnamefont {Carmichael}},\ and\
  \bibinfo {author} {\bibfnamefont {S.}~\bibnamefont {Hughes}},\ }\bibfield
  {title} {\bibinfo {title} {{Modeling quantum light-matter interactions in
  waveguide QED with retardation, nonlinear interactions, and a time-delayed
  feedback: Matrix product states versus a space-discretized waveguide
  model}},\ }\href {https://doi.org/10.1103/PhysRevResearch.3.023030}
  {\bibfield  {journal} {\bibinfo  {journal} {Phys. Rev. Res.}\ }\textbf
  {\bibinfo {volume} {3}},\ \bibinfo {pages} {023030} (\bibinfo {year}
  {2021})}\BibitemShut {NoStop}%
\bibitem [{\citenamefont {Mercurio}\ \emph {et~al.}(2022)\citenamefont
  {Mercurio}, \citenamefont {Macr{\ifmmode\grave{\imath}\else\`{\i}\fi}},
  \citenamefont {Gustin}, \citenamefont {Hughes}, \citenamefont {Savasta},\
  and\ \citenamefont {Nori}}]{Mercurio2022Apr}%
  \BibitemOpen
  \bibfield  {author} {\bibinfo {author} {\bibfnamefont {A.}~\bibnamefont
  {Mercurio}}, \bibinfo {author} {\bibfnamefont {V.}~\bibnamefont
  {Macr{\ifmmode\grave{\imath}\else\`{\i}\fi}}}, \bibinfo {author}
  {\bibfnamefont {C.}~\bibnamefont {Gustin}}, \bibinfo {author} {\bibfnamefont
  {S.}~\bibnamefont {Hughes}}, \bibinfo {author} {\bibfnamefont
  {S.}~\bibnamefont {Savasta}},\ and\ \bibinfo {author} {\bibfnamefont
  {F.}~\bibnamefont {Nori}},\ }\bibfield  {title} {\bibinfo {title} {{Regimes
  of cavity QED under incoherent excitation: From weak to deep strong
  coupling}},\ }\href {https://doi.org/10.1103/PhysRevResearch.4.023048}
  {\bibfield  {journal} {\bibinfo  {journal} {Phys. Rev. Res.}\ }\textbf
  {\bibinfo {volume} {4}},\ \bibinfo {pages} {023048} (\bibinfo {year}
  {2022})}\BibitemShut {NoStop}%
\bibitem [{\citenamefont {Stokes}\ and\ \citenamefont
  {Nazir}(2021{\natexlab{b}})}]{Stokes2021Feb}%
  \BibitemOpen
  \bibfield  {author} {\bibinfo {author} {\bibfnamefont {A.}~\bibnamefont
  {Stokes}}\ and\ \bibinfo {author} {\bibfnamefont {A.}~\bibnamefont {Nazir}},\
  }\bibfield  {title} {\bibinfo {title} {{Ultrastrong time-dependent
  light-matter interactions are gauge relative}},\ }\href
  {https://doi.org/10.1103/PhysRevResearch.3.013116} {\bibfield  {journal}
  {\bibinfo  {journal} {Phys. Rev. Res.}\ }\textbf {\bibinfo {volume} {3}},\
  \bibinfo {pages} {013116} (\bibinfo {year} {2021}{\natexlab{b}})}\BibitemShut
  {NoStop}%
\bibitem [{\citenamefont {Settineri}\ \emph {et~al.}(2021)\citenamefont
  {Settineri}, \citenamefont {Di~Stefano}, \citenamefont {Zueco}, \citenamefont
  {Hughes}, \citenamefont {Savasta},\ and\ \citenamefont
  {Nori}}]{PhysRevResearch.3.023079}%
  \BibitemOpen
  \bibfield  {author} {\bibinfo {author} {\bibfnamefont {A.}~\bibnamefont
  {Settineri}}, \bibinfo {author} {\bibfnamefont {O.}~\bibnamefont
  {Di~Stefano}}, \bibinfo {author} {\bibfnamefont {D.}~\bibnamefont {Zueco}},
  \bibinfo {author} {\bibfnamefont {S.}~\bibnamefont {Hughes}}, \bibinfo
  {author} {\bibfnamefont {S.}~\bibnamefont {Savasta}},\ and\ \bibinfo {author}
  {\bibfnamefont {F.}~\bibnamefont {Nori}},\ }\bibfield  {title} {\bibinfo
  {title} {Gauge freedom, quantum measurements, and time-dependent interactions
  in cavity qed},\ }\href {https://doi.org/10.1103/PhysRevResearch.3.023079}
  {\bibfield  {journal} {\bibinfo  {journal} {Phys. Rev. Research}\ }\textbf
  {\bibinfo {volume} {3}},\ \bibinfo {pages} {023079} (\bibinfo {year}
  {2021})}\BibitemShut {NoStop}%
\bibitem [{\citenamefont {Savasta}\ \emph
  {et~al.}(2021{\natexlab{b}})\citenamefont {Savasta}, \citenamefont
  {Di~Stefano},\ and\ \citenamefont {Nori}}]{Savasta2021Jan}%
  \BibitemOpen
  \bibfield  {author} {\bibinfo {author} {\bibfnamefont {S.}~\bibnamefont
  {Savasta}}, \bibinfo {author} {\bibfnamefont {O.}~\bibnamefont
  {Di~Stefano}},\ and\ \bibinfo {author} {\bibfnamefont {F.}~\bibnamefont
  {Nori}},\ }\bibfield  {title} {\bibinfo {title}
  {{Thomas{\textendash}Reiche{\textendash}Kuhn (TRK) sum rule for interacting
  photons}},\ }\href {https://doi.org/10.1515/nanoph-2020-0433} {\bibfield
  {journal} {\bibinfo  {journal} {Nanophotonics}\ }\textbf {\bibinfo {volume}
  {10}},\ \bibinfo {pages} {465} (\bibinfo {year}
  {2021}{\natexlab{b}})}\BibitemShut {NoStop}%
\bibitem [{\citenamefont {Stokes}\ and\ \citenamefont
  {Nazir}(2022)}]{Stokes2022Nov}%
  \BibitemOpen
  \bibfield  {author} {\bibinfo {author} {\bibfnamefont {A.}~\bibnamefont
  {Stokes}}\ and\ \bibinfo {author} {\bibfnamefont {A.}~\bibnamefont {Nazir}},\
  }\bibfield  {title} {\bibinfo {title} {{Implications of gauge freedom for
  nonrelativistic quantum electrodynamics}},\ }\href
  {https://doi.org/10.1103/RevModPhys.94.045003} {\bibfield  {journal}
  {\bibinfo  {journal} {Rev. Mod. Phys.}\ }\textbf {\bibinfo {volume} {94}},\
  \bibinfo {pages} {045003} (\bibinfo {year} {2022})}\BibitemShut {NoStop}%
\bibitem [{\citenamefont {Dalton}\ \emph {et~al.}(1996)\citenamefont {Dalton},
  \citenamefont {Guerra},\ and\ \citenamefont {Knight}}]{motu3}%
  \BibitemOpen
  \bibfield  {author} {\bibinfo {author} {\bibfnamefont {B.}~\bibnamefont
  {Dalton}}, \bibinfo {author} {\bibfnamefont {E.}~\bibnamefont {Guerra}},\
  and\ \bibinfo {author} {\bibfnamefont {P.}~\bibnamefont {Knight}},\
  }\bibfield  {title} {\bibinfo {title} {Field quantization in dielectric media
  and the generalized multipolar hamiltonian},\ }\href@noop {} {\bibfield
  {journal} {\bibinfo  {journal} {Phys. Rev. A}\ }\textbf {\bibinfo {volume}
  {54}},\ \bibinfo {pages} {2292} (\bibinfo {year} {1996})}\BibitemShut
  {NoStop}%
\bibitem [{\citenamefont {Dalton}\ and\ \citenamefont
  {Babiker}(1997)}]{Dalton1997Jul}%
  \BibitemOpen
  \bibfield  {author} {\bibinfo {author} {\bibfnamefont {B.~J.}\ \bibnamefont
  {Dalton}}\ and\ \bibinfo {author} {\bibfnamefont {M.}~\bibnamefont
  {Babiker}},\ }\bibfield  {title} {\bibinfo {title} {{Macroscopic quantization
  in quantum optics and cavity quantum electrodynamics: Interatomic
  interactions}},\ }\href {https://doi.org/10.1103/PhysRevA.56.905} {\bibfield
  {journal} {\bibinfo  {journal} {Phys. Rev. A}\ }\textbf {\bibinfo {volume}
  {56}},\ \bibinfo {pages} {905} (\bibinfo {year} {1997})}\BibitemShut
  {NoStop}%
\bibitem [{\citenamefont {Wubs}\ \emph {et~al.}(2003)\citenamefont {Wubs},
  \citenamefont {Suttorp},\ and\ \citenamefont {Lagendijk}}]{Wubs2003Jul}%
  \BibitemOpen
  \bibfield  {author} {\bibinfo {author} {\bibfnamefont {M.}~\bibnamefont
  {Wubs}}, \bibinfo {author} {\bibfnamefont {L.~G.}\ \bibnamefont {Suttorp}},\
  and\ \bibinfo {author} {\bibfnamefont {A.}~\bibnamefont {Lagendijk}},\
  }\bibfield  {title} {\bibinfo {title} {{Multipole interaction between atoms
  and their photonic environment}},\ }\href
  {https://doi.org/10.1103/PhysRevA.68.013822} {\bibfield  {journal} {\bibinfo
  {journal} {Phys. Rev. A}\ }\textbf {\bibinfo {volume} {68}},\ \bibinfo
  {pages} {013822} (\bibinfo {year} {2003})}\BibitemShut {NoStop}%
\bibitem [{\citenamefont {Cohen-Tannoudji}\ \emph {et~al.}(1997)\citenamefont
  {Cohen-Tannoudji}, \citenamefont {Dupont-Roc},\ and\ \citenamefont
  {Grynberg}}]{Cohen-Tannoudji1997Mar}%
  \BibitemOpen
  \bibfield  {author} {\bibinfo {author} {\bibfnamefont {C.}~\bibnamefont
  {Cohen-Tannoudji}}, \bibinfo {author} {\bibfnamefont {J.}~\bibnamefont
  {Dupont-Roc}},\ and\ \bibinfo {author} {\bibfnamefont {G.}~\bibnamefont
  {Grynberg}},\ }\href
  {https://www.amazon.com/Photons-Atoms-Introduction-Quantum-Electrodynamics/dp/0471184330}
  {\emph {\bibinfo {title} {{Photons and Atoms: Introduction to Quantum
  Electrodynamics}}}}\ (\bibinfo  {publisher} {Wiley-VCH},\ \bibinfo {year}
  {1997})\BibitemShut {NoStop}%
\bibitem [{\citenamefont {Sundermeyer}(1982)}]{sundermeyer1982constrained}%
  \BibitemOpen
  \bibfield  {author} {\bibinfo {author} {\bibfnamefont {K.}~\bibnamefont
  {Sundermeyer}},\ }\href@noop {} {\emph {\bibinfo {title} {Constrained
  dynamics with applications to Yang-Mills theory, general relativity,
  classical spin, dual string model}}}\ (\bibinfo {year} {1982})\BibitemShut
  {NoStop}%
\bibitem [{\citenamefont {Rouse}\ \emph {et~al.}(2021)\citenamefont {Rouse},
  \citenamefont {Lovett}, \citenamefont {Gauger},\ and\ \citenamefont
  {Westerberg}}]{Rouse2021Feb}%
  \BibitemOpen
  \bibfield  {author} {\bibinfo {author} {\bibfnamefont {D.~M.}\ \bibnamefont
  {Rouse}}, \bibinfo {author} {\bibfnamefont {B.~W.}\ \bibnamefont {Lovett}},
  \bibinfo {author} {\bibfnamefont {E.~M.}\ \bibnamefont {Gauger}},\ and\
  \bibinfo {author} {\bibfnamefont {N.}~\bibnamefont {Westerberg}},\ }\bibfield
   {title} {\bibinfo {title} {{Avoiding gauge ambiguities in cavity quantum
  electrodynamics}},\ }\href {https://doi.org/10.1038/s41598-021-83214-z}
  {\bibfield  {journal} {\bibinfo  {journal} {Sci. Rep.}\ }\textbf {\bibinfo
  {volume} {11}},\ \bibinfo {pages} {1} (\bibinfo {year} {2021})}\BibitemShut
  {NoStop}%
\bibitem [{\citenamefont {Savasta}\ and\ \citenamefont
  {Girlanda}(1995)}]{Savasta1995Nov}%
  \BibitemOpen
  \bibfield  {author} {\bibinfo {author} {\bibfnamefont {S.}~\bibnamefont
  {Savasta}}\ and\ \bibinfo {author} {\bibfnamefont {R.}~\bibnamefont
  {Girlanda}},\ }\bibfield  {title} {\bibinfo {title} {{The particle-photon
  interaction in systems described by model Hamiltonians in second
  quantization}},\ }\href {https://doi.org/10.1016/0038-1098(95)00242-1}
  {\bibfield  {journal} {\bibinfo  {journal} {Solid State Commun.}\ }\textbf
  {\bibinfo {volume} {96}},\ \bibinfo {pages} {517} (\bibinfo {year}
  {1995})}\BibitemShut {NoStop}%
\bibitem [{\citenamefont {Graf}\ and\ \citenamefont
  {Vogl}(1995)}]{Graf1995Feb}%
  \BibitemOpen
  \bibfield  {author} {\bibinfo {author} {\bibfnamefont {M.}~\bibnamefont
  {Graf}}\ and\ \bibinfo {author} {\bibfnamefont {P.}~\bibnamefont {Vogl}},\
  }\bibfield  {title} {\bibinfo {title} {{Electromagnetic fields and dielectric
  response in empirical tight-binding theory}},\ }\href
  {https://doi.org/10.1103/PhysRevB.51.4940} {\bibfield  {journal} {\bibinfo
  {journal} {Phys. Rev. B}\ }\textbf {\bibinfo {volume} {51}},\ \bibinfo
  {pages} {4940} (\bibinfo {year} {1995})}\BibitemShut {NoStop}%
\bibitem [{\citenamefont {Onodera}\ \emph {et~al.}(2022)\citenamefont
  {Onodera}, \citenamefont {Ng}, \citenamefont {Gustin}, \citenamefont
  {L{\ifmmode\ddot{o}\else\"{o}\fi}rch}, \citenamefont {Yamamura},
  \citenamefont {Hamerly}, \citenamefont {McMahon}, \citenamefont {Marandi},\
  and\ \citenamefont {Mabuchi}}]{Onodera2022Mar}%
  \BibitemOpen
  \bibfield  {author} {\bibinfo {author} {\bibfnamefont {T.}~\bibnamefont
  {Onodera}}, \bibinfo {author} {\bibfnamefont {E.}~\bibnamefont {Ng}},
  \bibinfo {author} {\bibfnamefont {C.}~\bibnamefont {Gustin}}, \bibinfo
  {author} {\bibfnamefont {N.}~\bibnamefont
  {L{\ifmmode\ddot{o}\else\"{o}\fi}rch}}, \bibinfo {author} {\bibfnamefont
  {A.}~\bibnamefont {Yamamura}}, \bibinfo {author} {\bibfnamefont
  {R.}~\bibnamefont {Hamerly}}, \bibinfo {author} {\bibfnamefont {P.~L.}\
  \bibnamefont {McMahon}}, \bibinfo {author} {\bibfnamefont {A.}~\bibnamefont
  {Marandi}},\ and\ \bibinfo {author} {\bibfnamefont {H.}~\bibnamefont
  {Mabuchi}},\ }\bibfield  {title} {\bibinfo {title} {{Nonlinear quantum
  behavior of ultrashort-pulse optical parametric oscillators}},\ }\href
  {https://doi.org/10.1103/PhysRevA.105.033508} {\bibfield  {journal} {\bibinfo
   {journal} {Phys. Rev. A}\ }\textbf {\bibinfo {volume} {105}},\ \bibinfo
  {pages} {033508} (\bibinfo {year} {2022})}\BibitemShut {NoStop}%
\bibitem [{\citenamefont {Dutra}\ and\ \citenamefont
  {Nienhuis}(2000)}]{dutra2000quantized}%
  \BibitemOpen
  \bibfield  {author} {\bibinfo {author} {\bibfnamefont {S.}~\bibnamefont
  {Dutra}}\ and\ \bibinfo {author} {\bibfnamefont {G.}~\bibnamefont
  {Nienhuis}},\ }\bibfield  {title} {\bibinfo {title} {Quantized mode of a
  leaky cavity},\ }\href@noop {} {\bibfield  {journal} {\bibinfo  {journal}
  {Physical Review A}\ }\textbf {\bibinfo {volume} {62}},\ \bibinfo {pages}
  {063805} (\bibinfo {year} {2000})}\BibitemShut {NoStop}%
\bibitem [{\citenamefont {Viviescas}\ and\ \citenamefont
  {Hackenbroich}(2003)}]{Viviescas2003Jan}%
  \BibitemOpen
  \bibfield  {author} {\bibinfo {author} {\bibfnamefont {C.}~\bibnamefont
  {Viviescas}}\ and\ \bibinfo {author} {\bibfnamefont {G.}~\bibnamefont
  {Hackenbroich}},\ }\bibfield  {title} {\bibinfo {title} {{Field quantization
  for open optical cavities}},\ }\href
  {https://doi.org/10.1103/PhysRevA.67.013805} {\bibfield  {journal} {\bibinfo
  {journal} {Phys. Rev. A}\ }\textbf {\bibinfo {volume} {67}},\ \bibinfo
  {pages} {013805} (\bibinfo {year} {2003})}\BibitemShut {NoStop}%
\bibitem [{\citenamefont {Gitman}\ and\ \citenamefont
  {Tyutin}(1990)}]{Gitman1990}%
  \BibitemOpen
  \bibfield  {author} {\bibinfo {author} {\bibfnamefont {D.~M.}\ \bibnamefont
  {Gitman}}\ and\ \bibinfo {author} {\bibfnamefont {I.~V.}\ \bibnamefont
  {Tyutin}},\ }\href {https://link.springer.com/book/10.1007/978-3-642-83938-2}
  {\emph {\bibinfo {title} {{Quantization of Fields with Constraints}}}}\
  (\bibinfo  {publisher} {Springer},\ \bibinfo {address} {Berlin, Germany},\
  \bibinfo {year} {1990})\BibitemShut {NoStop}%
\bibitem [{\citenamefont {Goldstein}(1980)}]{goldstein:mechanics}%
  \BibitemOpen
  \bibfield  {author} {\bibinfo {author} {\bibfnamefont {H.}~\bibnamefont
  {Goldstein}},\ }\href@noop {} {\emph {\bibinfo {title} {Classical
  Mechanics}}}\ (\bibinfo  {publisher} {Addison-Wesley},\ \bibinfo {year}
  {1980})\BibitemShut {NoStop}%
\bibitem [{\citenamefont {Garziano}\ \emph {et~al.}(2020)\citenamefont
  {Garziano}, \citenamefont {Settineri}, \citenamefont {Di~Stefano},
  \citenamefont {Savasta},\ and\ \citenamefont {Nori}}]{Garziano2020Aug}%
  \BibitemOpen
  \bibfield  {author} {\bibinfo {author} {\bibfnamefont {L.}~\bibnamefont
  {Garziano}}, \bibinfo {author} {\bibfnamefont {A.}~\bibnamefont {Settineri}},
  \bibinfo {author} {\bibfnamefont {O.}~\bibnamefont {Di~Stefano}}, \bibinfo
  {author} {\bibfnamefont {S.}~\bibnamefont {Savasta}},\ and\ \bibinfo {author}
  {\bibfnamefont {F.}~\bibnamefont {Nori}},\ }\bibfield  {title} {\bibinfo
  {title} {{Gauge invariance of the Dicke and Hopfield models}},\ }\href
  {https://doi.org/10.1103/PhysRevA.102.023718} {\bibfield  {journal} {\bibinfo
   {journal} {Phys. Rev. A}\ }\textbf {\bibinfo {volume} {102}},\ \bibinfo
  {pages} {023718} (\bibinfo {year} {2020})}\BibitemShut {NoStop}%
\bibitem [{\citenamefont {Gardiner}\ and\ \citenamefont
  {Zoller}(2004)}]{Gardiner2004Aug}%
  \BibitemOpen
  \bibfield  {author} {\bibinfo {author} {\bibfnamefont {C.}~\bibnamefont
  {Gardiner}}\ and\ \bibinfo {author} {\bibfnamefont {P.}~\bibnamefont
  {Zoller}},\ }\href
  {https://www.amazon.com/Quantum-Noise-Non-Markovian-Applications-Synergetics/dp/3540223010}
  {\emph {\bibinfo {title} {{Quantum Noise: A Handbook of Markovian and
  Non-Markovian Quantum Stochastic Methods with Applications to Quantum Optics
  (Springer Series in Synergetics)}}}}\ (\bibinfo  {publisher} {Springer},\
  \bibinfo {address} {Berlin, Germany},\ \bibinfo {year} {2004})\BibitemShut
  {NoStop}%
\bibitem [{\citenamefont
  {S{\ifmmode\acute{a}\else\'{a}\fi}nchez~Mu{\ifmmode\tilde{n}\else\~{n}\fi}oz}\
  \emph {et~al.}(2018)\citenamefont
  {S{\ifmmode\acute{a}\else\'{a}\fi}nchez~Mu{\ifmmode\tilde{n}\else\~{n}\fi}oz},
  \citenamefont {Nori},\ and\ \citenamefont
  {De~Liberato}}]{SanchezMunoz2018May}%
  \BibitemOpen
  \bibfield  {author} {\bibinfo {author} {\bibfnamefont {C.}~\bibnamefont
  {S{\ifmmode\acute{a}\else\'{a}\fi}nchez~Mu{\ifmmode\tilde{n}\else\~{n}\fi}oz}},
  \bibinfo {author} {\bibfnamefont {F.}~\bibnamefont {Nori}},\ and\ \bibinfo
  {author} {\bibfnamefont {S.}~\bibnamefont {De~Liberato}},\ }\bibfield
  {title} {\bibinfo {title} {{Resolution of superluminal signalling in
  non-perturbative cavity quantum electrodynamics}},\ }\href
  {https://doi.org/10.1038/s41467-018-04339-w} {\bibfield  {journal} {\bibinfo
  {journal} {Nat. Commun.}\ }\textbf {\bibinfo {volume} {9}},\ \bibinfo {pages}
  {1} (\bibinfo {year} {2018})}\BibitemShut {NoStop}%
\bibitem [{\citenamefont {Dirac}(1958)}]{Dirac1958Aug}%
  \BibitemOpen
  \bibfield  {author} {\bibinfo {author} {\bibfnamefont {P.~A.~M.}\
  \bibnamefont {Dirac}},\ }\bibfield  {title} {\bibinfo {title} {{Generalized
  Hamiltonian dynamics}},\ }\href {https://doi.org/10.1098/rspa.1958.0141}
  {\bibfield  {journal} {\bibinfo  {journal} {Proc. R. Soc. London A - Math.
  Phys. Sci.}\ }\textbf {\bibinfo {volume} {246}},\ \bibinfo {pages} {326}
  (\bibinfo {year} {1958})}\BibitemShut {NoStop}%
\bibitem [{\citenamefont {Scheel}\ and\ \citenamefont
  {Buhmann}(2009)}]{Scheel2009Feb}%
  \BibitemOpen
  \bibfield  {author} {\bibinfo {author} {\bibfnamefont {S.}~\bibnamefont
  {Scheel}}\ and\ \bibinfo {author} {\bibfnamefont {S.~Y.}\ \bibnamefont
  {Buhmann}},\ }\bibfield  {title} {\bibinfo {title} {{Macroscopic QED -
  concepts and applications}},\ }\bibfield  {journal} {\bibinfo  {journal}
  {arXiv}\ }\href {https://doi.org/10.48550/arXiv.0902.3586}
  {10.48550/arXiv.0902.3586} (\bibinfo {year} {2009}),\ \Eprint
  {https://arxiv.org/abs/0902.3586} {0902.3586} \BibitemShut {NoStop}%
\bibitem [{\citenamefont {Franke}\ \emph
  {et~al.}(2020{\natexlab{b}})\citenamefont {Franke}, \citenamefont {Ren},
  \citenamefont {Hughes},\ and\ \citenamefont
  {Richter}}]{franke2020fluctuation}%
  \BibitemOpen
  \bibfield  {author} {\bibinfo {author} {\bibfnamefont {S.}~\bibnamefont
  {Franke}}, \bibinfo {author} {\bibfnamefont {J.}~\bibnamefont {Ren}},
  \bibinfo {author} {\bibfnamefont {S.}~\bibnamefont {Hughes}},\ and\ \bibinfo
  {author} {\bibfnamefont {M.}~\bibnamefont {Richter}},\ }\bibfield  {title}
  {\bibinfo {title} {Fluctuation-dissipation theorem and fundamental photon
  commutation relations in lossy nanostructures using quasinormal modes},\
  }\href {https://doi.org/10.1103/PhysRevResearch.2.033332} {\bibfield
  {journal} {\bibinfo  {journal} {Phys. Rev. Research}\ }\textbf {\bibinfo
  {volume} {2}},\ \bibinfo {pages} {033332} (\bibinfo {year}
  {2020}{\natexlab{b}})}\BibitemShut {NoStop}%
\bibitem [{\citenamefont {Ren}\ \emph {et~al.}(2020)\citenamefont {Ren},
  \citenamefont {Franke}, \citenamefont {Knorr}, \citenamefont {Richter},\ and\
  \citenamefont {Hughes}}]{ren_near-field_2020}%
  \BibitemOpen
  \bibfield  {author} {\bibinfo {author} {\bibfnamefont {J.}~\bibnamefont
  {Ren}}, \bibinfo {author} {\bibfnamefont {S.}~\bibnamefont {Franke}},
  \bibinfo {author} {\bibfnamefont {A.}~\bibnamefont {Knorr}}, \bibinfo
  {author} {\bibfnamefont {M.}~\bibnamefont {Richter}},\ and\ \bibinfo {author}
  {\bibfnamefont {S.}~\bibnamefont {Hughes}},\ }\bibfield  {title} {\bibinfo
  {title} {Near-field to far-field transformations of optical quasinormal modes
  and efficient calculation of quantized quasinormal modes for open cavities
  and plasmonic resonators},\ }\href
  {https://doi.org/10.1103/PhysRevB.101.205402} {\bibfield  {journal} {\bibinfo
   {journal} {Physical Review B}\ }\textbf {\bibinfo {volume} {101}},\ \bibinfo
  {pages} {205402} (\bibinfo {year} {2020})}\BibitemShut {NoStop}%
\bibitem [{\citenamefont {Ge}\ \emph {et~al.}(2014)\citenamefont {Ge},
  \citenamefont {Kristensen}, \citenamefont {Young},\ and\ \citenamefont
  {Hughes}}]{ge_quasinormal_2014}%
  \BibitemOpen
  \bibfield  {author} {\bibinfo {author} {\bibfnamefont {R.-C.}\ \bibnamefont
  {Ge}}, \bibinfo {author} {\bibfnamefont {P.~T.}\ \bibnamefont {Kristensen}},
  \bibinfo {author} {\bibfnamefont {J.~F.}\ \bibnamefont {Young}},\ and\
  \bibinfo {author} {\bibfnamefont {S.}~\bibnamefont {Hughes}},\ }\bibfield
  {title} {\bibinfo {title} {Quasinormal mode approach to modelling
  light-emission and propagation in nanoplasmonics},\ }\href
  {https://doi.org/10.1088/1367-2630/16/11/113048} {\bibfield  {journal}
  {\bibinfo  {journal} {New Journal of Physics}\ }\textbf {\bibinfo {volume}
  {16}},\ \bibinfo {pages} {113048} (\bibinfo {year} {2014})}\BibitemShut
  {NoStop}%
\bibitem [{\citenamefont {Ge}\ \emph {et~al.}(2015)\citenamefont {Ge},
  \citenamefont {Young},\ and\ \citenamefont {Hughes}}]{Ge2015Mar}%
  \BibitemOpen
  \bibfield  {author} {\bibinfo {author} {\bibfnamefont {R.-C.}\ \bibnamefont
  {Ge}}, \bibinfo {author} {\bibfnamefont {J.~F.}\ \bibnamefont {Young}},\ and\
  \bibinfo {author} {\bibfnamefont {S.}~\bibnamefont {Hughes}},\ }\bibfield
  {title} {\bibinfo {title} {{Quasi-normal mode approach to the local-field
  problem in quantum optics}},\ }\href
  {https://doi.org/10.1364/OPTICA.2.000246} {\bibfield  {journal} {\bibinfo
  {journal} {Optica}\ }\textbf {\bibinfo {volume} {2}},\ \bibinfo {pages} {246}
  (\bibinfo {year} {2015})}\BibitemShut {NoStop}%
\bibitem [{\citenamefont {Glauber}\ and\ \citenamefont
  {Lewenstein}(1991)}]{Glauber1991Jan}%
  \BibitemOpen
  \bibfield  {author} {\bibinfo {author} {\bibfnamefont {R.~J.}\ \bibnamefont
  {Glauber}}\ and\ \bibinfo {author} {\bibfnamefont {M.}~\bibnamefont
  {Lewenstein}},\ }\bibfield  {title} {\bibinfo {title} {{Quantum optics of
  dielectric media}},\ }\href {https://doi.org/10.1103/PhysRevA.43.467}
  {\bibfield  {journal} {\bibinfo  {journal} {Phys. Rev. A}\ }\textbf {\bibinfo
  {volume} {43}},\ \bibinfo {pages} {467} (\bibinfo {year} {1991})}\BibitemShut
  {NoStop}%
\bibitem [{\citenamefont {de~Lasson}\ \emph {et~al.}(2015)\citenamefont
  {de~Lasson}, \citenamefont {Kristensen}, \citenamefont {M{\o}rk},\ and\
  \citenamefont {Gregersen}}]{deLasson}%
  \BibitemOpen
  \bibfield  {author} {\bibinfo {author} {\bibfnamefont {J.~R.}\ \bibnamefont
  {de~Lasson}}, \bibinfo {author} {\bibfnamefont {P.~T.}\ \bibnamefont
  {Kristensen}}, \bibinfo {author} {\bibfnamefont {J.}~\bibnamefont
  {M{\o}rk}},\ and\ \bibinfo {author} {\bibfnamefont {N.}~\bibnamefont
  {Gregersen}},\ }\bibfield  {title} {\bibinfo {title} {Semianalytical
  quasi-normal mode theory for the local density of states in coupled photonic
  crystal cavity-waveguide structures},\ }\href
  {https://doi.org/10.1364/OL.40.005790} {\bibfield  {journal} {\bibinfo
  {journal} {Opt. Lett.}\ }\textbf {\bibinfo {volume} {40}},\ \bibinfo {pages}
  {5790} (\bibinfo {year} {2015})}\BibitemShut {NoStop}%
\bibitem [{\citenamefont {Ren}\ \emph {et~al.}(2022)\citenamefont {Ren},
  \citenamefont {Franke},\ and\ \citenamefont {Hughes}}]{2108.10194}%
  \BibitemOpen
  \bibfield  {author} {\bibinfo {author} {\bibfnamefont {J.}~\bibnamefont
  {Ren}}, \bibinfo {author} {\bibfnamefont {S.}~\bibnamefont {Franke}},\ and\
  \bibinfo {author} {\bibfnamefont {S.}~\bibnamefont {Hughes}},\ }\bibfield
  {title} {\bibinfo {title} {Connecting classical and quantum mode theories for
  coupled lossy cavity resonators using quasinormal modes},\ }\href
  {https://doi.org/10.1021/acsphotonics.1c01274} {\bibfield  {journal}
  {\bibinfo  {journal} {{ACS} Photonics}\ }\textbf {\bibinfo {volume} {9}},\
  \bibinfo {pages} {138} (\bibinfo {year} {2022})}\BibitemShut {NoStop}%
\end{thebibliography}%
\end{document}